# Stability analysis of semiconductor manufacturing process with EWMA run-to-run controllers

Bing Ai[a], David Shan-Hill Wong[b], Shi-Shang Jang[b]

[a]*Department of Computer Science, University of Texas at Austin, Texas, USA*

[b]*Department of Chemical Engineering, National Tsing-Hua University, Hsin-Chu, Taiwan*

**ABSTRACT:**

In the semiconductor manufacturing batch processes, each step is a complicated physiochemical batch process; generally it is difficult to perform measurements on-line or carry out the measurement for each run, and hence there will be delays in the feedback of the system. The effect of the delay on the stability of the system is an important issue which needs to be understood. Based on the exponentially weighted moving average (EWMA) algorithm, we propose two kinds of controllers, EWMA-I and II controllers for single product process and mixed product process in semiconductor manufacturing in this paper. For the single product process, the stabilities of systems with both controllers which undergo different kinds of metrology delays are investigated. Necessary and sufficient conditions for the stochastic stability are established. Routh-Hurwitz criterion and Lyapunov's direct method are used to obtain the stability regions for the system with fixed metrology delay. By using Lyapunov's direct method, the stability region is established for the system with fixed sampling metrology and with stochastic metrology delay. We also extended the theorems of single product process to mixed product process. Based on the proposed theorems, some numerical examples are provided to illustrate the stability of the delay system.

## 1. Introduction

In the field of semiconductor manufacturing industry, run-to-run control is now widely accepted as a means of production fabrication facilities to improve the efficiency and wastage in production. It's a form of discrete process and machine control in which the product recipe with respect to a particular machine process is modified ex situ, i.e., between machine "runs", so as to minimize process drift, shift and variability [1]. Two of the most basic run-to-run control algorithms used today in the semiconductor manufacturing industry are exponentially weighted moving average (EWMA) algorithm and predictor-corrector controller (or double EWMA) algorithm. Both of these algorithms need the instantaneous information of the process output; however, in semiconductor manufacturing, each step is a complicated physiochemical batch process, and measurements are almost exclusively performed off-line, often slow, inconsistent, or skipped by operators. This caused delayed, inconsistent, and infrequent measurement of the process output. Therefore, one problem in the application of run-to-run control is how metrology delay would affect the stability of the process.

Time delay system has been extensively studied in the last few years. For the system with stochastic time delay, as pointed out in [2], many manufacturing process can be modeled by Markovian jump linear systems. And the results of optimal control, robust control and stability for such kind of system can be widely seen in the recent literatures, such as [3]-[7]. However, the optimization problem, the robust control as well as the stability problem for Markovian jump linear systems usually change into the problem of solving a set of linear matrix inequalities (LMIs), and detailed algorithms to solve the LMIs can be found in [8]-[11].

The pioneer work on the stability of EWMA run-to-run controller without delay was carried out by Ingolfsson and Sachs [12]. They found that the predicted process gain "must not be too small" related to the true process gain to grantee the stability of the process. R. Good *et al.* examined stability bounds for the discount factors of both single input single output (SISO) and multiple inputs multiple outputs (MIMO) double EWMA run-to-run controllers when there is plant-model mismatch and delay between product manufacturing and product metrology [13]. Their work showed that for the SISO system, the size of the stability region decreases as metrology delay and plant-model mismatch increase and as the discount factors decrease. They also pointed out that when metrology delay exists in the system, intentionally overestimating process gain could lead to instabilities, unlike the system without delay. On the other hand, for the MIMO system, increasing the discount factors does not necessarily increase stability when a large plant-model mismatch exists. A few years later, after their first work on stability analysis of double EWMA run-to-run controllers, Good *et al.* analyzed the stability of MIMO EWMA run-to-run controller with metrology delay by using the generalized Routh-Hurwitz stability criterion [15]. They derived the necessary and sufficient conditions for stability with metrology delays up to two runs, and developed a sufficient condition for the stability of the MIMO system with metrology delay longer than two runs. The sufficient condition they got is that if all of the eigenvalues of a model-mismatch matrix fall inside a circle with unit radius and centered at {1, 0} on the complex plane, then the closed-loop system is stable. M.F. Wu *et al.* analyzed the influences of metrology delay on both the transient and asymptotic properties of the product quality for the case when a linear system with an initial bias and a stochastic autoregressive moving average disturbance is under EWMA run-to-run control. They pointed out that metrology delay is only important for processes that experience nonstationary stochastic disturbance. Based on the study of numerical optimization results of the analytical closed-loop output response they developed guidelines to tuning the discount factor. All the aforementioned works on EWMA or double EWMA run-to-run control system are based on the assumption that the metrology delay is fixed. However, the semiconductor manufacturing industry characterized by physical and chemical environments makes measurement in many of these environments difficult or time-consuming, and this combined with the fact that many process tools are not designed for the addition of in situ sensor, resulted in measurement taken less frequently than every run, or at stochastic runs. To the best of authors' knowledge, up till now, in the field of EWMA run-to-run control, there is only one work available which discusses the issue of the system with stochastic metrology delay [16]. The pioneer work by B. Ai *et al.* in [16] assumes that the system subject to the stochastic metrology delay, one kind of EWMA controller, which we call EWMA-I controller in this paper, is proposed to reject the system disturbances. However, sometimes EWMA-I controller should not be updated if there is no new information available. So in this paper, we should modify EWMA-I controller. Also, in [16], the transition probability matrix is given, but in the real manufacturing process, the transition probability matrix cannot be known directly and should be calculated from the probability distribution of the data we received, hence we will discuss how to calculate the transition probability matrix from the specific probability distribution.

What we have discussed is based on the assumption that there is only one kind of product in the manufacturing line which is far from realistic. Because of the high capital costs associated with the process equipments, it is a common practice in today's semiconductor manufacturing to have many different products and processes run on each processing tool, i.e., high-mix manufacturing. The first work of the stability of the high-mix manufacturing is done by Y. Zheng *et al.* in [20] where they studied a model with two products manufactured on the same tool, and proposed two kinds of control method: "tool-based" and "product-based" approaches; and, they found that the "tool-based" approach is unstable when the plant is non-stationary and the plant-model mismatch parameters are different for different products, while the "product-based" approach is stable. However they made the mistake in deriving the output of the system for "product-based" control, see Lemma 2, equations (22) and (23) in [20] for details. Also the assumption that only two products are manufactured on the production line is far from realistic, therefore B. Ai *et al.* corrected and extended Y. Zheng's work [21]-[27] to a more complicated case, where they assumed that a number of different kinds of products are manufactured on the same tool with variable manufacturing cycles, and the campaign length and break length of each cycle are also variable. They found that for mixed product drifted process, if the break length of a product is large, then at the beginning runs of each cycle, the process output will far deviate from the target value. They proposed cycle resetting algorithm for discount factor (CR-EWMA) algorithm, and cycle forecasting EWMA (CF-EWMA) to reduce the large deviations as well as to achieve the minimum asymptotic variance control; they also proposed a discount factor resetting fault tolerant (RFT) approach and fault tolerant cycle forecasting EWMA (FTCF-EWMA) algorithm to handle the step fault. Although they had extended their models to more sophisticated conditions, the models are still not well matching the real manufacturing situations. Again to the best of authors' knowledge, up till now, in the field of semiconductor manufacturing, there is no work available which discusses the stability of mixed product process subject to stochastic metrology delay. So in this paper, we shall try to establish the model for mixed product process with stochastic metrology delay and extend the theorems obtained for single product process to mixed product process. The stability of mixed product process will also be analyzed.

For better of presentation, the remainder of the paper is organized as follows: From Section 2 to Section 4, we will focus on single product process, i.e., in Section 2, two kinds of EWMA controllers, i.e., EWMA-I controller and EWMA-II controller, are proposed for single product process. In Section 3, we will discuss how to obtain the transition probability matrix from a specific probability distribution. Numerical examples are provided in Section 4 to obtain the stability regions for the systems subject to different metrology delays. In Section 5, we will discuss mixed product process, and the theorems obtained for single product process are extended to the mixed product process. The conclusion remarks are presented in Section 6.

## 2. EWMA Run-To-Run Controllers in Single Product Process

In semiconductor manufacturing, the same products are usually manufactured on the same tool, i.e., single product process. The most widely used algorithm in this process is EWMA algorithm which needs the information of the output of the system for feedback. However, outputs sometimes are not available timely because the measurement is usually time-consuming, so there will be delays in the system. In this Section, we will first discuss EWMA controller without metrology delay, and then we will propose two kinds of EWMA controllers for the system with metrology delay. The stability of the controllers will be examined in the last part of this Section.

### 2.1. EWMA Run-to-Run Controller without Metrology Delay

A typical EWMA run-to-run observer assumes a static linear model between control variable $Y_t$, and manipulated variable $u_t$, i.e.,

$$Y_t = \beta u_t + a_t \tag{1}$$

where $\beta$ is the process gain between the process input and output, and $a_t$ is the instantaneous disturbance at run $t$.

Given the predicted model of the process

$$\hat{Y}_t = bu_t + \hat{a}_t \tag{2}$$

where $b$ is the model gain, and $\hat{a}_t$ is the estimated offset at run $t$ for the process.

When information of the current run is available without delay, an EWMA update of the offset is given by:

$$\hat{a}_{t+1} = \omega(Y_t - bu_t) + (1-\omega)\hat{a}_t \tag{3}$$

where $\omega$ is a discount factor between zero and one.

A control law is used to determine the control recipe for the next run, i.e.,

$$u_t = \frac{T - \hat{a}_t}{b} \tag{4}$$

where $T$ is the desired target. Without loss of generality, in this paper, we assume $T = 0$. **Fig. 1** shows the structural diagram of the above algorithm.

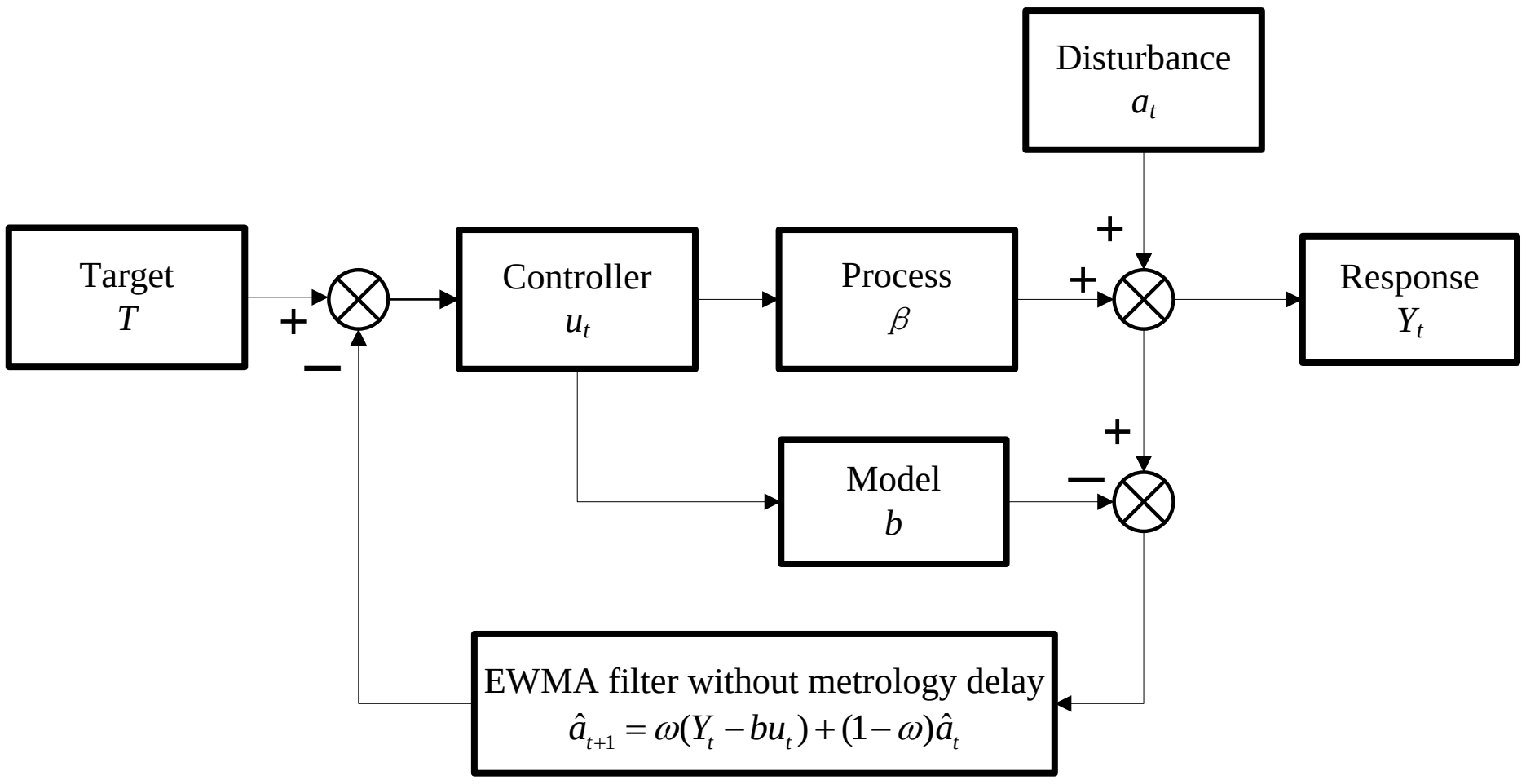

**Fig. 1**. Structural diagram of EWMA run-to-run controller without metrology delay

## 2.2. EWMA Run-to-Run Controllers with Metrology Delay

As mentioned previously, in an actual manufacturing plant, measurement delay is a common phenomenon which constitutes a stochastic process. When the measurement delay happens, in time output of the system is no longer available, so the original EWMA filter described in (3) does not hold. Other EWMA filter should be proposed to reject the disturbance of the system. **Fig. 2** is the general structural diagram for EWMA run-to-run control with metrology delay.

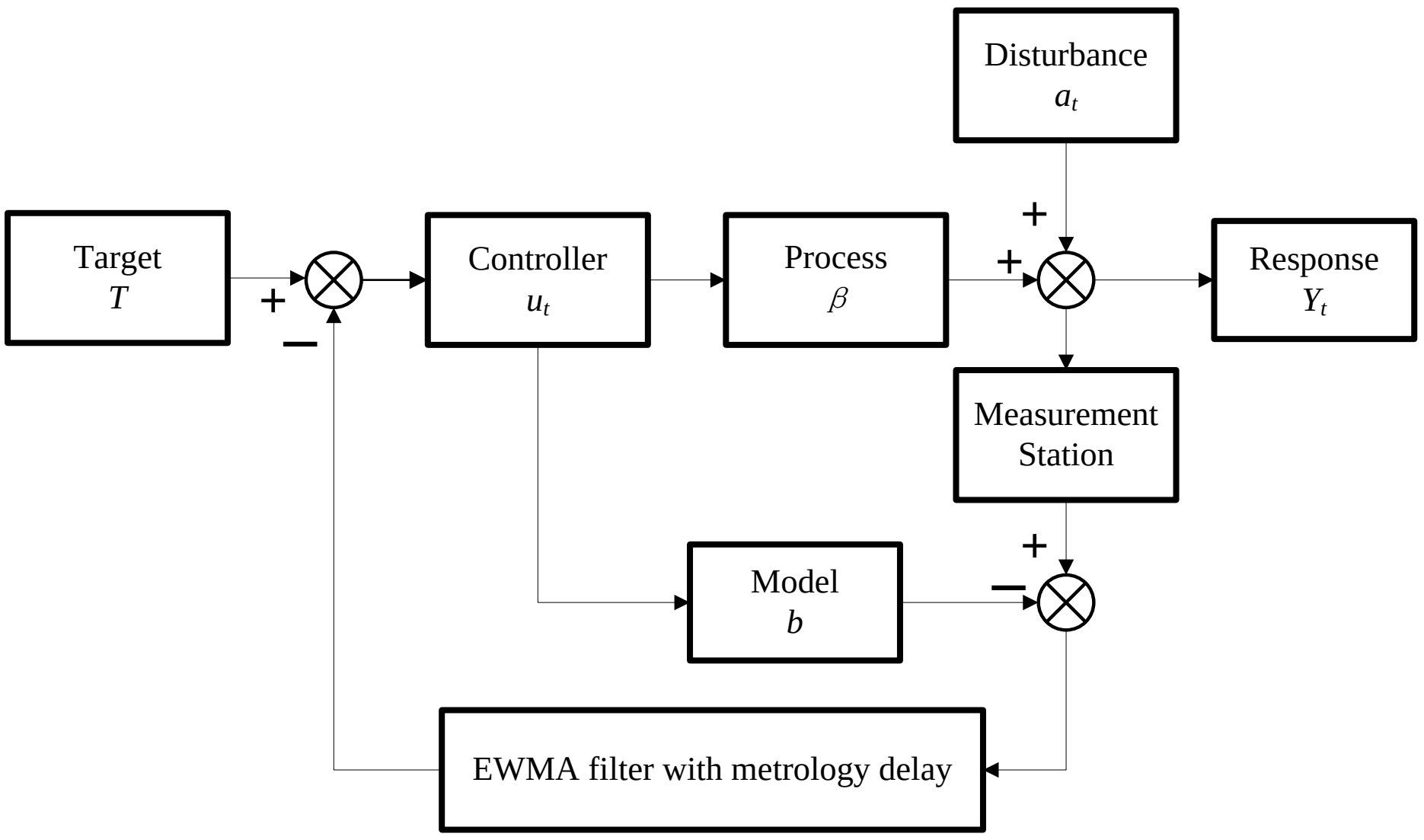

**Fig. 2**. General structural diagram for EWMA run-to-run control with metrology delay

Suppose that there is a stochastic metrology delay $\tau_t$ at run $t$, then at this time, two kinds of EWMA run-to-run controllers, EWMA-I and II controllers, can be used to estimate the disturbance:

For EWMA-I controller, which was first proposed by B. Ai *et al.* in [16], the control action is also chosen as (4), and the EWMA filter is

$$\hat{a}_{t+1} = \omega(Y_{t-\tau_t} - bu_{t-\tau_t}) + (1-\omega)\hat{a}_t \tag{5}$$

Combining (1), (2), (4) and (5), we rewrite EWMA-I controller in the form of (6),

$$\hat{a}_{t+1} = (1-\omega)\hat{a}_t + \omega(1-\xi)\hat{a}_{t-\tau_t} \tag{6}$$

where $\xi = \beta / b$ represents the plant-model mismatch of the process gain. From Equations (5) and (6) we know that the EWMA-I controller is always updating its state $\hat{a}_t$ regardless of the availability of the output $Y_t$, i.e., in each recursion, $\hat{a}_t$ is updated by the available output which may not be the latest output of the system. However, in some processes, if the newest output is not available, the controller will not update its state, and we name this controller as EWMA-II controller, i.e.,

$$\hat{a}_{t+1} = \begin{cases} (1-\omega)\hat{a}_t + \omega(1-\xi)\hat{a}_{t-\tau_t}, & if\ \tau_{t-1} \geq \tau_t; \\ \hat{a}_t, & otherwise. \end{cases} \tag{7}$$

For example, at the $t^{th}$ production run, we have output of $Y_{t-3}$ for feedback, i.e., $\tau_t = 3$: if $\tau_{t+1} > 3$, i.e., no new output is available at the $(t+1)^{th}$ production run, then the state of the system should not be updated; however, if $\tau_{t+1} = 2$, then at the $(t+1)^{th}$ production run, the latest output of the system is $Y_{t-1}$, and at this time the state of the system should be updated by using $Y_{t-1}$.

***Remark 1:*** For the fixed metrology delay, EWMA-I and II controllers are equivalent because both of the controllers are in the form of

$$\hat{a}_{t+1} = \omega(Y_{t-f} - bu_{t-f}) + (1-\omega)\hat{a}_t \tag{8}$$

for fixed $f$ runs metrology delay.

## 2.3. Stability of EWMA Run-to-Run Controllers

Stability is a fundamental requirement for any system. An unstable control scheme should not be implemented. In other words, the control scheme should lead the process under control to a stable state. In this subsection, we shall examine the stability of EWMA run-to-run controllers subject to fixed and stochastic metrology delay.

### 2.3.1. Fixed Metrology Delay

For the system with fixed $f$ runs metrology lags, combining (1), (2), (4) and (8), we have the output of closed-loop system,

$$Y_t = \frac{1-(1-\omega)B-\omega B^{f+1}}{1-(1-\omega)B-\omega(1-\xi)B^{f+1}} a_t \tag{9}$$

where $B$ is the backshift operator. Equation (9) yields the characteristic equation of the closed-loop system $G(B) = 1-(1-\omega)B-\omega(1-\xi)B^{f+1} = 0$; then the system described by (1), (2), (4) and (8) is stable if and only if the roots of $G(B)$ lie outside the unit cycle. With the well-known bilinear transformation

$$B = \frac{1-W}{1+W} \tag{10}$$

the outside of the unit cycle is mapped to the open left half-plane, and then the Routh-Hurwitz criterion will be used on (11) to find the stability boundaries of systems with metrology delays of $f$ runs.

$$G(W) = 1-(1-\omega)\frac{1-W}{1+W}-\omega(1-\xi)\left(\frac{1-W}{1+W}\right)^{f+1} = 0 \tag{11}$$

***Remark 2:*** The Jury's test can also be implemented directly on $G(B)$ to check whether or not its roots lie outside the unit cycle.

### 2.3.2. Stochastic Metrology Delay

For the system with stochastic metrology delay, neither Routh-Hurwitz criterion nor Jury's test is valid in obtaining stability region. In the following, we try to find other ways to get the stability region for such kind of system.

Augment the state variable $\hat{a}_t$ at run $t$ into a vector as

$$\boldsymbol{X}_t = \left(\hat{a}_t\ \hat{a}_{t-1}\ \cdots \hat{a}_{t-\tau_t} \cdots\ \hat{a}_{t-\tau_{\max}}\right)^T \tag{12}$$

where $\tau_{\max}$ is the maximum delay possible, and the update equation in (6) or (7) can be rewritten as

$$\boldsymbol{X}_{t+1} = \Xi(\tau_{t-1}, \tau_t)\boldsymbol{X}_t \tag{13}$$

where $\Xi(\tau_{t-1}, \tau_t)$ is determined by the actual type of controller and the maximum delay of the system.

Combining (6) or (7) with (12), we can write $\Xi(\tau_{t-1}, \tau_t)$ for systems of any metrology delay with either EWMA-I or II controller.

For example, if the controller of the system is EWMA-I controller, then the system matrix $\Xi(\tau_{t-1}, \tau_t)$ is relevant to $\tau_t$, but irrelevant to $\tau_{t-1}$:if the system is delay free, i.e., $\tau_t=0$, and $\Xi(\tau_{t-1}, \tau_t) = \Xi(0,0) = 1-\xi\omega$; if the maximum delay of the system is 1, if $\tau_t=0$, then $\Xi(\tau_{t-1}, \tau_t) = \Xi(0,0) = \Xi(1,0) = \begin{pmatrix} 1-\xi\omega & 0 \\ 1 & 0 \end{pmatrix}$, and if $\tau_t=1$, then $\Xi(\tau_{t-1}, \tau_t) = \Xi(0,1) = \Xi(1,1) = \begin{pmatrix} 1-\omega & \omega(1-\xi) \\ 1 & 0 \end{pmatrix}$.

However, for the system with EWMA-II controller, the system matrix of the system is relevant to both $\tau_{t-1}$ and $\tau_t$: if the system is delay free, then $\Xi(0,0) = 1-\xi\omega$; if the maximum delay of the system is 1, then $\Xi(0,0) = \begin{pmatrix} 1-\xi\omega & 0 \\ 1 & 0 \end{pmatrix}$, $\Xi(0,1) = \begin{pmatrix} 1 & 0 \\ 1 & 0 \end{pmatrix}$ $\Xi(1,0) = \begin{pmatrix} 1-\xi\omega & 0 \\ 1 & 0 \end{pmatrix}$ and $\Xi(1,1) = \begin{pmatrix} 1-\omega & \omega(1-\xi) \\ 1 & 0 \end{pmatrix}$; if the maximum delay of the system is 2, then $\Xi(0,0) = \begin{pmatrix} 1-\xi\omega & 0 & 0 \\ 1 & 0 & 0 \\ 0 & 1 & 0 \end{pmatrix}$,

$$\Xi(0,1)=\begin{pmatrix}1&0&0\\1&0&0\\0&1&0\end{pmatrix},\quad \Xi(1,0)=\begin{pmatrix}1-\xi\omega&0&0\\1&0&0\\0&1&0\end{pmatrix},\quad \Xi(1,1)=\begin{pmatrix}1-\omega&\omega(1-\xi)&0\\1&0&0\\0&1&0\end{pmatrix},$$

$$\Xi(1,2)=\begin{pmatrix}1&0&0\\1&0&0\\0&1&0\end{pmatrix},\quad \Xi(2,0)=\begin{pmatrix}1-\xi\omega&0&0\\1&0&0\\0&1&0\end{pmatrix},\quad \Xi(2,1)=\begin{pmatrix}1-\omega&\omega(1-\xi)&0\\1&0&0\\0&1&0\end{pmatrix},$$

$$\Xi(2,2)=\begin{pmatrix}1-\omega&0&\omega(1-\xi)\\1&0&0\\0&1&0\end{pmatrix}.$$

***Remark 3:*** Through the augmentation technique, it is clear that the stochastic metrology delay system described by (1), (2), (4) and (6) or described by (1), (2), (4) and (7) is transformed into a delay-free discrete-time jump linear system modeled by a homogeneous Markov chains as is expressed by (13).

### 2.3.3. Stability Criterion of the System with Variable Delay

In this paper, we will use the following definition for stochastically stable.

***Definition 1:*** The system in (13) is stochastically stable, if for every finite $X_0$ and initial mode $\tau_0 \in S$, and $S=\{0,1,\cdots,\tau_{\max}\}$ being the set of all possible delays, the following condition

$$E\left(\sum_{t=0}^{\infty}\|\boldsymbol{X}_t\|^2 \,|\, \boldsymbol{X}_0,\tau_0\right)<\infty \tag{14}$$

is satisfied.

Let transition probabilities matrix between metrology delays be $P=[p_{ij}]$, and $p_{ij}$ is defined by

$$p_{ij}=\text{Prob}\left(\tau_{t+1}=j \,|\, \tau_t=i\right) \tag{15}$$

where $i,j\in\{0,1,\ldots,\infty\}$, then, $0\le p_{ij}\le 1$ and $\sum_{j=0}^{\infty}p_{ij}=1$. The following theorem gives sufficient and necessary condition to guarantee the stochastic stability of system (13).

***Theorem 1:*** System (13) is stochastically stable if and only if there exists a positive-definite matrix

$\boldsymbol{Q}(\tau_{t-1},\tau_t) > 0$ for $\forall \tau_{t-1},\tau_t \in \{0,1,\cdots,\tau_{\max}\}$, satisfying the following matrix inequalities:

$$\boldsymbol{L}(\tau_{t-2},\tau_{t-1}) = \sum_{\tau_t=0}^{\tau_{\max}} p_{\tau_{t-1},\tau_t} \Xi(\tau_{t-1},\tau_t)^T \boldsymbol{Q}(\tau_{t-1},\tau_t) \Xi(\tau_{t-1},\tau_t) - \boldsymbol{Q}(\tau_{t-2},\tau_{t-1}) < 0 \qquad (16)$$

***Proof:***

*Sufficiency:*

Construct the stochastic Lyapunov function $\boldsymbol{V}\left(\boldsymbol{X}_t,\tau_{t-2},\tau_{t-1}\right)$ as follows:

$$\boldsymbol{V}(\boldsymbol{X}_t,\tau_{t-2},\tau_{t-1}) = \boldsymbol{X}_t^T \boldsymbol{Q}(\tau_{t-2},\tau_{t-1}) \boldsymbol{X}_t \qquad (17)$$

then

$$\begin{aligned}
& E\left[\Delta \boldsymbol{V}\left(\boldsymbol{X}_t,\tau_{t-2},\tau_{t-1}\right)\right] \\
&= E\left[\boldsymbol{V}\left(\boldsymbol{X}_{t+1},\tau_{t-1},\tau_t\right) - \boldsymbol{V}\left(\boldsymbol{X}_t,\tau_{t-2},\tau_{t-1}\right)\right] \\
&= E\left[\boldsymbol{V}\left(\boldsymbol{X}_{t+1},\tau_{t-1},\tau_t\right)\right] - E\left[\boldsymbol{V}\left(\boldsymbol{X}_t,\tau_{t-2},\tau_{t-1}\right)\right] \\
&= E\left[\boldsymbol{V}\left(\boldsymbol{X}_{t+1},\tau_{t-1},\tau_t \middle| \boldsymbol{X}_t,\tau_{t-2},\tau_{t-1}\right)\right] - \boldsymbol{V}\left(\boldsymbol{X}_t,\tau_{t-2},\tau_{t-1}\right) \\
&= \sum_{\tau_t=0}^{\tau_{\max}} p_{\tau_{t-1},\tau_t} \boldsymbol{X}_{t+1}^T \boldsymbol{Q}\left(\tau_{t-1},\tau_t\right) \boldsymbol{X}_{t+1} - \boldsymbol{X}_t^T \boldsymbol{Q}\left(\tau_{t-2},\tau_{t-1}\right) \boldsymbol{X}_t \\
&= \sum_{\tau_t=0}^{\tau_{\max}} p_{\tau_{t-1},\tau_t} \boldsymbol{X}_t^T \Xi\left(\tau_{t-1},\tau_t\right)^T \boldsymbol{Q}\left(\tau_{t-1},\tau_t\right) \Xi\left(\tau_{t-1},\tau_t\right) \boldsymbol{X}_t - \boldsymbol{X}_t^T \boldsymbol{Q}\left(\tau_{t-2},\tau_{t-1}\right) \boldsymbol{X}_t \\
&= \boldsymbol{X}_t^T \left[\sum_{\tau_t=0}^{\tau_{\max}} p_{\tau_{t-1},\tau_t} \Xi\left(\tau_{t-1},\tau_t\right)^T \boldsymbol{Q}\left(\tau_{t-1},\tau_t\right) \Xi\left(\tau_{t-1},\tau_t\right) \boldsymbol{X}_t - \boldsymbol{Q}\left(\tau_{t-2},\tau_{t-1}\right) \boldsymbol{X}_t\right] \\
&= \boldsymbol{X}_t^T \boldsymbol{L}\left(\tau_{t-2},\tau_{t-1}\right) \boldsymbol{X}_t = -\boldsymbol{X}_t^T \left(-\boldsymbol{L}\left(\tau_{t-2},\tau_{t-1}\right)\right) \boldsymbol{X}_t \\
&\le -\lambda_{\min}\left(\tau_{t-2},\tau_{t-1}\right) \boldsymbol{X}_t^T \boldsymbol{X}_t \le -\Gamma_{\min} \left\|\boldsymbol{X}_t\right\|^2
\end{aligned} \qquad (18)$$

where $\lambda_{\min}\left(\tau_{t-2},\tau_{t-1}\right)$ is the minimum eigenvalue of $-\boldsymbol{L}\left(\tau_{t-2},\tau_{t-1}\right)$ and $\Gamma_{\min} = \inf_{\tau_{t-2},\tau_{t-1}\in S} \left[\lambda_{\min}\left(\tau_{t-2},\tau_{t-1}\right)\right]$ is the minimum value of $\lambda_{\min}\left(\tau_{t-2},\tau_{t-1}\right)$ for all possible combinations of $\tau_{t-2}$ and $\tau_{t-1}$.

Given the inequation (18), we can derive the following recursive relation

$$\begin{aligned}
& E\left[\boldsymbol{V}\left(\boldsymbol{X}_2,\tau_0,\tau_1\right)\right] - E\left[\boldsymbol{V}\left(\boldsymbol{X}_1,\tau_0\right)\right] \le -\beta \left\|\boldsymbol{X}_1\right\|^2 \\
& E\left[\boldsymbol{V}\left(\boldsymbol{X}_3,\tau_1,\tau_2\right)\right] - E\left[\boldsymbol{V}\left(\boldsymbol{X}_2,\tau_0,\tau_1\right)\right] \le -\beta \left\|\boldsymbol{X}_2\right\|^2 \\
& \vdots \\
& E\left[\boldsymbol{V}\left(\boldsymbol{X}_{T+1},\tau_{T-1},\tau_T\right)\right] - E\left[\boldsymbol{V}\left(\boldsymbol{X}_T,\tau_{T-2},\tau_{T-1}\right)\right] \le -\beta \left\|\boldsymbol{X}_T\right\|^2
\end{aligned} \qquad (19)$$

Hence

$$E\left[\boldsymbol{V}\left(\boldsymbol{X}_{T+1},\tau_{T-1},\tau_T\right)\right]-\boldsymbol{V}\left(\boldsymbol{X}_0,\tau_0\right)\le-\beta\sum_{t=1}^{T}\left\|\boldsymbol{X}_t\right\|^2 \tag{20}$$

i.e.,

$$\begin{aligned}\sum_{t=1}^{T}\left\|\boldsymbol{X}_t\right\|^2 &\le\frac{1}{\beta}E\left[\boldsymbol{V}(\boldsymbol{X}_0,\tau_0)-\boldsymbol{V}(\boldsymbol{X}_{T+1},\tau_{T-1},\tau_T)\right]\\ &\le\frac{1}{\beta}\boldsymbol{V}(\boldsymbol{X}_0,\tau_0)\end{aligned}$$

which implies

$$\begin{aligned}\lim_{T\to\infty}E\left[\sum_{t=1}^{T}\left\|\boldsymbol{X}_t\right\|^2|\boldsymbol{X}_0,\tau_0\right]&\le\frac{1}{\beta}E\left[\boldsymbol{V}(\boldsymbol{X}_0,\tau_0)\right]\\ &=\frac{1}{\beta}\boldsymbol{X}_0^TQ(\tau_0)\boldsymbol{X}_0\\ &<\infty\end{aligned}$$

i.e., system (13) is stochastically stable.

*Necessity*:

Define

$$E\left[\boldsymbol{X}_t^T\tilde{\boldsymbol{Q}}\left(T\text{-}t,\tau_{t-2},\tau_{t-1}\right)\boldsymbol{X}_t\right]\equiv E\left[\sum_{k=t}^{T}\boldsymbol{X}_k^T\boldsymbol{R}\left(\tau_{k-2},\tau_{k-1}\right)\boldsymbol{X}_k\left|\boldsymbol{X}_t,\tau_{t-2},\tau_{t-1}\right.\right] \tag{21}$$

with $\boldsymbol{R}\left(\tau_{k-2},\tau_{k-1}\right)$ is a set of positive definite matrix. Let $r_{ij}=\boldsymbol{R}\left(\tau_{k-2}=i,\tau_{k-1}=j\right)$ and given that $\left(x_{k-i}-x_{k-j}\right)^2\ge0\Rightarrow x_{k-i}^2+x_{k-j}^2\ge2x_{k-i}x_{k-j}$, we have

$$\begin{aligned}\boldsymbol{X}_k^T\boldsymbol{R}\left(\tau_{k-2},\tau_{k-1}\right)\boldsymbol{X}_k&=\sum_{i=0}^{\tau_{\max}}\sum_{j=0}^{\tau_{\max}}x_{k-i}r_{ij}x_{k-j}=\sum_{j=0}^{\tau_{\max}}x_{k\text{-}j}^2r_{jj}+2\sum_{\substack{i=0\\ i\ne j}}^{\tau_{\max}}\sum_{j=0}^{\tau_{\max}}x_{k-i}r_{ij}x_{k-j}\\ &\le\sum_{j=0}^{\tau_{\max}}x_{k\text{-}j}^2r_{jj}+2\sum_{\substack{i=0\\ i\ne j}}^{\tau_{\max}}\sum_{j=0}^{\tau_{\max}}r_{ij}\left(x_{k-i}^2+x_{k-j}^2\right)=\sum_{j=0}^{\tau_{\max}}c_jx_{k-j}^2\le c_{\max}\left\|\boldsymbol{X}_k\right\|^2\end{aligned} \tag{22}$$

with $c_{\max}=\max\left(c_0,\cdots,c_{\tau_{\max}}\right)$. Hence

$$E\left[\boldsymbol{X}_t^{\mathrm{T}}\tilde{\boldsymbol{Q}}\left(T-t,\tau_{t-2},\tau_{t-1}\right)\boldsymbol{X}_t\right]\le E\left[\sum_{k=t}^{T}c_{\max}\left\|\boldsymbol{X}_k\right\|^2|\boldsymbol{X}_t,\tau_{t-2},\tau_{t-1}\right]$$

Since the system is stochastic stable, we have

$$\lim_{T\to\infty}E\left(\sum_{t=1}^{T}\left\|\boldsymbol{X}_t\right\|^2|\boldsymbol{X}_0,\tau_0\right)<\infty \tag{23}$$

$E\left[ X_t^T \tilde{Q}\left(T-t,\tau_{t-2},\tau_{t-1}\right) X_t \right]$ will be bounded and its asymptotic value is given by:

$$E\left[ X_t^T Q\left(\tau_{t-2},\tau_{t-1}\right) X_t \right] = \lim_{T\to\infty} E[\sum_{k=t}^{T} X_k^T R(\tau_{k-2},\tau_{k-1}) X_k \,|X_t,\tau_{t-2},\tau_{t-1}] \tag{24}$$

Similarly, we can write

$$\begin{aligned} E[X_{t+1}^T Q\left(\tau_{t-1},\tau_t\right) X_{t+1} |X_t,\tau_{t-2},\tau_{t-1}] &= \lim_{T\to\infty} E[X_{t+1}^T \tilde{Q}\left(T-t-1,\tau_{t-1},\tau_t\right) X_{t+1}] \\ &= \lim_{T\to\infty} E[\sum_{k=t+1}^{T} X_k^T R\left(\tau_{k-2},\tau_{k-1}\right) X_k \,|X_t,\tau_{t-2},\tau_{t-1}] \end{aligned} \tag{25}$$

Subtracting (24) from (25), we have

$$\begin{aligned} & E[X_{t+1}^T Q\left(\tau_{t-1},\tau_t\right) X_{t+1} |X_t,\tau_{t-2},\tau_{t-1}] - E[X_t^T Q\left(\tau_{t-2},\tau_{t-1}\right) X_t] \\ &= -X_t^T R(\tau_{t-2},\tau_{t-1}) X_t \\ &= \sum_{\tau_t=0}^{\tau_{\max}} p_{\tau_{t-1},\tau_t} X_t^T \Xi(\tau_{t-1},\tau_t)^T Q(\tau_{t-1},\tau_t) \Xi(\tau_{t-1},\tau_t) X_t - X_t^T Q(\tau_{t-2},\tau_{t-1}) X_t \end{aligned} \tag{26}$$

Hence we have

$$\begin{aligned} -R(\tau_{t-2},\tau_{t-1}) &= \sum_{\tau_t=0}^{\tau_{\max}} p_{\tau_{t-1},\tau_t} \Xi(\tau_{t-1},\tau_t)^T Q(\tau_{t-1},\tau_t) \Xi(\tau_{t-1},\tau_t) - Q(\tau_{t-2},\tau_{t-1}) \\ &= L(\tau_{t-2},\tau_{t-1}) < 0 \end{aligned} \tag{27}$$

To sum those up, ***Theorem 1*** holds.■

Since $L(\tau_{t-2},\tau_{t-1})$ are nonlinear in the system matrix, it is difficult to check whether (16) is feasible or not. To this end, we have the equivalent condition for (16), i.e.,

***Proposition 1:*** The matrix inequality $L(\tau_{t-2},\tau_{t-1}) < 0$ in (16) is equivalent to the following matrix inequality

$$\begin{pmatrix} -Q(\tau_{t-2},\tau_{t-1}) & \bar{\Xi}(\tau_{t-1})^T V(\tau_{t-1}) \hat{Q}(\tau_{t-1}) \\ \hat{Q}(\tau_{t-1})^T V(\tau_{t-1})^T \bar{\Xi}(\tau_{t-1}) & -\hat{Q}(\tau_{t-1}) \end{pmatrix} < 0 \tag{28}$$

where $\bar{\Xi}(\tau_{t-1}) = \left( \Xi(\tau_{t-1},0)^T \quad \Xi(\tau_{t-1},1)^T \quad \cdots \quad \Xi(\tau_{t-1},\tau_{\max})^T \right)^T$,

$\hat{Q}(\tau_{t-1}) = diag\left\{ Q(\tau_{t-1},0), Q(\tau_{t-1},1), \cdots, Q(\tau_{t-1},\tau_{\max}) \right\}$,

$V(\tau_{t-1}) = diag\left\{ \sqrt{p_{\tau_{t-1},0}} I, \sqrt{p_{\tau_{t-1},1}} I, \cdots, \sqrt{p_{\tau_{t-1},\tau_{\max}}} I \right\}$, and $I$ is an identity matrix with a proper dimension.

Proof:

In order to prove *Proposition 1*, we need to use Schur complement [28], which says that for symmetric matrix $M=\begin{bmatrix} M_{11} & M_{12} \\ M_{12}^T & M_{22}\end{bmatrix}$, the following three conditions are equivalent:

1. $M<0$;
2. $M_{11}<0, M_{22}-M_{12}^T M_{11}^{-1} M_{12}<0$;
3. $M_{22}<0, M_{11}-M_{12}M_{22}^{-1}M_{12}^T<0$.

Based on the Schur complement, in the following, we will establish the relationship between $L(\tau_{t-2},\tau_{t-1})<0$ and *Proposition 1*:

$$
\begin{aligned}
&L(\tau_{t-2},\tau_{t-1})\\
&=\sum_{\tau_t=0}^{\tau_{\max}} p_{\tau_{t-1},\tau_t}\Xi(\tau_{t-1},\tau_t)^T Q(\tau_{t-1},\tau_t)\Xi(\tau_{t-1},\tau_t)-Q(\tau_{t-2},\tau_{t-1})\\
&=\Xi(\tau_{t-1},0)^T p_{\tau_{t-1},0}Q(\tau_{t-1},0)\Xi(\tau_{t-1},0)+\cdots+\Xi(\tau_{t-1},\tau_{\max})^T p_{\tau_{t-1},\tau_{\max}}Q(\tau_{t-1},\tau_{\max})\Xi(\tau_{t-1},\tau_{\max})-Q(\tau_{t-2},\tau_{t-1})\\
&=\begin{pmatrix}\Xi(\tau_{t-1},0)^T & \cdots & \Xi(\tau_{t-1},\tau_{\max})^T\end{pmatrix}\begin{pmatrix} p_{\tau_{t-1},0}Q(\tau_{t-1},0) & & \\ & \ddots & \\ & & p_{\tau_{t-1},\tau_{\max}}Q(\tau_{t-1},\tau_{\max})\end{pmatrix}\begin{pmatrix}\Xi(\tau_{t-1},0)\\ \vdots \\ \Xi(\tau_{t-1},\tau_{\max})\end{pmatrix}-Q(\tau_{t-2},\tau_{t-1})\\
&=\begin{pmatrix}\Xi(\tau_{t-1},0)^T & \cdots & \Xi(\tau_{t-1},\tau_{\max})^T\end{pmatrix}\begin{pmatrix} \sqrt{p_{\tau_{t-1},0}}Q(\tau_{t-1},0) & & \\ & \ddots & \\ & & \sqrt{p_{\tau_{t-1},\tau_{\max}}}Q(\tau_{t-1},\tau_{\max})\end{pmatrix}\begin{pmatrix} \sqrt{p_{\tau_{t-1},0}}I & & \\ & \ddots & \\ & & \sqrt{p_{\tau_{t-1},\tau_{\max}}}I\end{pmatrix}\begin{pmatrix}\Xi(\tau_{t-1},0)\\ \vdots \\ \Xi(\tau_{t-1},\tau_{\max})\end{pmatrix}-Q(\tau_{t-2},\tau_{t-1})\\
&=\underbrace{\begin{pmatrix}\Xi(\tau_{t-1},0)^T & \cdots & \Xi(\tau_{t-1},\tau_{\max})^T\end{pmatrix}}_{\bar\Xi(\tau_{t-1})^T}\underbrace{\begin{pmatrix} \sqrt{p_{\tau_{t-1},0}}I & & \\ & \ddots & \\ & & \sqrt{p_{\tau_{t-1},\tau_{\max}}}I\end{pmatrix}}_{V(\tau_{t-1})}\underbrace{\begin{pmatrix} Q(\tau_{t-1},0) & & \\ & \ddots & \\ & & Q(\tau_{t-1},\tau_{\max})\end{pmatrix}}_{\hat Q(\tau_{t-1})}\underbrace{\begin{pmatrix} \sqrt{p_{\tau_{t-1},0}}I & & \\ & \ddots & \\ & & \sqrt{p_{\tau_{t-1},\tau_{\max}}}I\end{pmatrix}}_{V(\tau_{t-1})^T}\underbrace{\begin{pmatrix}\Xi(\tau_{t-1},0)\\ \vdots \\ \Xi(\tau_{t-1},\tau_{\max})\end{pmatrix}}_{\bar\Xi(\tau_{t-1})}-Q(\tau_{t-2},\tau_{t-1})\\
&\equiv\bar\Xi(\tau_{t-1})^T V(\tau_{t-1})\hat Q(\tau_{t-1})V(\tau_{t-1})^T\bar\Xi(\tau_{t-1})-Q(\tau_{t-2},\tau_{t-1})<0
\end{aligned}
$$

we take $M_{11}=-Q(\tau_{t-2},\tau_{t-1})$, $M_{12}=\bar\Xi(\tau_{t-1})^T V(\tau_{t-1})\hat Q(\tau_{t-1})$ and $M_{22}=-\hat Q(\tau_{t-1})$, then the third condition of Schur complement is satisfied, and therefore $M=\begin{bmatrix} M_{11} & M_{12} \\ M_{12}^T & M_{22}\end{bmatrix}=\begin{pmatrix} -Q(\tau_{t-2},\tau_{t-1}) & \bar\Xi(\tau_{t-1})^T V(\tau_{t-1})\hat Q(\tau_{t-1}) \\ \hat Q(\tau_{t-1})^T V(\tau_{t-1})^T\bar\Xi(\tau_{t-1}) & -\hat Q(\tau_{t-1})\end{pmatrix}<0$, i.e., *Proposition 1* holds.■

***Remark 4:*** the necessary and sufficiency conditions of stochastic stable for the system with EWMA run-to-run controllers (i.e., EWMA-I controller in this paper) subject to stochastic metrology delay were first proposed by B. Ai *et al.* in [16]; however they made mistakes in establishing the relationship between transition probability matrix and the system matrix (please see Theorem in [16] for detail), and therefore the proposition (also see [16]) they obtained for solving the matrix inequality is wrong. In fact, (28) is a set of linear matrix inequalities, we can use Matlab Robust Control Toolbox to solve them.

## 3. Transition Probability Matrix for Single Product Process

Since the metrology delay is a stochastic variable instead of being fixed in an actual manufacturing plant, how to establish the relationship between the metrology lags will be studied in this Section. Firstly, we will show the process of obtaining actual metrology delay sequence from the original metrology delay sequence, and then we will derive the transition probability matrix from a specific probability distribution. How to calculate the average delay of the system will be discussed in the last part of this Section.

### 3.1. Original and Observed Metrology Delay

The process of sampling, measurement and reporting metrology results is a stochastic process so that the "original" or "actual" metrology delay is a sequence of stochastic variable generated by a probability function. Let $OMD_t$, the "original" metrology delay of the system at the run $t$, be a random number generated by a specific probability distribution. Hence the information of this run will be available at run $OMD_t + t$. Let $LRA_t$ be the index of the latest run at which metrology results are available at the run $t$, we have:

$$LRA_t = \arg\min_{k=[1,\ldots,t-1]} \{OMD_k + k\} \quad (29)$$

Furthermore, if the latest run at which metrology results are available is more recent than that of the last run, then controller will be notified of this new delay and metrology, otherwise the delay will be increased by 1:

$$\tau_t = t - LRA_t \quad (30)$$

**Table 1** illustrated an example of the process of calculating resampled delay numbers from original delay numbers. The information of the first production run is available immediately, hence the delay is zero. At the 2nd production run, the newest data available is still the data of the 1st run, hence the delay is 1. At the 3rd run, the information of the 2nd run was received; hence the delay remained to be 1. At the 6th run, two previous runs (the 4th and 5th) are available; only the latest (the 5th) will be used, hence the delay is 1. Since the 6th and the 8th runs were not sampled, and the 7th run was reported only after long delay, the data of the 5th production run remained the latest information available for 7th and the 8th runs. Hence the corresponding delays for 7th and the 8th runs were 2 and 3 accordingly.

**Table 1:** The relation between original metrology delay and actual metrology delay by the controller

| Run Number $t$ | Original Metrology Delay $OMD_t$ | Run when metrology available $OMD_t + t$ | Latest Run Available $LRA_t$ | Actual Delay in EWMA Controller $\tau_t$ |
|---|---|---|---|---|
| 1 | 0 | 1 | 1 | 0 |
| 2 | 1 | 3 | 1 | 1 |
| 3 | 2 | 5 | 2 | 1 |
| 4 | 2 | 6 | 2 | 2 |
| 5 | 1 | 6 | 3 | 2 |
| 6 | NM[1] | NM[1] | 5 | 1 |
| 7 | 5 | 12 | 5 | 2 |
| 8 | NM[1] | NM[1] | 5 | 3 |
| [1] run is not sampled | | | | |

From the previous analysis we conclude that the actual metrology delay $\tau_t$ can increase at most 1 at each run, i.e., $\tau_{t+1} \le \tau_t + 1$ and $\mathrm{Prob}(\tau_{t+1} \ge \tau_t + 2) = 0$. Since $\tau_{t+1}$ is only affected by $\tau_t$, $\{\tau_t\}$ is a Markov Chain. If the constraint conditions for metrology delay are considered, then the structure of the transition probability matrix will be

$$P = [p_{ij}]$$

$$= \begin{bmatrix} p_{00} & p_{01} & 0 & 0 & \cdots & 0 & 0 & 0 & 0 & \cdots \\ p_{10} & p_{11} & p_{12} & 0 & \cdots & 0 & 0 & 0 & 0 & \cdots \\ \vdots & \vdots & \vdots & \vdots & \ddots & \vdots & \vdots & \vdots & \vdots & \ddots \\ p_{\tau-2,0} & p_{\tau-2,1} & p_{\tau-2,2} & p_{\tau-2,3} & \cdots & 0 & 0 & 0 & 0 & \cdots \\ p_{\tau-1,0} & p_{\tau-1,1} & p_{\tau-1,2} & p_{\tau-1,3} & \cdots & p_{\tau-1,\tau} & 0 & 0 & 0 & \cdots \\ p_{\tau 0} & p_{\tau 1} & p_{\tau 2} & p_{\tau 3} & \cdots & p_{\tau\tau} & p_{\tau,\tau+1} & 0 & 0 & \cdots \\ p_{\tau+1,0} & p_{\tau+1,1} & p_{\tau+1,2} & p_{\tau+1,3} & \cdots & p_{\tau+1,\tau} & p_{\tau+1,\tau+1} & p_{\tau+1,\tau+2} & 0 & \cdots \\ \vdots & \vdots & \vdots & \vdots & \ddots & \vdots & \vdots & \vdots & \vdots & \ddots \end{bmatrix}_{\infty\times\infty} \quad (31)$$

Each row represents the transition probabilities from a fixed state to all the states, the diagonal elements are the probabilities of metrology sequences with equal delays, the elements below the diagonal indicate shorter delays, and the elements above the diagonal are the probabilities of encountering longer delays. **Fig. 3** illustrates three states transition diagram. From the figure, we can see that it can jump from $\tau = 1$ and $\tau = 2$ to any states, while it cannot jump from $\tau = 0$ to $\tau = 2$.

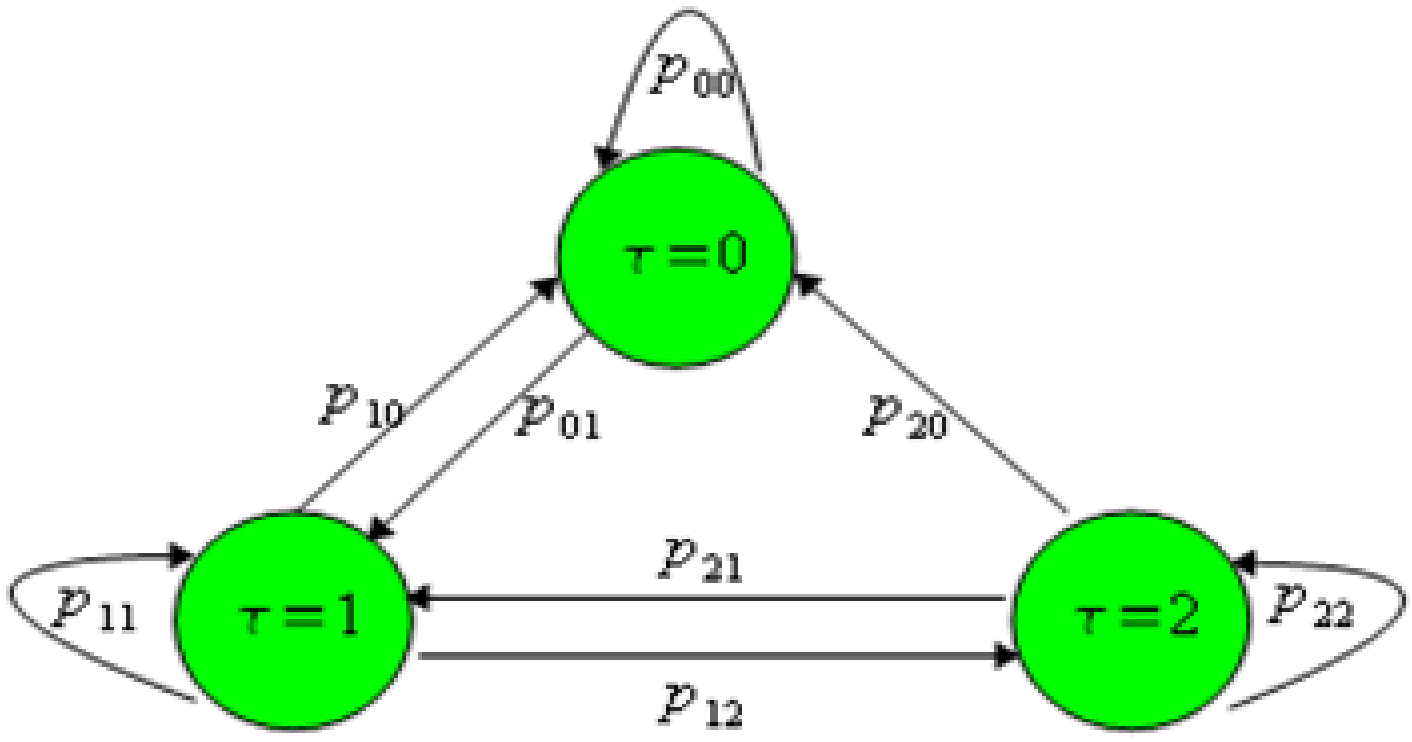

**Fig. 3**. Three states transition diagram

### 3.2. Calculation of Transition Probability Matrix from a Specific Probability Distribution

Let $\eta_j \equiv \text{Prob}(OMD_t = j)$ be the probabilities of a metrology being reported after $j$ runs; and $p_{NM}$ be the probabilities of a run not sampled at all. The following theorem calculates the probabilities of observing $\tau_{t+1} = j$ and $\tau_t = i$, i.e., $p_{ij}$ given $\eta_j$ and $p_{NM}$:

***Theorem 2:*** The transition probability matrix $p_{ij}$ can be calculated from the probability distribution of the metrology delay, $\eta_j$ and $p_{NM}$ by:

$$p_{ij} = p(\tau_{t+1} = j \mid \tau_t = i) = \begin{cases} 0, & i+1 < j; \\ p_{NM} + (1-p_{NM}) \cdot \sum_{k=i+1}^{\infty} \eta_k, & j = i+1; \\ (1-p_{NM}) \cdot \eta_j, & 0 \le j \le i. \end{cases} \quad (32)$$

***Proof:***

If $\tau_t = i$, the latest data available at run $t$ is $t-i$, i.e., the data of run $t-i$ is sampled with metrology delay of $i$ runs. Furthermore, the data of runs $t-i+1,\cdots,t$ are either not sampled, or are reported with metrology delays $OMD$ longer than or equal to $i,\cdots,1$ runs respectively, i.e.:

$$\begin{aligned}\mathrm{Prob}(\tau_t = i) = (1-p_{NM})\eta_i \bullet \left( p_{NM} + (1-p_{NM}) \bullet \sum_{k=i}^{\infty} \eta_k \right) \bullet \left( p_{NM} + (1-p_{NM}) \bullet \sum_{k=i-1}^{\infty} \eta_k \right) \bullet \cdots \\ \bullet \left( p_{NM} + (1-p_{NM}) \bullet \sum_{k=2}^{\infty} \eta_k \right) \bullet \left( p_{NM} + (1-p_{NM}) \bullet \sum_{k=1}^{\infty} \eta_k \right)\end{aligned} \quad (33)$$

(A) If $\tau_t = i$ and $\tau_{t+1} = j \le i$, the $OMDs$ of runs between $t-i+1$ and $t-j$ has to be longer than or equal to $i,\cdots,j+1$ because even if the $OMD$ of $t-i+1$ is $i$ and arrived at $t+1$ as did the metrology of $t-j+1$, it will be disregarded as $t-j+1$ is a newer run. Furthermore, the data of runs $t-j+2,\cdots,t-1$ are either not sampled, or are reported with metrology delays $OMD$ longer than or equal to $j,\cdots,1$ runs respectively. Therefore:

$$\begin{aligned}&\mathrm{Prob}(\tau_{t+1} = j \cap \tau_t = i) \\ &= (1-p_{NM})\eta_i \bullet \left( p_{NM} + (1-p_{NM}) \bullet \sum_{k=i}^{\infty} \eta_k \right) \bullet \left( p_{NM} + (1-p_{NM}) \bullet \sum_{k=i-1}^{\infty} \eta_k \right) \bullet \cdots \\ &\bullet \left( p_{NM} + (1-p_{NM}) \bullet \sum_{k=j+1}^{\infty} \eta_k \right) \bullet (1-p_{NM})\eta_j \bullet \left( p_{NM} + (1-p_{NM}) \bullet \sum_{k=j}^{\infty} \eta_k \right) \\ &\bullet \left( p_{NM} + (1-p_{NM}) \bullet \sum_{k=j-1}^{\infty} \eta_k \right) \bullet \cdots \\ &\bullet \left( p_{NM} + (1-p_{NM}) \bullet \sum_{k=2}^{\infty} \eta_k \right) \bullet \left( p_{NM} + (1-p_{NM}) \bullet \sum_{k=1}^{\infty} \eta_k \right)\end{aligned} \quad (34)$$

$$p_{ij}=\text{Prob}\left(\tau_{t+1}=j\middle|\tau_t=i\right)=\frac{\text{Prob}\left(\tau_{t+1}=j\cap\tau_t=i\right)}{\text{Prob}\left(\tau_t=i\right)}$$

$$=\frac{(1-p_{NM})\eta_i}{(1-p_{NM})\eta_i}\bullet\frac{p_{NM}+(1-p_{NM})\bullet\sum_{k=i}^{\infty}\eta_k}{p_{NM}+(1-p_{NM})\bullet\sum_{k=i}^{\infty}\eta_k}\bullet\frac{p_{NM}+(1-p_{NM})\bullet\sum_{k=i-1}^{\infty}\eta_k}{p_{NM}+(1-p_{NM})\bullet\sum_{k=i-1}^{\infty}\eta_k}\bullet\cdots$$

$$\bullet\frac{p_{NM}+(1-p_{NM})\bullet\sum_{k=j+1}^{\infty}\eta_k}{p_{NM}+(1-p_{NM})\bullet\sum_{k=j+1}^{\infty}\eta_k}\bullet\frac{(1-p_{NM})\eta_j}{p_{NM}+(1-p_{NM})\bullet\sum_{k=j}^{\infty}\eta_k}\bullet\frac{p_{NM}+(1-p_{NM})\bullet\sum_{k=j}^{\infty}\eta_k}{p_{NM}+(1-p_{NM})\bullet\sum_{k=j-1}^{\infty}\eta_k}$$

$$\bullet\frac{p_{NM}+(1-p_{NM})\bullet\sum_{k=j-1}^{\infty}\eta_k}{p_{NM}+(1-p_{NM})\bullet\sum_{k=j-2}^{\infty}\eta_k}\bullet\cdots\bullet\frac{p_{NM}+(1-p_{NM})\bullet\sum_{k=2}^{\infty}\eta_k}{p_{NM}+(1-p_{NM})\bullet\sum_{k=1}^{\infty}\eta_k}\bullet\frac{p_{NM}+(1-p_{NM})\bullet\sum_{k=1}^{\infty}\eta_k}{1}$$

$$=(1-p_{NM})\eta_j \tag{35}$$

(B) If $\tau_t=i$ and $\tau_{t+1}=j=i+1$, the *OMDs* of runs between $t-i+1,\cdots t,t+1$ and $t$ has to be longer than or equal to $i+1,\cdots 2,1$. Therefore:

$$\text{Prob}\left(\tau_{t+1}=i+1\cap\tau_t=i\right)$$

$$=(1-p_{NM})\eta_i\bullet\left(p_{NM}+(1-p_{NM})\bullet\sum_{k=i+1}^{\infty}\eta_k\right)\bullet\left(p_{NM}+(1-p_{NM})\bullet\sum_{k=i}^{\infty}\eta_k\right)\bullet\cdots \tag{36}$$

$$\bullet\left(p_{NM}+(1-p_{NM})\bullet\sum_{k=2}^{\infty}\eta_k\right)\bullet\left(p_{NM}+(1-p_{NM})\bullet\sum_{k=1}^{\infty}\eta_k\right)$$

$$p_{i,i+1}=\text{Prob}\left(\tau_{t+1}=i+1\middle|\tau_t=i\right)=\frac{\text{Prob}\left(\tau_{t+1}=i+1\cap\tau_t=i\right)}{\text{Prob}\left(\tau_t=i\right)}$$

$$=\frac{(1-p_{NM})\eta_i}{(1-p_{NM})\eta_i}\bullet\frac{p_{NM}+(1-p_{NM})\bullet\sum_{k=i+1}^{\infty}\eta_k}{p_{NM}+(1-p_{NM})\bullet\sum_{k=i}^{\infty}\eta_k}\bullet\frac{p_{NM}+(1-p_{NM})\bullet\sum_{k=i}^{\infty}\eta_k}{p_{NM}+(1-p_{NM})\bullet\sum_{k=i-1}^{\infty}\eta_k}\bullet\cdots$$

$$\bullet\frac{p_{NM}+(1-p_{NM})\bullet\sum_{k=2}^{\infty}\eta_k}{p_{NM}+(1-p_{NM})\bullet\sum_{k=1}^{\infty}\eta_k}\bullet\frac{p_{NM}+(1-p_{NM})\bullet\sum_{k=1}^{\infty}\eta_k}{1} \tag{37}$$

$$=p_{NM}+(1-p_{NM})\bullet\sum_{k=i+1}^{\infty}\eta_k$$

(C) $j>i+1$

Analysis in subsection 3.1 showed that the observed delay can increase by 1 but no more, i.e. $p_{ij}=0$ if $j>i+1$.

To sum those up, we have Theorem 2.■

***Remark 5:*** The summation of the transition probabilities from a fixed state to all the states is 1:

$$\sum_{j=0}^{\infty} p_{ij}=1 \tag{38}$$

***Proof:***

$$\begin{aligned}\sum_{j=0}^{\infty} p_{ij} &= \sum_{j=0}^{i} p_{ij} + \sum_{j=i+1}^{\infty} p_{ij} \\ &= \sum_{k=0}^{i}(1-p_{NM})\cdot\eta_k + p_{NM} + (1-p_{NM})\cdot\sum_{k=i+1}^{\infty}\eta_k \\ &= p_{NM} + (1-p_{NM})\cdot\sum_{k=0}^{\infty}\eta_k \\ &= 1\end{aligned} \tag{39}$$

***Remark 6:*** The calculation of transition probability matrix from a given probability distribution was first derived by Z.X. Yu *et al.* for the problem of data-packet transfer in network control system (NCS) [17]-[18]; Y. Zheng *et al.* extended this work to a more general case which assumes the data-packet will be lost sometimes; however, the theorems obtained by Yu and Zheng *et al.* are wrong (see **Theorem 1** in [17]-[19] for details).

### 3.3. Asymptomatic Probability and Average Delay

For any homogeneous Markov chain, from the relationship between absolute probability distribution and initial probability distribution, we have the following equation:

$$\overline{P}_j=\sum_{i=0}^{\infty}\overline{P}_i\cdot p_{ij} \tag{40}$$

with the constraint:

$$\sum_{j=0}^{\infty}\overline{P}_j=1 \tag{41}$$

where $\overline{P}_j$ is the average probability at state $j$.

The average delay $E(\tau)$ can be obtained by taking the expectation of states in the state-set. i.e.,

$$E(\tau) = \sum_{j=0}^{\infty} j \bullet \bar{P}_j \tag{42}$$

If $\eta_i = 0$ for $i > \tau_p$, the above equations can be truncated to a set of finite delay states $\{0,1,\cdots \tau_p\}$ with

$$\tilde{P}_j = \sum_{j=0}^{\tau_p} \tilde{P}_i \cdot \tilde{p}_{ij} \tag{43}$$

with

$$\tilde{p}_{ij} = \begin{cases} p_{ij} & 0 \le i < \tau_p, 0 \le j \le \tau_p; \\ \eta_j & i = \tau_p, 0 \le j \le \tau_p. \end{cases} \tag{44}$$

***Proof:***

If $\eta_i = 0$ for $i > \tau_p$, all rows with $i > \tau_p$ can be truncated since $\text{Prob}(\tau_t = i) = 0$ and hence $p_{ij}$ is undefined. Furthermore, if we truncate all columns of probability transition matrix from the column $j = \tau_p + 1$ onwards, the reduced transition probability matrix must be normalized accordingly

$$\tilde{p}_{ij} = \frac{p_{ij}}{1 - \sum_{j=\tau_p+1}^{\infty} p_{ij}} \tag{45}$$

However, $p_{ij} = 0$ for all $i < \tau_p$, $j \ge \tau_p + 1$, therefore, we have $\tilde{p}_{ij} = p_{ij}$ for $0 \le i \le \tau_p$; $0 \le j \le \tau_p$. When $i = \tau_p$, we have

$$p_{\tau_p,\tau_p+1} = p_{NM} + (1 - p_{NM}) \bullet \sum_{j=\tau_p+1}^{\infty} \eta_j = p_{NM} \tag{46}$$

thus

$$\tilde{p}_{\tau_p,j} = \frac{p_{\tau_p,j}}{1 - p_{NM}} = \frac{(1 - p_{NM}) \bullet \eta_j}{(1 - p_{NM})} = \eta_j \tag{47}$$

Therefore (44) holds.■

The asymptotic average delay is given by:

$$E(\tilde{\tau}) = \sum_{j=0}^{\tau_p} j \bullet \tilde{P}_j \tag{48}$$

The value of $\tau_p$ can be chosen so that changes in $E(\tilde{\tau})$ is negligible with increase of $\tau_p$.

## 4. Numerical Examples for Single Product Process

In this Section, we will investigate the stability of systems with EWMA-I and II controllers subject to different kinds of metrology delays. In subsection 4.1, the stability regions for systems with fixed metrology delay will be obtained; we will consider the stability regions for the systems with EWMA-I and II controllers with measurement taken in a particular sampling interval in subsection 4.2; the stability regions for systems with EWMA-I and II controllers subject to stochastic metrology delay will be discussed in subsection 4.3.

### 4.1. Stability Analysis for Systems with Fixed Metrology Delay

***Theorem 1*** which is obtained in subsection 2.3.2 is based on Lyapunov's direct method. In this subsection, we will compare the stability regions obtained by Lyapunov's direct method and Routh-Hurwitz criterion (or Jury's test) for the systems with fixed metrology delays.

**Fig. 4 - Fig. 13** are the simulation results for systems without metrology delay and with fixed one to fixed nine runs metrology delay. From the figures, it is clear that for the same metrology delay system, the stability region will be the same despite of which kind of controller, EWMA-I controller or EWMA-II controller, is adopted. This result coincides with the theoretic result obtained in subsection 2.3. Also Lyapunov's direct method and Routh-Hurwitz criterion arrive at the same stability regions for the same systems.

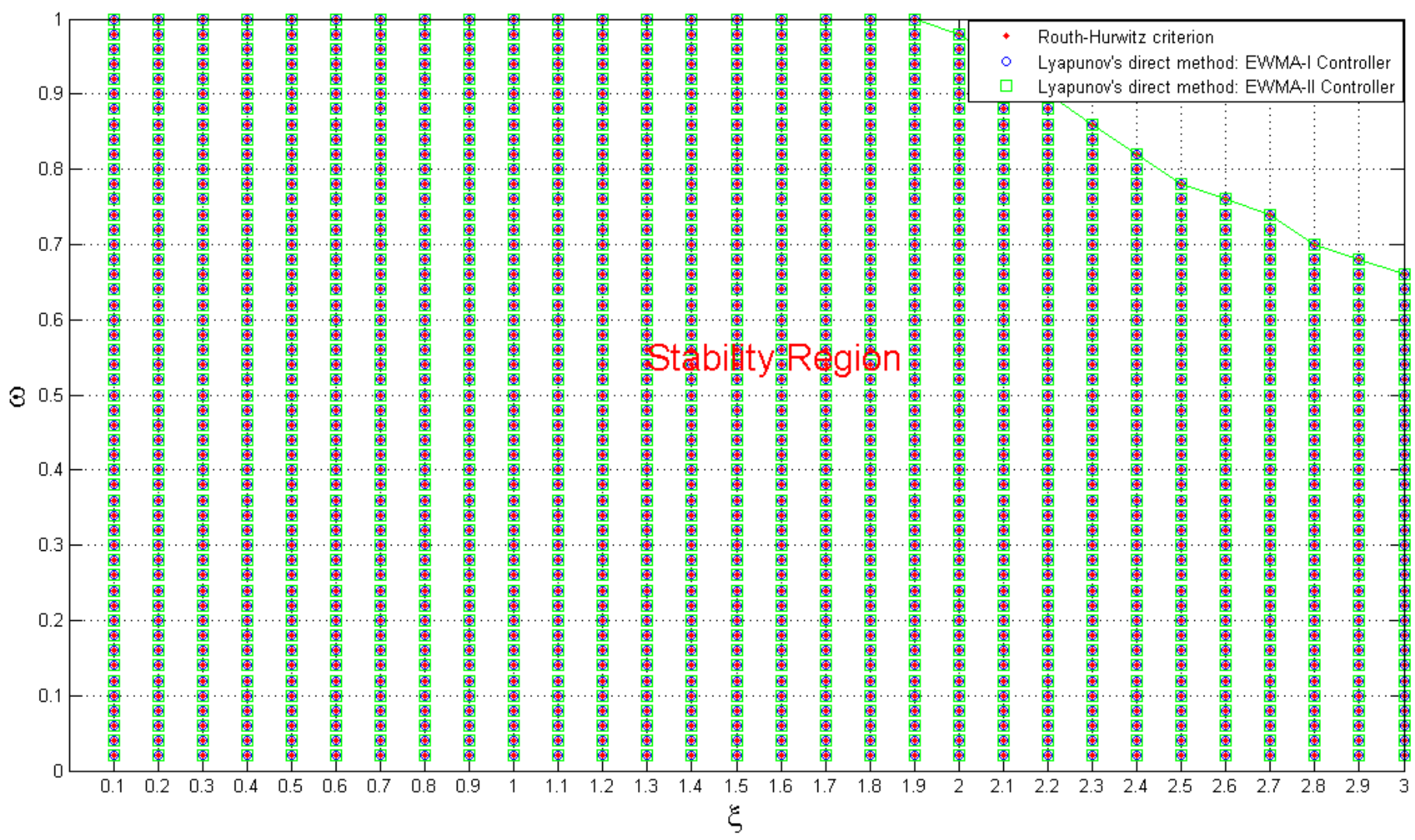

**Fig. 4.** Stability regions for systems without metrology delay

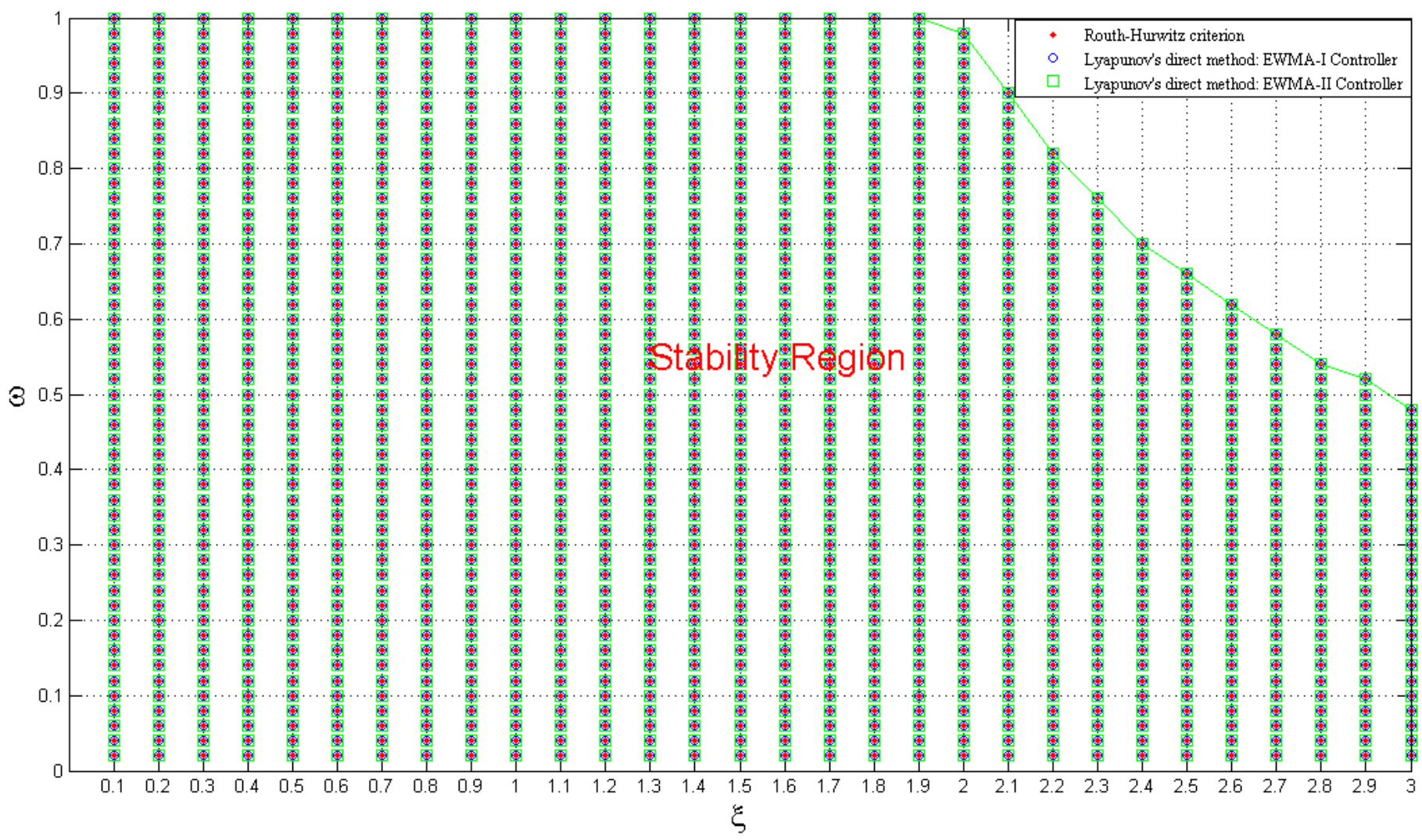

**Fig. 5.** Stability regions for systems with fixed one run metrology delay

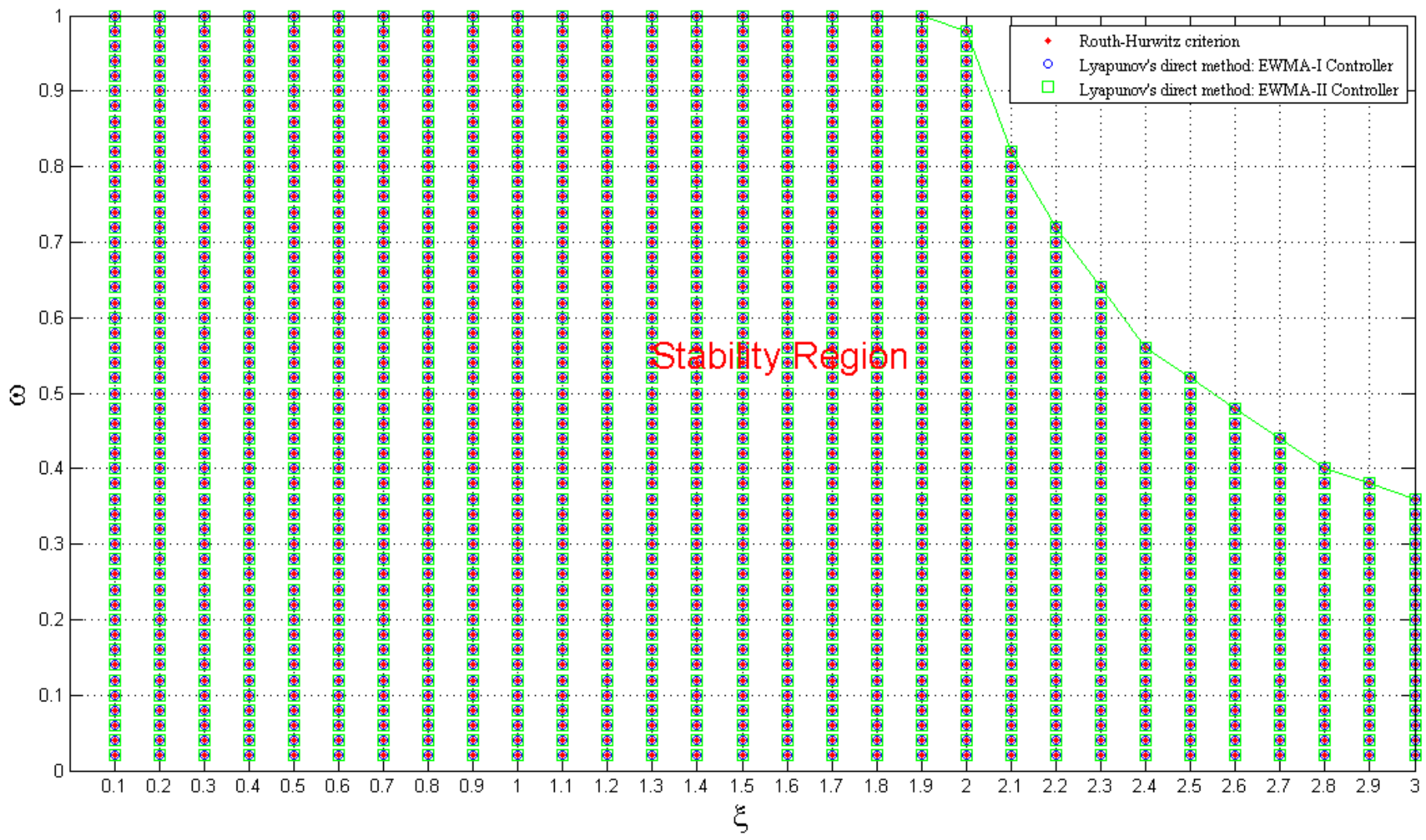

**Fig. 6.** Stability regions for systems with fixed two runs metrology delay

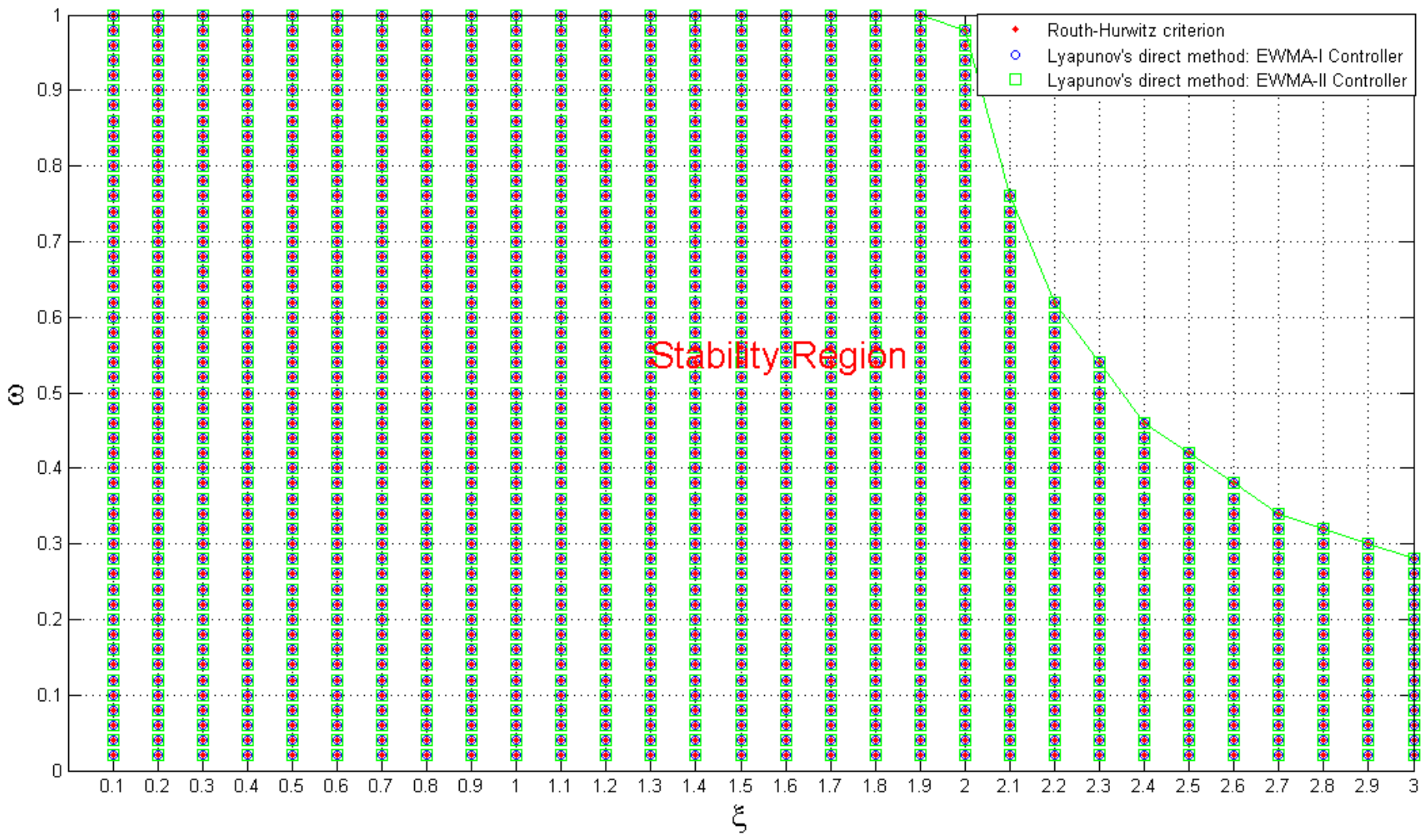

**Fig. 7.** Stability regions for systems with fixed three runs metrology delay

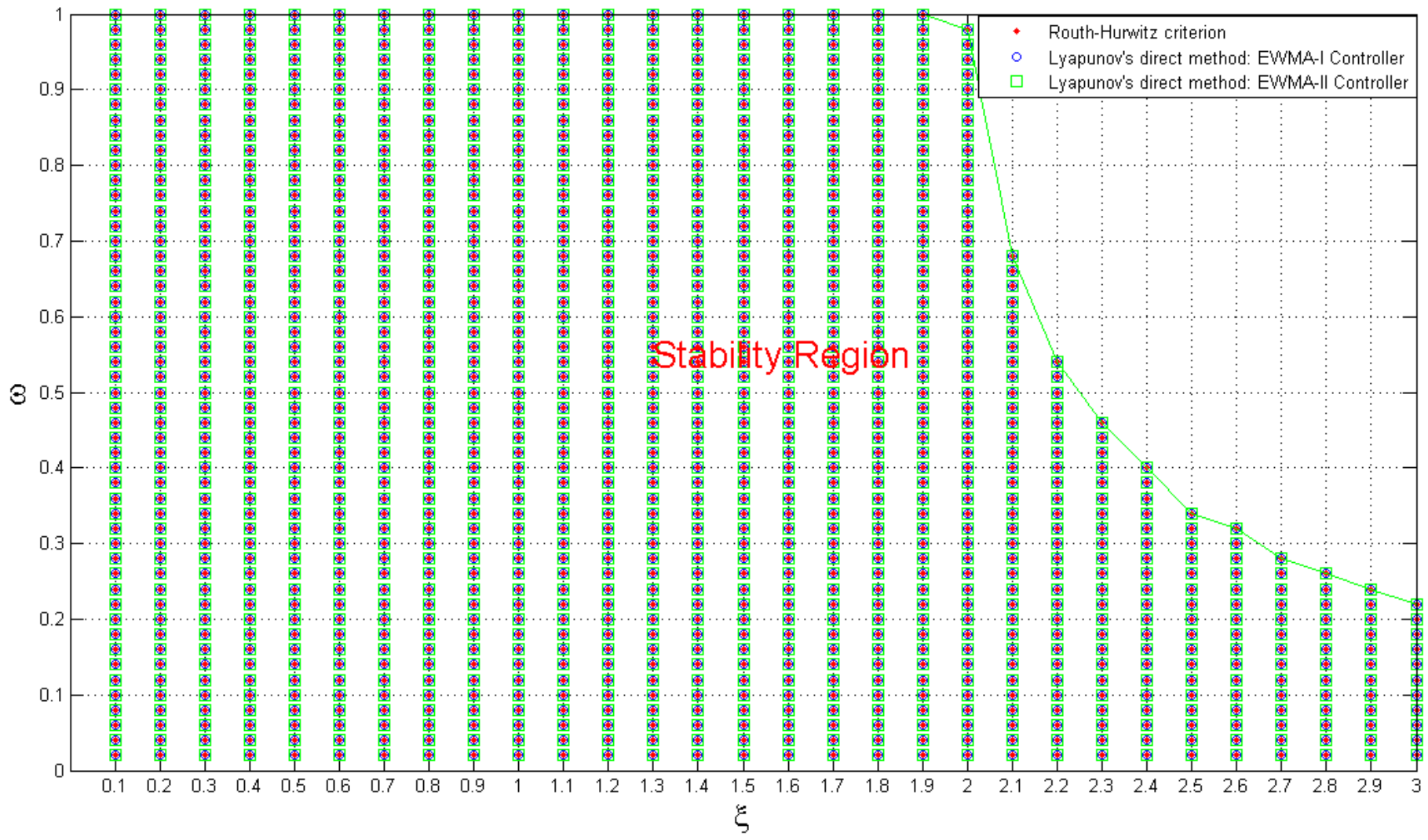

**Fig. 8.** Stability regions for systems with fixed four runs metrology delay

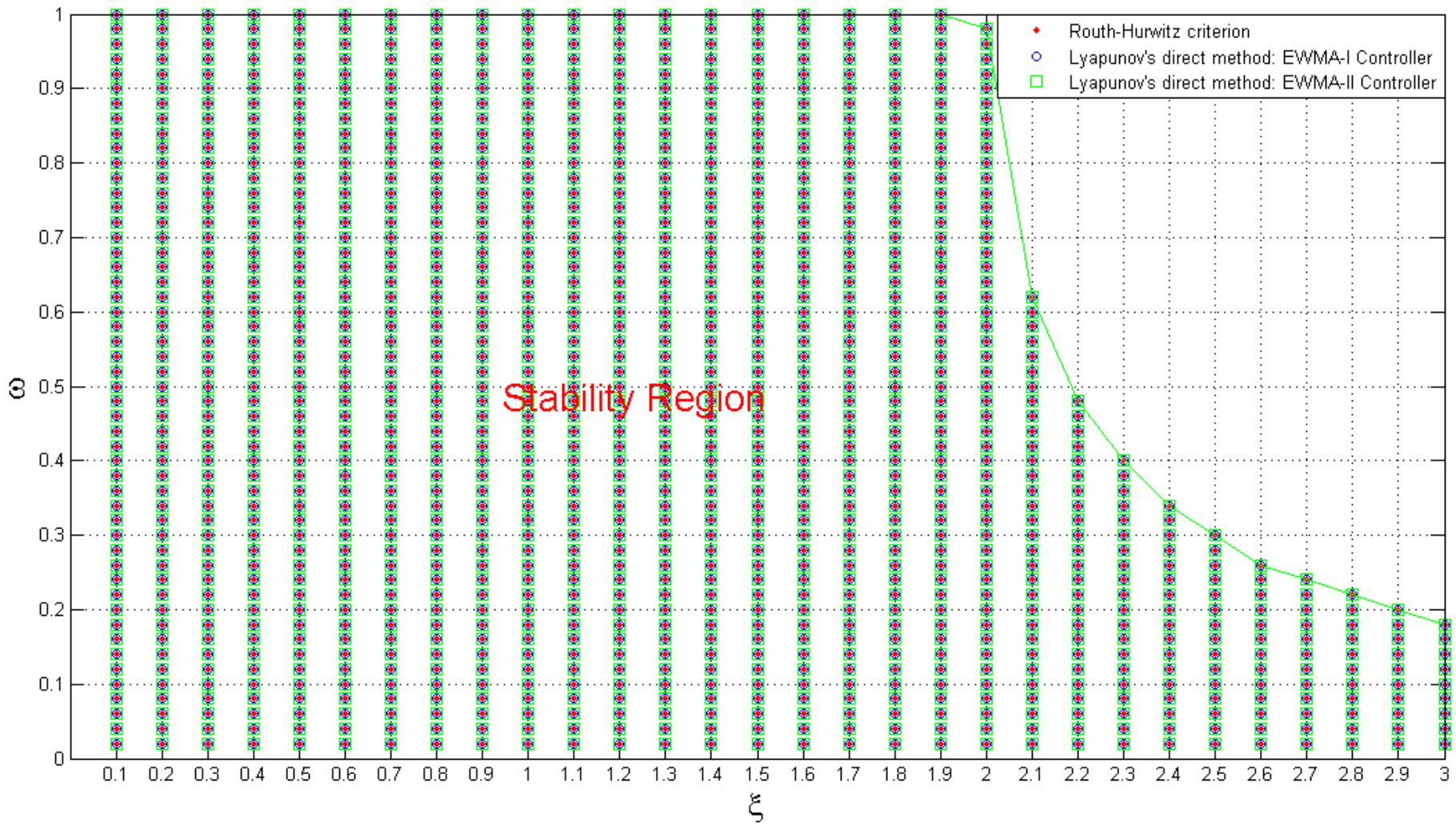

**Fig. 9.** Stability regions for systems with fixed five runs metrology delay

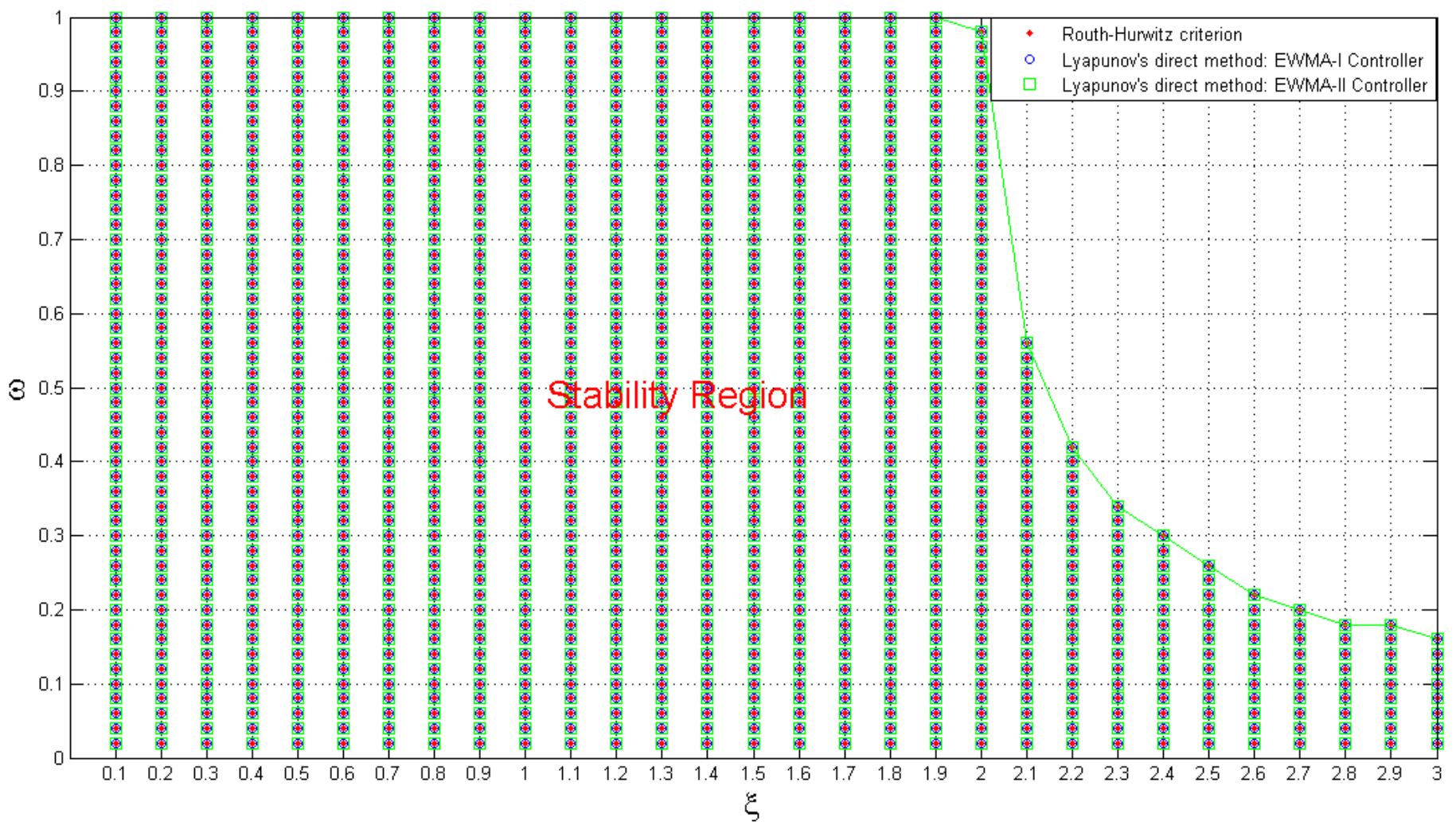

**Fig. 10.** Stability regions for systems with fixed six runs metrology delay

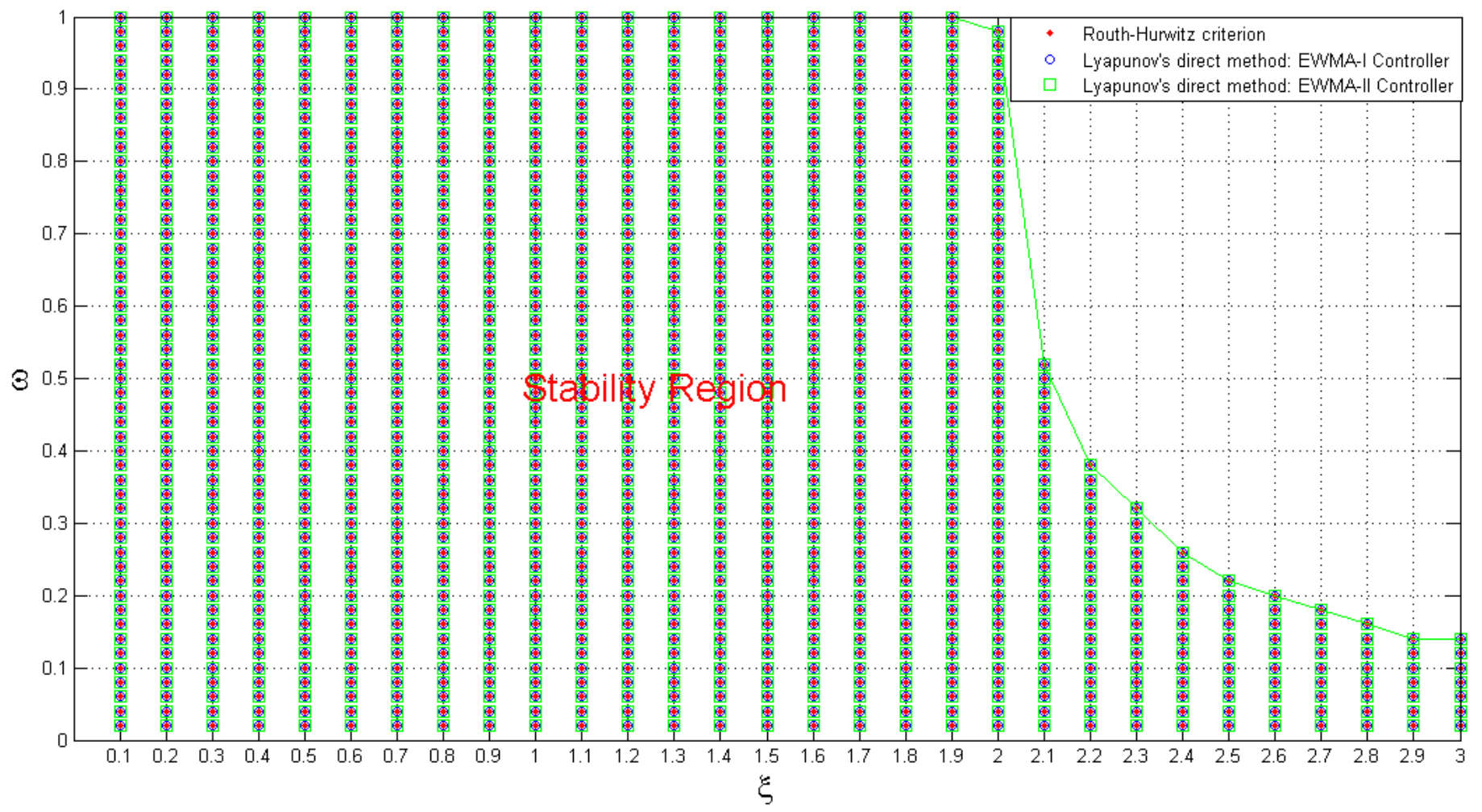

**Fig. 11.** Stability regions for systems with fixed seven runs metrology delay

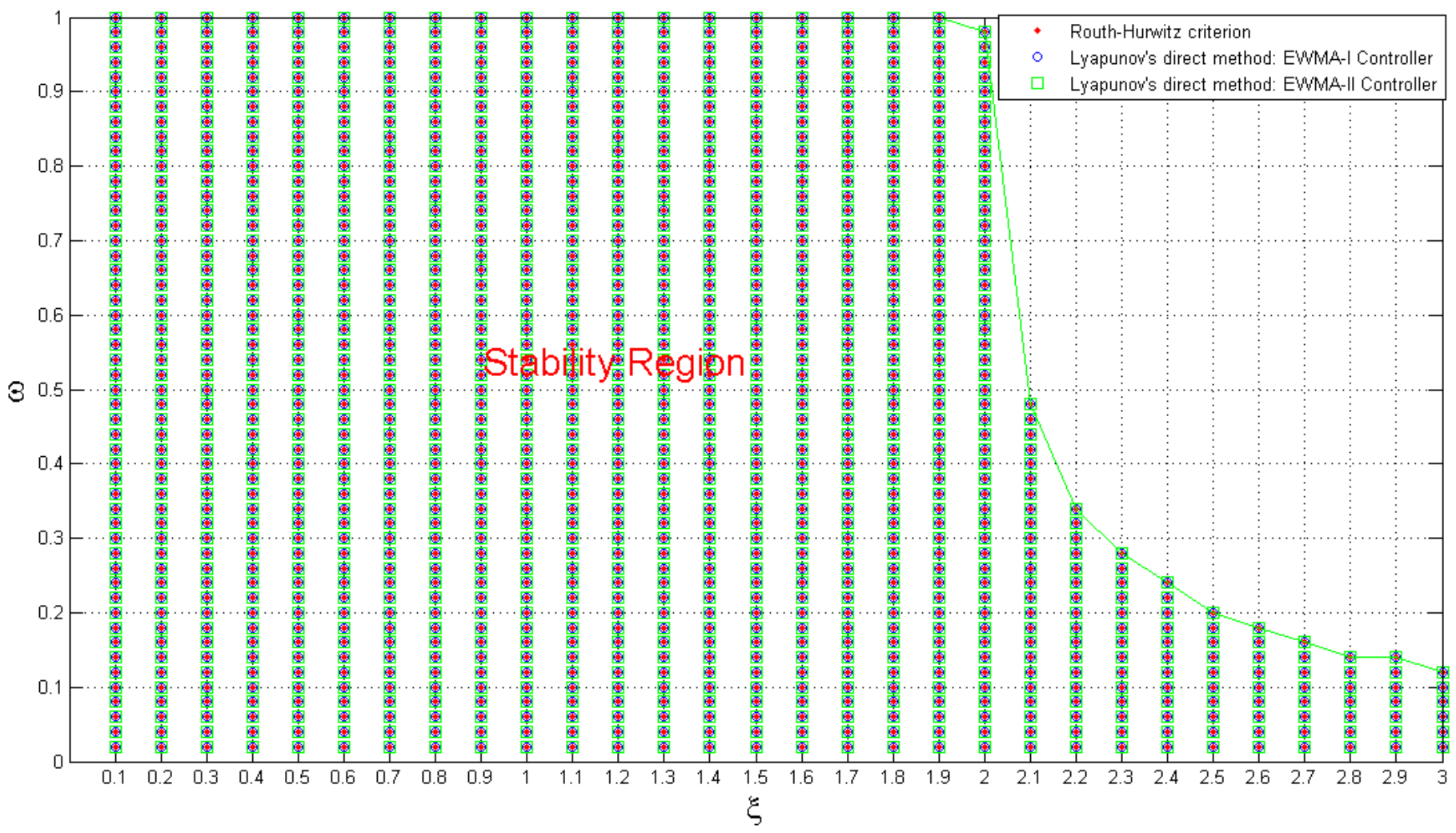

**Fig. 12.** Stability regions for systems with fixed eight runs metrology delay

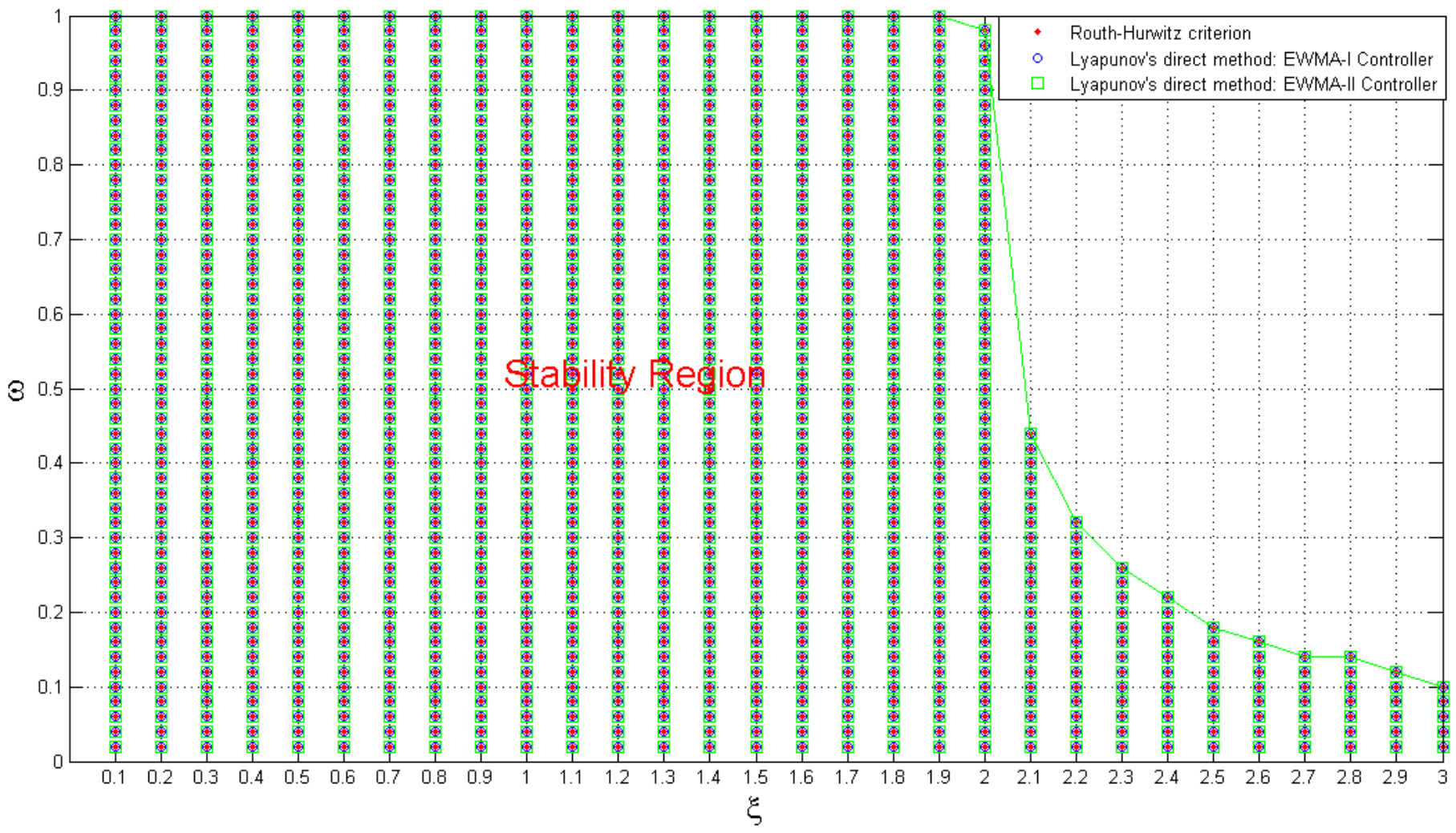

**Fig. 13.** Stability regions for systems with fixed nine runs metrology delay

**Fig. 14** is the comparison figure which shows the stability regions of systems with different fixed metrology delays. From the figure, it is noticed that although the systems have different fixed metrology delays, they have similarities in the stability region: when the estimated process gain is greater than half of the true process gain ($\xi < 2$), the system is guaranteed closed-loop stable for any discount factor $\omega$ between 0 to 1; however, if $\xi \geq 2$, $\omega$ is decreasing in $\xi$ to keep the closed-loop system stable. In addition, the size of stability region will shrink with the increase of metrology delay.

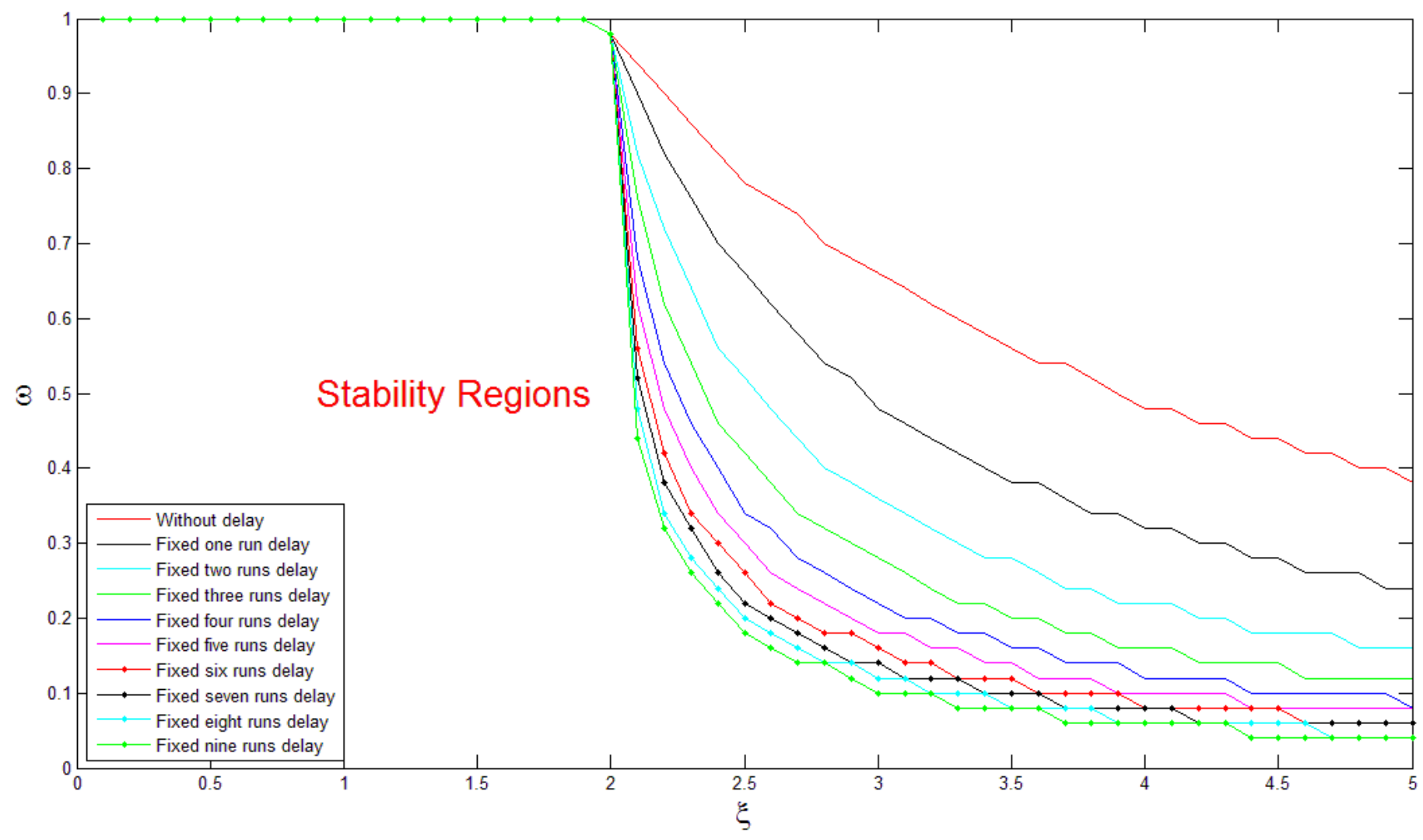

**Fig. 14.** Comparison of stability regions for systems with different fixed runs delays

### 4.2. Stability Analysis for Systems with Fixed Sampling Metrology

In semiconductor manufacturing, in situ measurements are usually taken at the particular sampling interval; thus, there is a need to analyze the stability of the system with such kind of metrology delay. Taking one run as the sampling interval for example, if at the $t^{th}$ production run, we have taken the metrology, then at the $(t+1)^{th}$ production run, we will use the information we got at the $t^{th}$ production run, i.e., the system will be of one run delay, and therefore $p_{01}=1$. On the other hand, if at the $t^{th}$ production run, no metrology is taken, then at the $(t+1)^{th}$ production run, a new metrology will be taken, i.e., the system is delay free, and $p_{10}=1$. Based on the above analysis, we know that the transition probability matrix is $P=\begin{bmatrix}0 & 1\\ 1 & 0\end{bmatrix}$. The same analysis can be done for the system with other sampling metrology intervals.

And the transition probability matrix is $P=\begin{bmatrix}0&1&0\\0&0&1\\1&0&0\end{bmatrix}$ and $P=\begin{bmatrix}0&1&0&0\\0&0&1&0\\0&0&0&1\\1&0&0&0\end{bmatrix}$ for the sampling interval two and three respectively.

**Fig. 15** shows the simulation results for systems with different fixed sampling metrology intervals. From the figure, we notice that when the model mismatch $\xi < 2$, the system with either controller is guaranteed closed-loop stable for any discount factor $\omega$ between 0 to 1. For the system with EWMA-I controller, the larger the sampling metrology interval is, the smaller the size of stability region will be; while for the system with EWMA-II controller, the stability region is irrelevant to the length of sampling intervals and this result does not coincide with our intuition.

In fact, for the fixed sampling metrology delay, EWMA-II controller (7) is equivalent to EWMA-III controller, $x_{t+1}=\begin{cases}(1-\xi\omega)x_t, & if\ \tau_t=0;\\ x_t, & otherwise.\end{cases}$. If the metrology is taken in the form of fixed sampling, then the stability condition for system with EWMA-III controller is the same as that of the system without metrology delay. There are two reasons for this result: first, the data used to update the controller are the measured data which are free of delay, and this happens with probability 1; second, when delay appears, EWMA-III controller will not update.

For the system of fixed sampling metrology, if the sampling interval is $d$, then from (40) and (42), it is easy to get that the average metrology delay $E(\tau)=\frac{d}{2}$ and $\bar{P}_0=\bar{P}_1=\cdots=\bar{P}_d=\frac{1}{d+1}$. From **Fig. 15**, we know that it is wrong to use ceil(E(τ))[1] as the delay for the system with EWMA-I controller to get the stability region, while if we do so for the system with EWMA-II controller, we will obtain conservative stability regions.

[1] ceil(A) rounds the elements of A to the nearest integers greater than or equal to A.

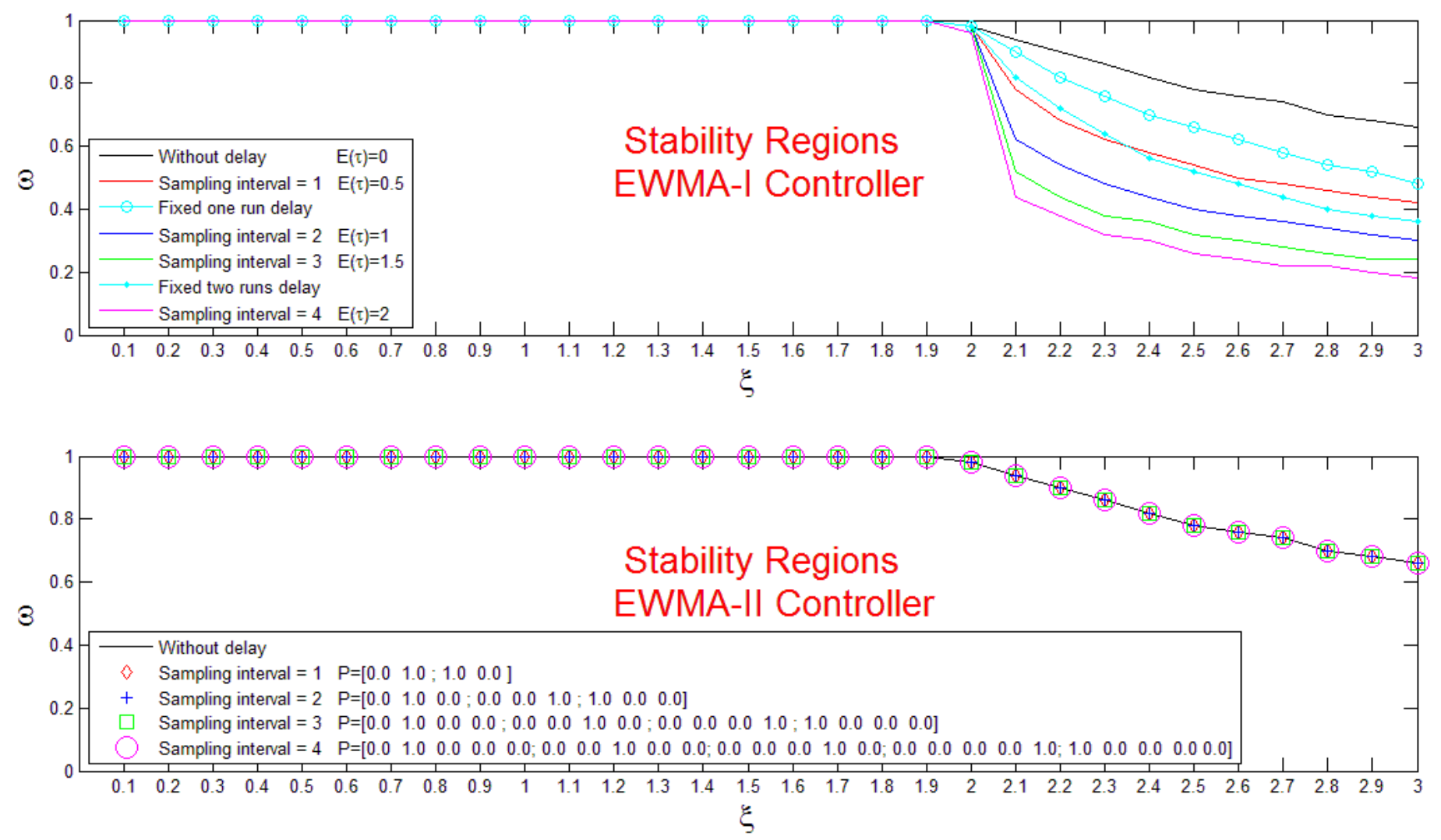

**Fig. 15.** Stability regions for delay systems with different sampling intervals

## 4.3. Stability Analysis for Systems with Stochastic Metrology Delay

In this subsection, we will discuss the stability of the system with EWMA-I and II controllers. We first assume that the transition probability matrix is unknown, but can be calculated from Poisson distribution (for other probability distributions the calculations are also straightforward); and then we will find out the stability regions for the system with this kind of metrology delay. The conclusion for stability of systems with EWMA-I and II controllers will be given in the last part of this subsection.

### 4.3.1. Transition Probability Matrix Calculated from Poisson distribution

In the real manufacturing process, we sometimes hardly do the measurement for each run, or there is no need to obtain the data of every run. We call these cases as the sampling metrology. In this subsection, we first give an example with the sampling metrology to verify ***Theorem 2***. Then we will discuss how to make an approximation of the average delay of the system, i.e., how to choose a proper $\tau_p$, to accelerate the computation. And the stability analysis will be done on the system with transition probability matrix calculated from Poisson distribution.

We assume that the metrology delay of the real manufacturing process follows Poisson distribution with mean parameter $\lambda = 1$, and the probability of the current run not being measured is $p_{NM}$ which is chosen from 0 to 0.9 in 0.1 increments. Based on the calculation process discussed in **Table 1**, 50000 Poisson random numbers are generated, as the original measurement delays of the system, to get the resampled delay numbers. The simulations are done 50 times for each $p_{NM}$ to obtain the transition probability matrix $P$ and the average delay of the system $E(\tau)$.

During the simulations, $P$ is truncated into a $3 \times 3$ matrix in each calculation[2]. **Fig. 16** is the simulation results for $P$ and $E(\tau)$. Each element of transition probability matrix calculated by ***Theorem 2*** is denoted by $p_{Calculated_{ij}}$, and $p_{Observed_{ij}}$ is computed from the simulations. From the figure, we know the differences between $p_{Calculated_{ij}}$ and $p_{Observed_{ij}}$ are very small, less than 3%. Also the relationships between $E(\tau)_{Calculated}$, calculated from (48), and $E(\tau)_{Observed}$, computed by simulation, are linear with slope 1. These simulation results verify ***Theorem 2***.

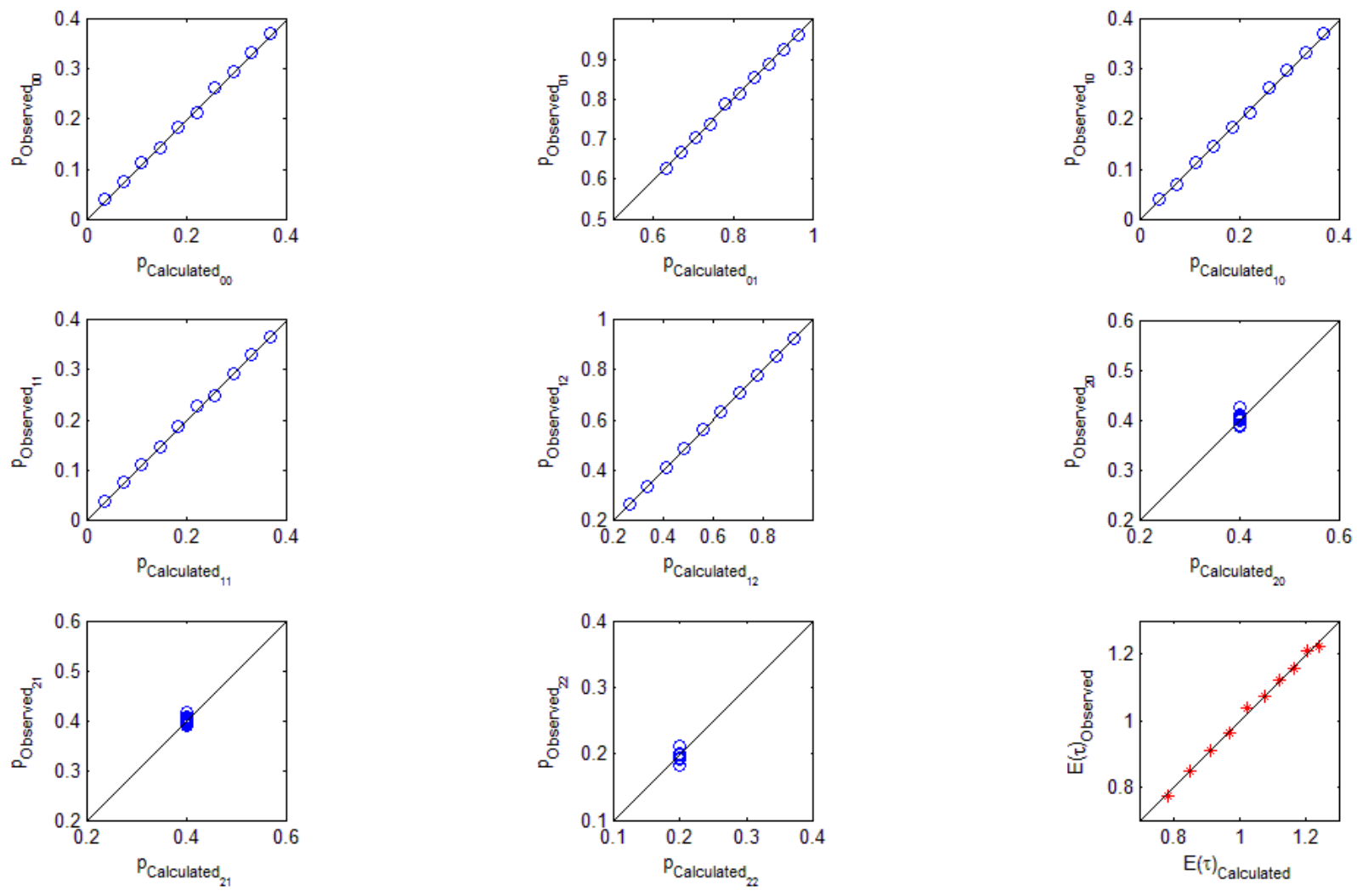

Fig. 16. Simulation test of $P$ and $E(\tau)$

**Fig. 17** shows the relationship between truncation of the transition probability matrix and the average delay of the system. The computation in this simulation is based on (32) and (48). From the figure, we know the minimum $\tau$, denoted as $\tau_{\min}$, is increasing with the increase of $p_{NM}$, to make the average delay of the system converge to its limitation. For example, if $p_{NM}=0$, i.e., every data of the system will be measured for feedback, $\tau_{\min}$ should be 4; and $\tau_{\min}=31$ for $p_{NM}=0.7$.

[2] The truncation will not affect the result, because both $P_{Calculated}$ and $P_{Observed}$ are normalized at the same time, hence $E(\tau)_{Calculated}$ and $E(\tau)_{Observed}$ are normalized at the same time.

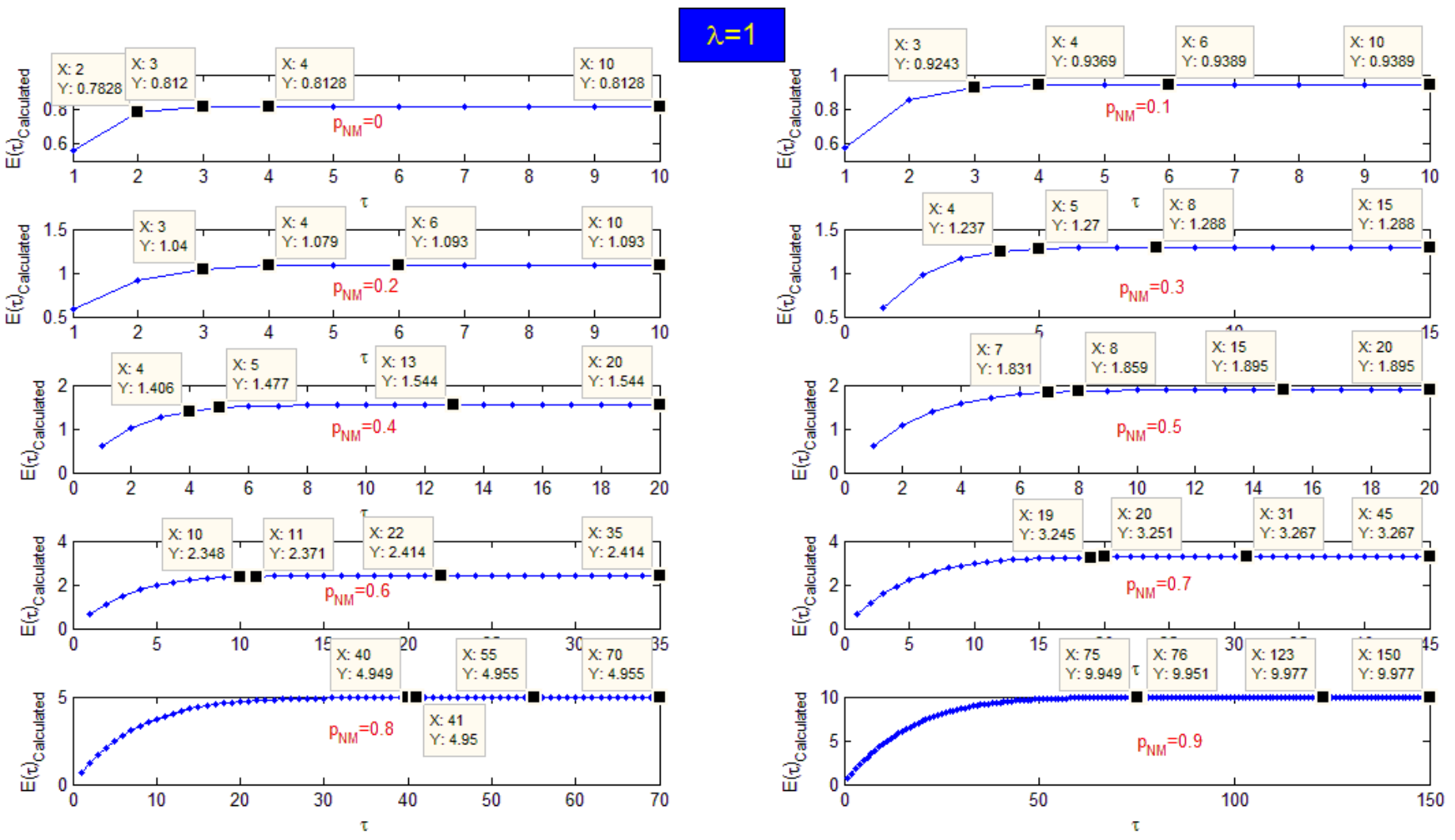

**Fig. 17.** Relationship between trancation of the transition probability matrix and average delay of the system

**Fig. 18** shows the relationship between $p_{NM}$ and the average delay of the system. It is clear that the relationship between $p_{NM}$ and $E(\tau)$ is nonlinear, and the nonlinearity becomes strong especially when $p_{NM}$ approaches to 1.

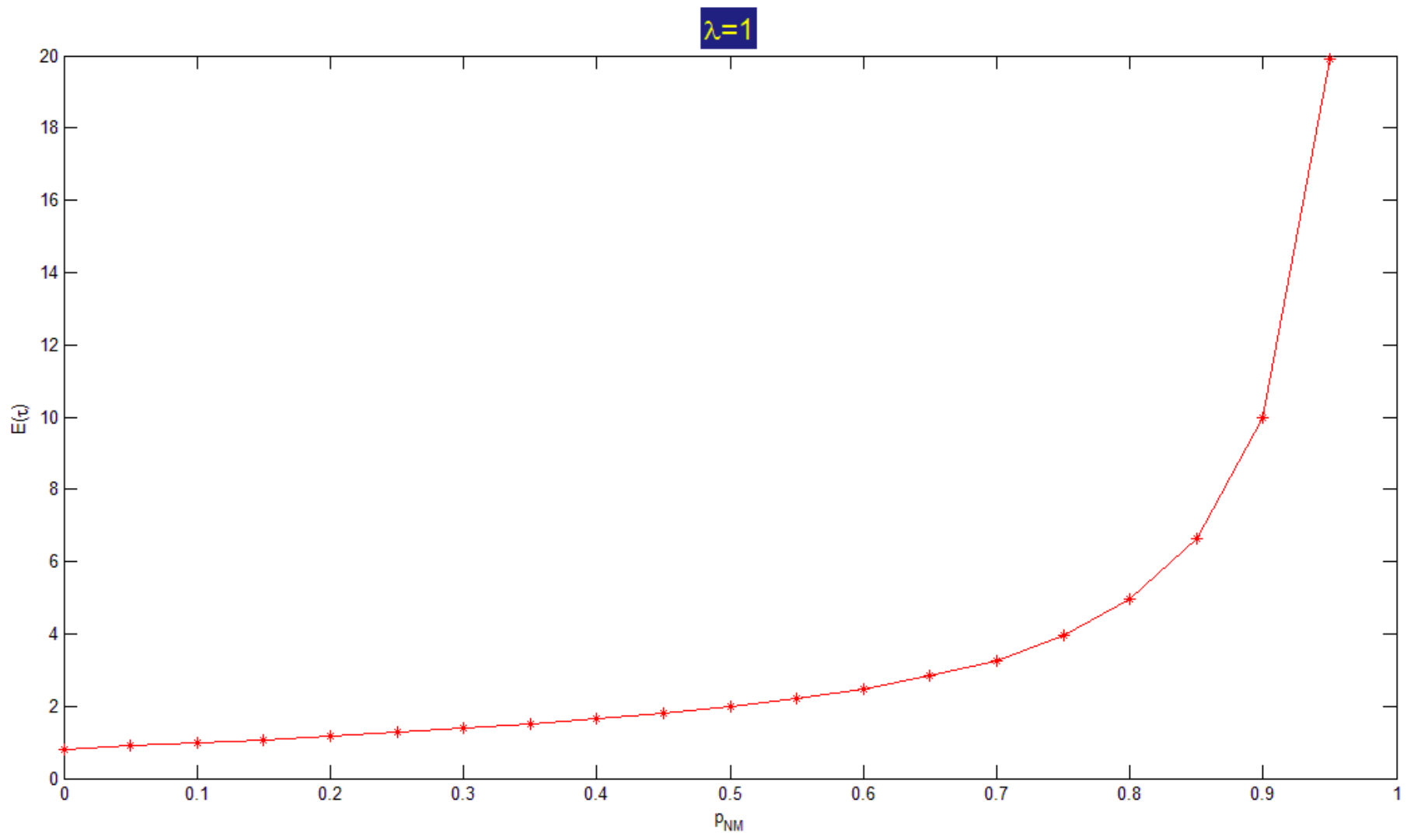

**Fig. 18.** Relationship between $p_{NM}$ and $E(\tau)$

***Remark 7:*** If $\tau_p$ is greater than 5, then it's time consuming (to solve LMIs) to get stability regions for the delay system (more than 13 hours for $\tau_p = 8$; more than 3 days for $\tau_p = 9$, and the LMIs are unsolvable if $\tau_p \geq 10$ by MATLAB), so we shall make some approximation on the average delay of the system to choose a reasonable average delay, $\tau_p$. **Table 2** gives the reasonable $\tau_p$ for each $p_{NM}$.

**Table 2:** The approximation of $E(\tau)_{Calculated}$ to $E(\tau)_{Approximated}$

| $p_{NM}$ | $(\tau_{\min}, E(\tau))$ | $E(\tau)_{Approximated}$ | $(\tau_p, E(\tau)_{Calculated})$ |
|---|---|---|---|
| 0 | (4, 0.8128) | 0.813 | (4, 0.8128) |
| 0.1 | (6, 0.9389) | 0.94 | (4, 0.9369) |
| 0.2 | (6, 1.093) | 1.1 | (4, 1.079) |
| 0.3 | (8, 1.288) | 1.3 | (5, 1.27) |
| 0.4 | (13, 1.544) | 1.5 | (5, 1.477) |
| 0.5 | (15, 1.895) | 1.9 | (8, 1.859) |
| 0.6 | (22, 2.414) | 2.4 | (11, 2.371) |
| 0.7 | (31, 3.267) | 3.3 | (20, 3.251) |
| 0.8 | (55, 4.955) | 5 | (41, 4.95) |
| 0.9 | (123, 9.977) | 10 | (76, 9.951) |

**Fig. 19** is the comparison figure of stability regions which are obtained by using $\tau_{\min}$ and $\tau_p$ as the maximum delay of the system. From the figure we can see that the stability regions got by using $\tau_{\min}$ and $\tau_p$ as the delays of the system are the same. So a proper $\tau_p$ can be used instead of $\tau_{\min}$ to accelerate the numerical computation.

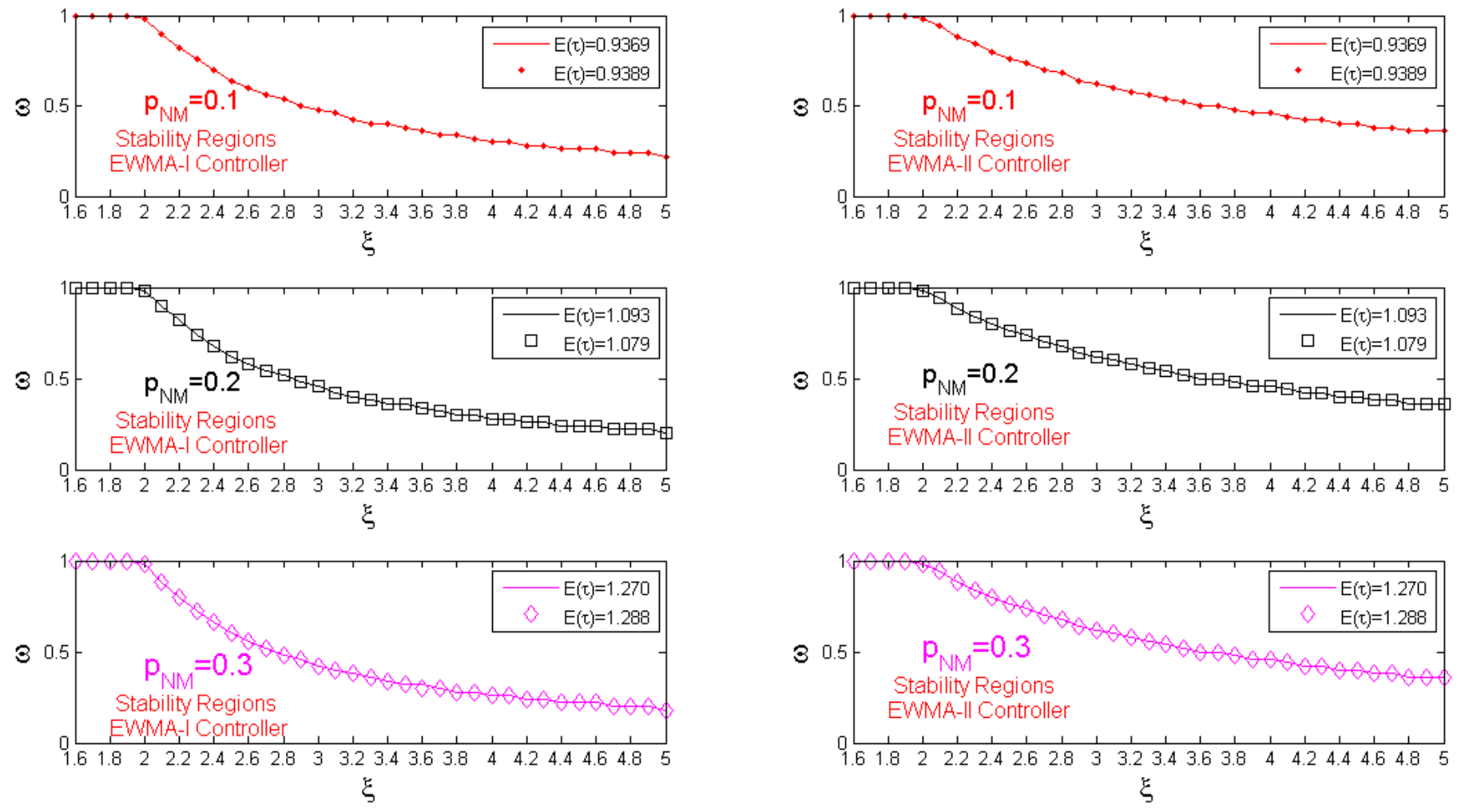

Fig. 19. Comparison of stability regions obtained by using $\tau_{\min}$ and $\tau_p$ as the maximum delay of the system

### 4.3.2. Stability Regions for Single Product Proce

**Fig. 20** shows the stability regions of stochastic metrology delay systems with EWMA-I and II controllers. Since the measurement delay of the system follows Poisson distribution with parameters $\lambda = 1$, $\eta_0 = 0.3679$, $\eta_1 = 0.3679$, $\eta_2 = 0.1809$, $\eta_3 = 0.0613$, $\eta_4 = 0.0153$, $\eta_5 = 0.0031$, $\eta_6 = 0.0005$, $\eta_7 = 0.0001$, and $\eta_k \approx 0$ for $k \geq 8$. The $p_{NM}$ is from 0 to 0.5 in 0.1 increment. The proper $\tau_p$ is chosen from **Table 2**, the transition probability matrix and the average delay of the system are calculated by (32) and (48) for each $p_{NM}$. For instance, if $p_{NM}=0$, then we should truncate the probability matrix $P$ into a $5\times 5$ (i.e., $\tau_p=4$) matrix

$$P=\begin{bmatrix} 0.3679 & 0.6321 & 0 & 0 & 0 \\ 0.3679 & 0.3679 & 0.2642 & 0 & 0 \\ 0.3679 & 0.3679 & 0.1839 & 0.0803 & 0 \\ 0.3679 & 0.3679 & 0.1839 & 0.0613 & 0.0190 \\ 0.3679 & 0.3679 & 0.1839 & 0.0613 & 0.0153 \end{bmatrix}$$

to obtain the convergent average delay, $E(\tau)=0.8128$, of the system; if $p_{NM}=0.3$, $P$ should be a matrix of $6\times 6$ (i.e., $\tau_p=5$),

$$P=\begin{bmatrix} 0.2575 & 0.7425 & 0 & 0 & 0 & 0 \\ 0.2575 & 0.2575 & 0.4850 & 0 & 0 & 0 \\ 0.2575 & 0.2575 & 0.1288 & 0.3562 & 0 & 0 \\ 0.2575 & 0.2575 & 0.1288 & 0.0429 & 0.3133 & 0 \\ 0.2575 & 0.2575 & 0.1288 & 0.0429 & 0.0107 & 0.3026 \\ 0.2575 & 0.2575 & 0.1288 & 0.0429 & 0.0107 & 0.0021 \end{bmatrix}$$

and $E(\tau)=1.2695$.

In **Fig. 20(a)**, i.e., the system with EWMA-I controller, it is clear that with the increase of $p_{NM}$, the stability regions decrease. For the stochastic case, the increasing of $E(\tau)$ will deteriorate the stability region. However, it is sometimes wrong to use $\text{ceil}(E(\tau))$ as the delay of the system to get the stability region, which can be seen from the stability regions obtained when $p_{NM}=0.1$ or 0.5.

For the system with EWMA-II controller, the simulation results in **Fig. 20(b)**, play tricks with our intuition: First, the increasing of $p_{NM}$ (or $E(\tau)$) has minor effects on the stability of the system; second, the stability regions lie between those without delay and those with fixed one run delay despite of what the $p_{NM}$ (or $E(\tau)$) is, in other words, $p_{NM}$ (or $E(\tau)$) conveys little information about stability of the system with EWMA-II controller. Using $\text{ceil}(E(\tau))$ as the delay of the system to get the stability regions will lead to conservative stability regions for the system, especially when $E(\tau)$ has a large value.

Comparing **Fig. 20(a)** and **(b)**, we can also note that the system with EWMA-I or II controller is stable for any transition probability matrices ( $p_{NM}$ or $E(\tau)$) if $\xi < 2$, when $\omega$ takes value between 0 to 1. Also, for the same stochastic metrology delay, the system with EWMA-II controller has a bigger stability region compared with the system with EWMA-I controller.

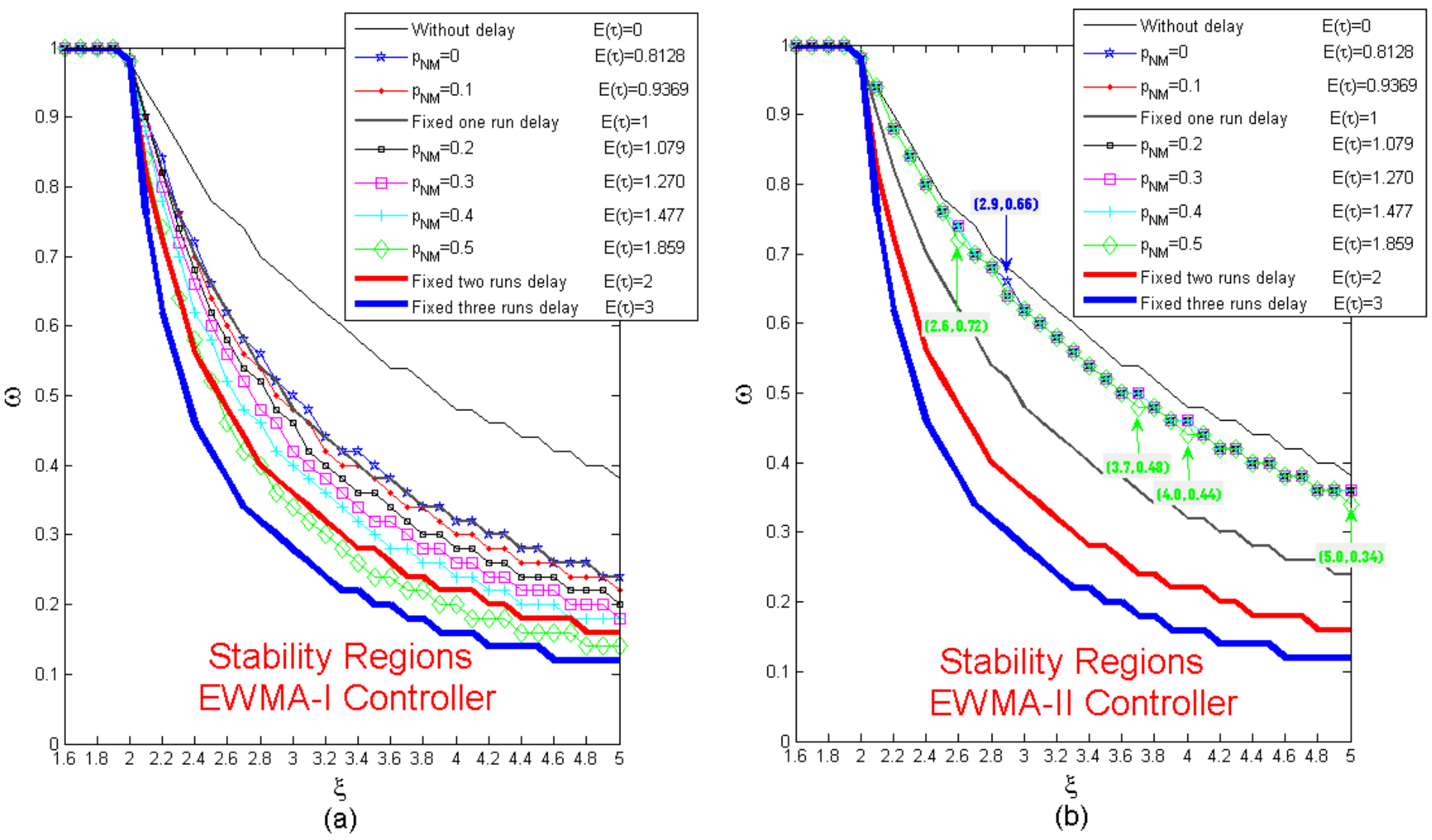

**Fig. 20.** Stability regions for systems with *P* calculated from Poisson distribution

From **Table 2**, it is clear that although the approximations on $E(\tau)$ are made, we still cannot get the stability regions for systems when $p_{NM} \geq 0.6$ because the LMIs are unsolvable by Matlab for large $\tau_p$. However, we can affirm that the stability regions for systems with EWMA-II controller lie between the stability regions of delay free system and those of the system with fixed one run delay. For the system with EWMA-I controller, the stability region may lies around the stability region obtained by using $\text{ceil}(E(\tau))$ as the fixed delay.

As an example to illustrate our affirmation, **Fig. 21** gives the responses of state variables of both controllers under the same stochastic metrology delay which follows Poisson distribution with $\lambda=1$ , $p_{NM} = 0.9$ and the corresponding

$E(\tau)_{Approximated} = 10$.

**Fig. 21(a)** and **Fig. 21(b)** are the responses of state variables of system with EWMA-I controller under different $(\xi,\omega)$: in **Fig. 21(a)**, $(\xi,\omega)$ is chosen as $(2.6, 0.34)$ under which the system with fixed nine runs delay is unstable, and from the same figure, we can see that the state variable is divergent; while in **Fig. 21(b)**, $(\xi,\omega)$ is $(2.6, 0.14)$ which guarantees the stability of the system with fixed ten runs delay, also from the same figure, we notice that the state variable converges to 0. And the results obtained from those two figures convince us that we can use the stability region of fixed ten runs delay system to approximate the stability region of the system with EWMA-I controller under this kind of stochastic metrology delay.

**Fig. 21(c)** and **Fig. 21(d)** are the responses of state variables of system with EWMA-II controller with different $(\xi,\omega)$: The $(\xi,\omega)$ in **Fig. 21(c)** is $(2.6, 0.78)$ which will cause the delay free system to be unstable, and therefore if this pair of $(\xi,\omega)$ is used for the system with EWMA-II controller, the system is unstable by all means (the state variable divergent, see **Fig. 21(c)**); in **Fig. 21(d)**, $(\xi,\omega)$ is chosen such that the system with fixed one run metrology delay is stable, and from the same figure we note that the state variable of the system with EWMA-II controller is convergent. Also the combination of those two figures tells us that if the delay follows Poisson distribution, no matter what $p_{NM}$ is, the stability region of the system with EWMA-II controller always falls between the stability region of delay free system and the stability region of the system with fixed one run delay.

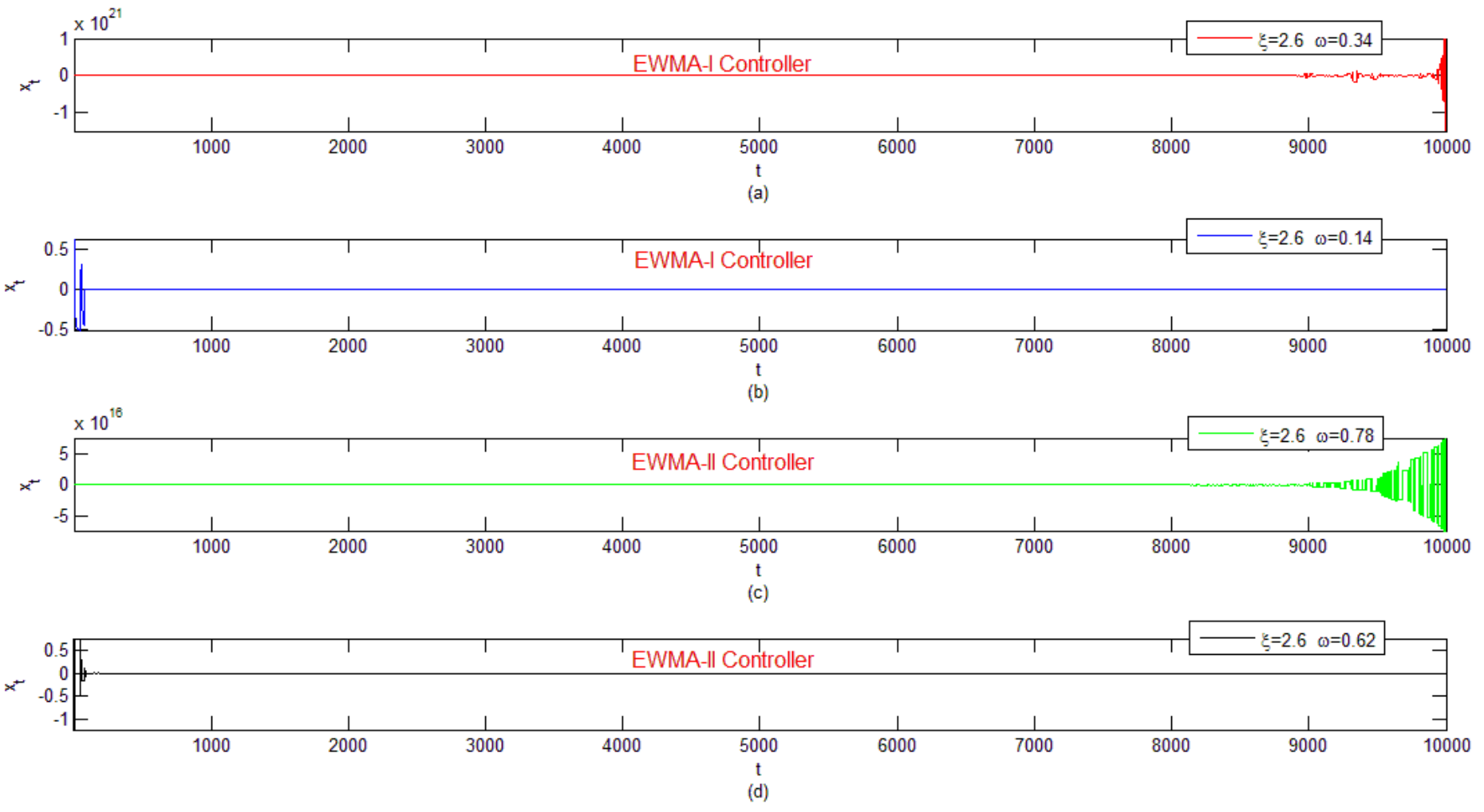

Fig. 21. Responses of state variables when $p_{NM}=0.9$

It is worth mentioning that for other kinds of metrology delay, the stability region can be got by the theorems proposed in Section 2.

***Remark 8:*** The simulation results in **Fig. 20(b)** and **Fig. 21(d)** tell us that if the transition probability matrix of the system is calculated from the Poisson distribution, then the stability regions of the system with EWMA-II controller lie between the regions of systems without delay and those with fixed one run metrology delay. However, we cannot affirm that the stability region of system with EWMA-II controller, for any transition probability matrix, lies between the regions of delay free system and those of fixed one run delay system. The following examples are cases in point to illustrate this affirmation.

**Fig. 22** is the simulation results of the stability regions for systems with EWMA-I and II controllers with different transition probability matrices which are randomly chosen as $P=\begin{bmatrix}0.8 & 0.2 & 0.0\\ 0.5 & 0.3 & 0.2\\ 0.6 & 0.3 & 0.1\end{bmatrix}$, $P=\begin{bmatrix}0.2 & 0.8 & 0.0\\ 0.2 & 0.3 & 0.5\\ 0.1 & 0.3 & 0.6\end{bmatrix}$, $P=\begin{bmatrix}0.1 & 0.9 & 0.0\\ 0.1 & 0.2 & 0.7\\ 0.1 & 0.2 & 0.7\end{bmatrix}$, and $P=\begin{bmatrix}0.1 & 0.9 & 0.0\\ 0.1 & 0.1 & 0.8\\ 0.0 & 0.2 & 0.8\end{bmatrix}$; the corresponding average delays are $E(\tau)=0.3291$, $E(\tau)=1.3176$, $E(\tau)=1.5300$, and $E(\tau)=1.7609$ respectively. From **Fig. 22(a)**, it is obvious that with the increase of average delay, $E(\tau)$, the stability regions of the system with EWMA-I controller are decreasing. While this result holds for systems with EWMA-II controller only provided that the systems are subject to stochastic metrology delay. Using $\text{ceil}(E(\tau))$ as the delay of the system will lead to conservative stability regions for both controllers, but it is much more conservative to do this for EWMA-II controller.

For the same transition probability matrix (the same average delay), the stability regions for the systems with EWMA-I and II controllers are compared in **Fig. 23**. From the figure, we can find that the system with EWMA-II controller has a bigger stability region compared with the system with EWMA-I controller for the same transition probability matrix. Also the system with either controller is guaranteed stable for $\xi < 2$ and $0 < \omega \le 1$.

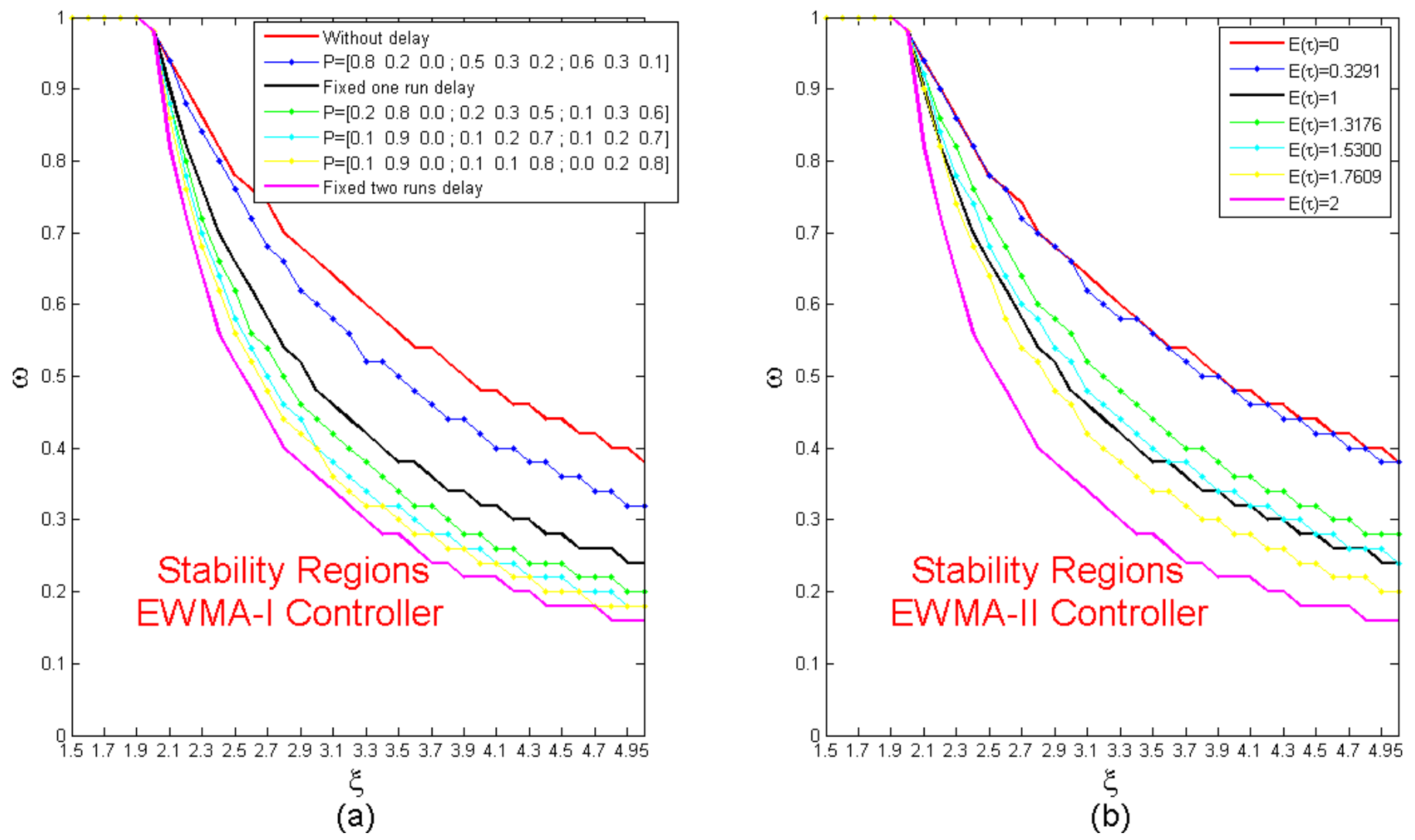

**Fig. 22.** Stability regions for systems with different transition probability matrix

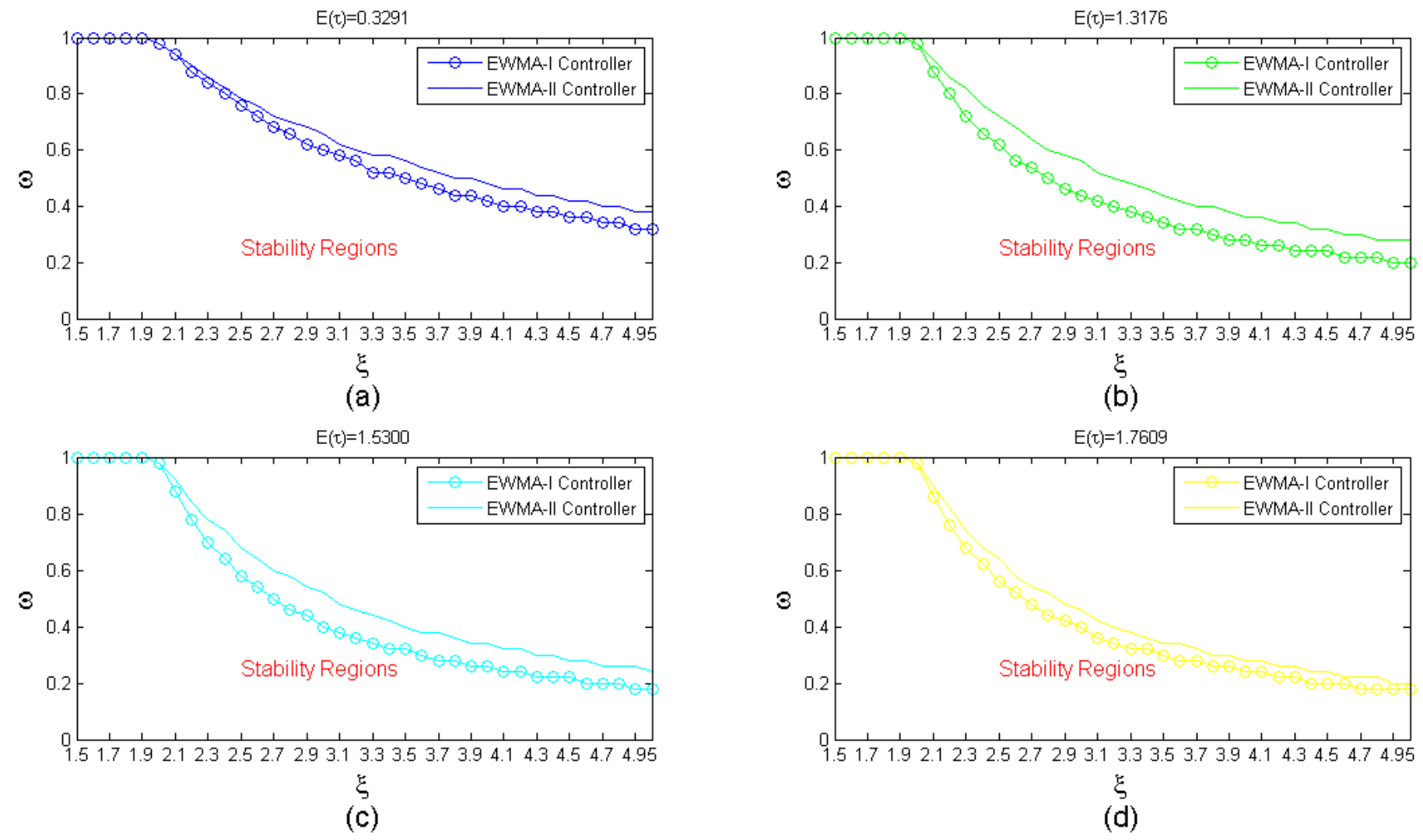

**Fig. 23.** Comparison of stability regions for systems with EWMA-I and II controllers with different transition probability matrix

### 4.3.3. Conclusions for Systems with Stochastic Metrology Delay

Comparing the results obtained in subsection 4.3.2, we can infer the following conclusions:

1. For the system with stochastic metrology delay, we have the delayed data of the system, and these delayed data deteriorate the stability of the system.
2. For stochastic case, the stability region of the system with EWMA-I controller is decreasing with the increase of average delay; however one cannot always expect to use average delay as a benchmark to determine the stability region of the system with EWMA-II controller.
3. Using $\text{ceil}(E(\tau))$ as the delay of the system will lead to conservative stability regions for systems with EWMA-II controller, but it is sometimes wrong to do this for systems with EWMA-I controller.
4. The system with either controller is stable for any $\omega$ between 0 to 1 if $\xi < 2$.

# 5. EWMA Run-to-Run Controllers in Mixed Product Process

What we have discussed in the previous Sections are for the single product process. However, it is a common practice in today's semiconductor manufacturing to have many different products and processes run on each processing tool, i.e., high-mix manufacturing. The engineers are interested in the stabilities of each kind of product on the specific tool, i.e., mixed product process.

In this section, we will extend the theorems obtained for single product process to analyze mixed product process. We first establish the system model for mixed product process, and then we will discuss the transition probability matrix for the same kind of product. The numerical examples for this kind of process are provided in the last part of this section.

## 5.1. System Model

Suppose that a number of products are randomly manufactured on a specific tool. Taking product $N$ as an example, the run number of product $N$ is denoted as $n$ and the corresponding run number of the total process is expressed as $t_n$. Fig. 24 gives an example of mixed product process.

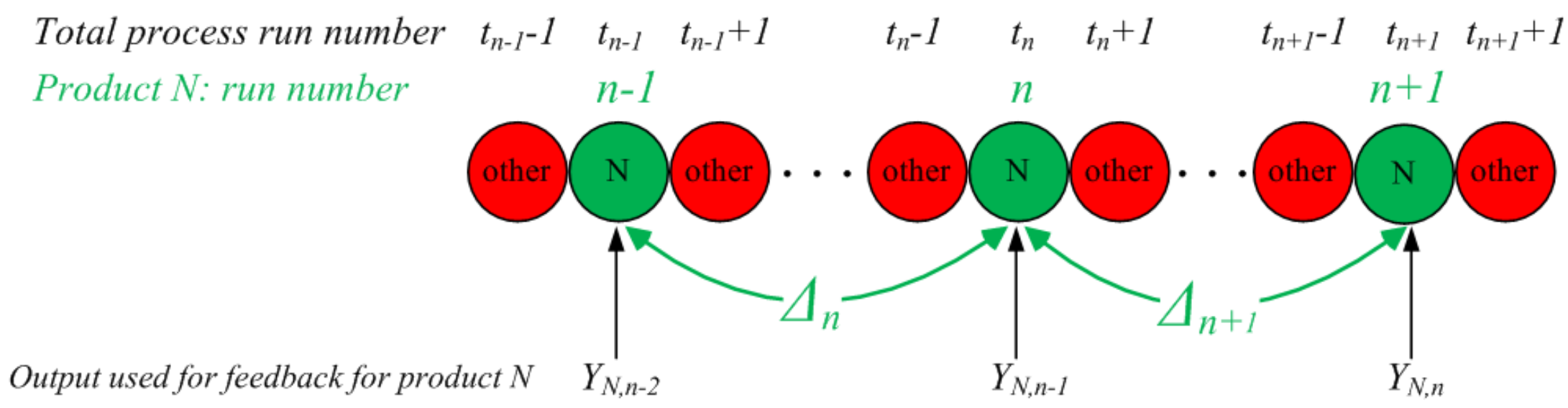

**Fig. 24.** An example of mixed product process

Assume that the input-output relationship for product $N$ on the specific tool is linear with slope $\beta_N$, then the process model is:

$$Y_{t_n} = \beta_N u_{t_n} + a_{t_n} \tag{49}$$

Or

$$Y_{N,n} = \beta_N u_{N,n} + a_{N,n} \tag{50}$$

where $Y_{t_n} = Y_{N,n}$ is the output of product $N$ at the $t_n{}^{th}$ production run, and $a_{t_n} = a_{N,n}$ is the instantaneous disturbance at run $t_n$.

The predicted model for product $N$ is

$$\hat{Y}_{t_n} = b_N u_{t_n} + \hat{a}_{t_n} \tag{51}$$

Or

$$\hat{Y}_{N,n} = b_N u_{N,n} + \hat{a}_{N,n} \tag{52}$$

where $b_N$ and $\hat{a}_{t_n}$ (or $\hat{a}_{N,n}$) are model gain and offset parameters estimated for product $N$.

We also suppose that there is a stochastic metrology delay $\tau_{t_n} = \tau_{N,n} \in S$ at run $t_n$, two kinds of EWMA run-to-run controllers, EWMA-I and II controllers, can be used to estimate the disturbance:

For EWMA-I controller, the control action is also chosen as

$$u_{t_n} = \frac{T_N - \hat{a}_{t_n}}{b_N} \quad or \quad u_{N,n} = \frac{T_N - \hat{a}_{N,n}}{b_N} \tag{53}$$

where $T_N$ is the target for product $N$. And the EWMA filter with discount factor $\omega_N$ for product $N$ is

$$\begin{gathered} \hat{a}_{t_{n+1}} = \omega_N (Y_{t_n - \tau_{t_n}} - b_N u_{t_n - \tau_{t_n}}) + (1 - \omega_N) \hat{a}_{t_n - \tau_{t_n}} \\ or \\ \hat{a}_{N,n+1} = \omega_N (Y_{N,n\text{-}\tau_{N,n}} - b_N u_{N,n}) + (1 - \omega_N) \hat{a}_{N,n\text{-}\tau_{N,n}} \end{gathered} \tag{54}$$

Combining (51)-(54), we rewrite EWMA-I controller in the form of (55),

$$\begin{gathered} \hat{a}_{t_{n+1}} = (1 - \omega_N) \hat{a}_{t_n} + \omega_N (1 - \xi_N) \hat{a}_{t_n - \tau_{t_n}} \\ or \\ \hat{a}_{N,n+1} = (1 - \omega_N) \hat{a}_{N,n} + \omega_N (1 - \xi_N) \hat{a}_{N,n\text{-}\tau_{N,n}} \end{gathered} \tag{55}$$

where $\xi_N$ is the plant-model mismatch for product $N$.

For the same reason as in the single product process, in the mixed product process, EWMA-II controller should also be considered, i.e.,

$$\hat{a}_{t_{n+1}} = \begin{cases} (1-\omega_N)\hat{a}_{t_n} + \omega_N(1-\xi_N)\hat{a}_{t_n-\tau_{t_n}}, & if\ \tau_{t_{n-1}} \geq \tau_{t_n}; \\ \hat{a}_{t_n}, & otherwise. \end{cases}$$

$$or \tag{56}$$

$$\hat{a}_{N,n+1} = \begin{cases} (1-\omega_N)\hat{a}_{N,n} + \omega_N(1-\xi_N)\hat{a}_{N,n-\tau_{N,n}}, & if\ \tau_{N,n\text{-}1} \geq \tau_{N,n}; \\ \hat{a}_{N,n}, & otherwise. \end{cases}$$

Augmenting the state variables at run $t_n$, as what we have done in subsection 2.3, the delay system in (55) or (56) can be changed into a delay free system

$$\boldsymbol{X}_{t_{n+1}} = \Xi_N(\tau_{t_{n-1}}, \tau_{t_n})\boldsymbol{X}_{t_n}$$

$$or \tag{57}$$

$$\boldsymbol{X}_{N,n+1} = \Xi_N(\tau_{N,n\text{-}1}, \tau_{N,n})\boldsymbol{X}_{N,n}$$

where $\Xi_N(\tau_{t_{n-1}}, \tau_{t_n})$ and $\Xi_N(\tau_{N,n\text{-}1}, \tau_{N,n})$ are determined by which kind of EWMA controller the system has adopted. If we do the same analysis as that of subsection 2.3.2, $\Xi_N(\tau_{t_{n-1}}, \tau_{t_n})$ and $\Xi_N(\tau_{N,n\text{-}1}, \tau_{N,n})$ are easy to write out. The stability problem of (57) can also be solved by ***Theorem 1*** in subsection 2.3.2.

***Remark 9:*** If only the same products, say product $N$, are manufactured on the same tool, then $t_n = n = t$, $t_{n+1} = n+1 = t+1$, and the model of product $N$ in mixed product process becomes the same as that of single product process.

It is worth pointing out that similar analysis can be done for other products.

## 5.2. Transition Probability Matrix

In order to get the stability regions for mixed product process, we have to know the transition probability matrix of the metrology delay for each kind of product.

We also take product $N$ for example. The probability of producing this product in the total process is $q_N$. For the total process, we use $\eta_i$ as the probability of the data at time $t$ being delayed by $i$ runs and the probability of current run not being measured is $p_{NM}$. The distance between current run, $t_n$, and the last run, $t_{n-1}$, for product $N$, is $\Delta_n = t_n - t_{n-1}$. Then for any production run, the probability of the distance between adjacent product $N$ being $r$ is

$$P(\Delta = r) = (1-q_N)^{r-1} q_N \tag{58}$$

Define $\eta_i'$ as the probability that the time delay of the current run for product $N$ is $i$, $\eta_i' = P(\tau_n = i)$, which can be estimated by the following theorem:

***Theorem 3:*** For product $N$, the probability that the data of current run being delayed by $i$ runs is

$$\eta_i' = \sum_{k=i}^{\infty} \eta_k \cdot b(k-1, i-1, q_N) \tag{59}$$

for $i \geq 1$, and $\eta_0' = \eta_0$ .Where $b(k,i,q_N) = C_{k-1}^{i-1}(1-q_N)^{k-i} \cdot q_N{}^{i-1}$ is the binomial probability of producing $i-1$ product $N$ in $k-1$ runs.

***Proof:***

(A) If the current run, run $n$, is measured, then

$$\eta_0' = p(\tau_n = 0) = \eta_0 \tag{60}$$

(B) If the measurement delay of run $n$ is 1, then

$$\begin{aligned}
\eta_1' &= p(\tau_n = 1) \\
&= \eta_1 \cdot P(\Delta_{n+1} \geq 1) + \eta_2 \cdot P(\Delta_{n+1} \geq 2) + \cdots + \eta_d \cdot P(\Delta_{n+1} \geq d) + \cdots \\
&= \sum_{k=1}^{\infty} \eta_k \cdot P(\Delta_{n+1} \geq k) \\
&= \sum_{k=1}^{\infty} \eta_k \cdot \left( \sum_{l=k}^{\infty} P(\Delta_{n+1} = l) \right) \\
&= \sum_{k=1}^{\infty} \eta_k \cdot \left( \sum_{l=k}^{\infty} (1-q_N)^{l-1} \cdot q_N \right) \\
&= \sum_{k=1}^{\infty} \eta_k \cdot (1-q_N)^{k-1} \\
&\triangleq \sum_{k=1}^{\infty} \eta_k \cdot b(k-1, 0, q_N)
\end{aligned} \tag{61}$$

(C) If the measurement delay of run $n$ is 2, then we have

$$
\begin{aligned}
\eta_2' &= p\left(\tau_n = 2\right) \\
&= \eta_2 \cdot P\left(\Delta_{n+1} < 2 \cap \Delta_{n+1} + \Delta_{n+2} \geq 2\right) + \eta_3 \cdot P\left(\Delta_{n+1} < 3 \cap \Delta_{n+1} + \Delta_{n+2} \geq 3\right) \\
&\quad + \cdots + \eta_d \cdot P\left(\Delta_{n+1} < d \cap \Delta_{n+1} + \Delta_{n+2} \geq d\right) + \cdots \\
&= \sum_{k=2}^{\infty} \eta_k \cdot P\left(\Delta_{n+1} < k \cap \Delta_{n+1} + \Delta_{n+2} \geq k\right) \\
&= \sum_{k=2}^{\infty} \eta_k \cdot \left( \sum_{l=k}^{\infty} P\left(\Delta_{n+1} < l \cap \Delta_{n+1} + \Delta_{n+2} = l\right) \right) \\
&= \sum_{k=2}^{\infty} \eta_k \cdot \sum_{l=k}^{\infty} C_{k-1}^{1} \left(1 - q_N\right)^{k-2} \cdot q_N \cdot \left(1 - q_N\right)^{l-k} \cdot q_N \\
&= \sum_{k=2}^{\infty} \eta_k \cdot \left( \sum_{l=k}^{\infty} C_{k-1}^{1} \left(1 - q_N\right)^{k-2} \cdot q_N \cdot \left(1 - q_N\right)^{l-k} \cdot q_N \right) \\
&= \sum_{k=2}^{\infty} \eta_k \cdot \left( C_{k-1}^{1} \left(1 - q_N\right)^{k-2} \cdot q_N \right) \\
&\triangleq \sum_{k=2}^{\infty} \eta_k \cdot b\left(k-1, 1, q_N\right)
\end{aligned}
\tag{62}
$$

(D) If the measurement delay of run $n$ is 3, then we have

$$
\begin{aligned}
\eta_3' &= p\left(\tau_n = 3\right) \\
&= \eta_3 \cdot P\left(\Delta_{n+1} + \Delta_{n+2} < 3 \cap \Delta_{n+1} + \Delta_{n+2} + \Delta_{n+3} \geq 3\right) \\
&\quad + \eta_4 \cdot P\left(\Delta_{n+1} + \Delta_{n+2} < 4 \cap \Delta_{n+1} + \Delta_{n+2} + \Delta_{n+3} \geq 4\right) \\
&\quad + \cdots + \eta_d \cdot P\left(\Delta_{n+1} + \Delta_{n+2} < d \cap \Delta_{n+1} + \Delta_{n+2} + \Delta_{n+3} \geq d\right) + \cdots \\
&= \sum_{k=3}^{\infty} \eta_k \cdot P\left(\Delta_{n+1} + \Delta_{n+2} < k \cap \Delta_{n+1} + \Delta_{n+2} + \Delta_{n+3} \geq k\right) \\
&= \sum_{k=3}^{\infty} \eta_k \cdot \left( \sum_{l=k}^{\infty} P\left(\Delta_{n+1} + \Delta_{n+2} < l \cap \Delta_{n+1} + \Delta_{n+2} + \Delta_{n+3} = l\right) \right) \\
&= \sum_{k=3}^{\infty} \eta_k \cdot \left( \sum_{l=k}^{\infty} C_{k-1}^{2} \left(1 - q_N\right)^{k-3} \cdot q_N^2 \cdot \left(1 - q_N\right)^{l-k} \cdot q_N \right) \\
&= \sum_{k=3}^{\infty} \eta_k \cdot \left( C_{k-1}^{2} \left(1 - q_N\right)^{k-3} \cdot q_N^2 \right) \\
&\triangleq \sum_{k=3}^{\infty} \eta_k \cdot b\left(k-1, 2, q_N\right)
\end{aligned}
\tag{63}
$$

(E) If the measurement delay of run $n$ is $i$, then

$$
\begin{aligned}
\eta_i' &= p(\tau_n = i) \\
&= \eta_i \cdot P(\Delta_{n+1} + \Delta_{n+2} + \cdots + \Delta_{n+i-1} < i \cap \Delta_{n+1} + \Delta_{n+2} + \cdots + \Delta_{n+i-1} + \Delta_{n+i} \geq i) \\
&\quad + \eta_{i+1} \cdot P(\Delta_{n+1} + \Delta_{n+2} + \cdots + \Delta_{n+i-1} < i+1 \cap \Delta_{n+1} + \Delta_{n+2} + \cdots + \Delta_{n+i-1} + \Delta_{n+i} \geq i+1) \\
&\quad + \cdots \\
&\quad + \eta_d \cdot P(\Delta_{n+1} + \Delta_{n+2} + \cdots + \Delta_{n+i-1} < d \cap \Delta_{n+1} + \Delta_{n+2} + \cdots + \Delta_{n+i-1} + \Delta_{n+i} \geq d) + \cdots \\
&= \sum_{k=i}^{\infty} \eta_k \cdot P(\Delta_{n+1} + \Delta_{n+2} + \cdots + \Delta_{n+i-1} < k \cap \Delta_{n+1} + \Delta_{n+2} + \cdots + \Delta_{n+i-1} + \Delta_{n+i} \geq k) \\
&= \sum_{k=i}^{\infty} \eta_k \cdot \left( \sum_{l=k}^{\infty} P(\Delta_{n+1} + \Delta_{n+2} + \cdots + \Delta_{n+i-1} < l \cap \Delta_{n+1} + \Delta_{n+2} + \cdots + \Delta_{n+i-1} + \Delta_{n+i} = l) \right) \\
&= \sum_{k=i}^{\infty} \eta_k \cdot \left( \sum_{l=k}^{\infty} C_{k-1}^{i-1} (1-q_N)^{k-i} \cdot q_N^{i-1} \cdot (1-q_N)^{l-k} \cdot q_N \right) \\
&= \sum_{k=2}^{\infty} \eta_k \cdot \left( C_{k-1}^{i-1} (1-q_N)^{k-i} \cdot q_N^{i-1} \right) \\
&\triangleq \sum_{k=i}^{\infty} \eta_k \cdot b(k-1, i-1, q_N)
\end{aligned}
\tag{64}
$$

And these constitute the proof.■

***Remark 10***: If $q_N = 1$, the tool only produce product $N$, i.e., the single product process; then from (59), we know $\eta_i' = \eta_i$.

***Remark 11:*** The probabilities of the current run being delayed or without delay can be normalized:

$$
\begin{aligned}
\sum_{i=0}^{\infty} \eta_i' &= \eta_0 + \sum_{i=1}^{\infty} \eta_i' = \eta_0 + \sum_{i=1}^{\infty} \sum_{k=i}^{\infty} \eta_k \cdot b(k-1, i-1, q_N) \\
&= \eta_0 + \sum_{k=1}^{\infty} \eta_k \cdot b(k-1, 0, q_N) + \sum_{k=2}^{\infty} \eta_k \cdot b(k-1, 1, q_N) + \cdots + \sum_{k=d+1}^{\infty} \eta_k \cdot b(k-1, d, q_N) + \cdots \\
&= \eta_0 + \eta_1 \cdot b(0, 0, q_N) + \sum_{k=2}^{\infty} \eta_k \cdot b(k-1, 1, q_N) + \cdots + \sum_{k=d+1}^{\infty} \eta_k \cdot b(k-1, d, q_N) + \cdots \\
&= \eta_0 + \eta_1 + \cdots + \eta_{d+1} + \cdots \\
&= 1
\end{aligned}
$$

***Remark 12:*** For product $N$, since we have derived the probabilities of the metrology delay for different runs, i.e., $\eta_i'$, by the same analysis as we do in Section 3, the transition probability matrix of metrology delay can be easily obtained as

$$p_{ij} = p(\tau_{t+1} = j \mid \tau_t = i) = \begin{cases} 0, & i+1 < j; \\ p_{NM} + (1-p_{NM}) \cdot \sum_{k=i+1}^{\infty} \eta'_k, & j = i+1; \\ (1-p_{NM}) \cdot \eta'_j, & 0 \le j \le i. \end{cases} \quad (65)$$

***Remark 13:*** The average delay for product $N$, $E(\tau)_N$ can be calculated by combining (41), (42) and (65).

***Remark 14:*** ***Theorem 3*** can be applied to other products.

### 5.3. Numerical Examples

In this subsection, we will first take Poisson distribution as the measurement delay of products as an example to test the validity of the theorems we have gotten in Section 5 for mixed product process. How to choose a proper $\tau$ to get the convergent average delay will also be discussed. Then, we will establish the stability results for mixed product process with the measurement delay following Poisson distribution. It is worth mentioning that for other probability distributions, by using the theorems in this paper, the stability regions will be obtained by redoing the computations.

#### 5.3.1. Transition Probability Matrix Calculated from Poisson distribution

Suppose that two kinds of products, product $M$ and $N$, are randomly manufactured on the same tool with the probability $q_M = 0.3$ and $q_N = 0.7$. We assume that the metrology delay of the real manufacturing process follows Poisson distribution with mean parameter $\lambda = 1$, and the probability of the current run not be measured is $p_{NM}$ which is chosen from 0 to 0.9 in 0.1 increments. Based on the calculation process discussed in **Table 1** 100000 Poisson random numbers are generated, as the original measurement delays for both products, to get the resampled delay numbers. The simulations are done 20 times for each $p_{NM}$ to obtain the transition probability matrix, $P_M$ for product $M$ and $P_N$ for product $N$, as well as the average delay $E(\tau)_M$ for product $M$ and $E(\tau)_N$ for product $N$. During the simulations, $P_M$ and $P_N$ are truncated into $3\times 3$ matrices in each calculation. **Fig. 25** and **Fig. 26** are the simulated results for each element of $P_M$ and $P_N$, as well as $E(\tau)_M$ and $E(\tau)_N$. From the figure, it is clear that for the same product, both the transition probability matrix and the average delay of the system which are calculated from the simulations are close to those obtained from ***Theorem 2*** and ***Theorem 3***, and this verifies ***Theorem 2*** and ***Theorem 3***.

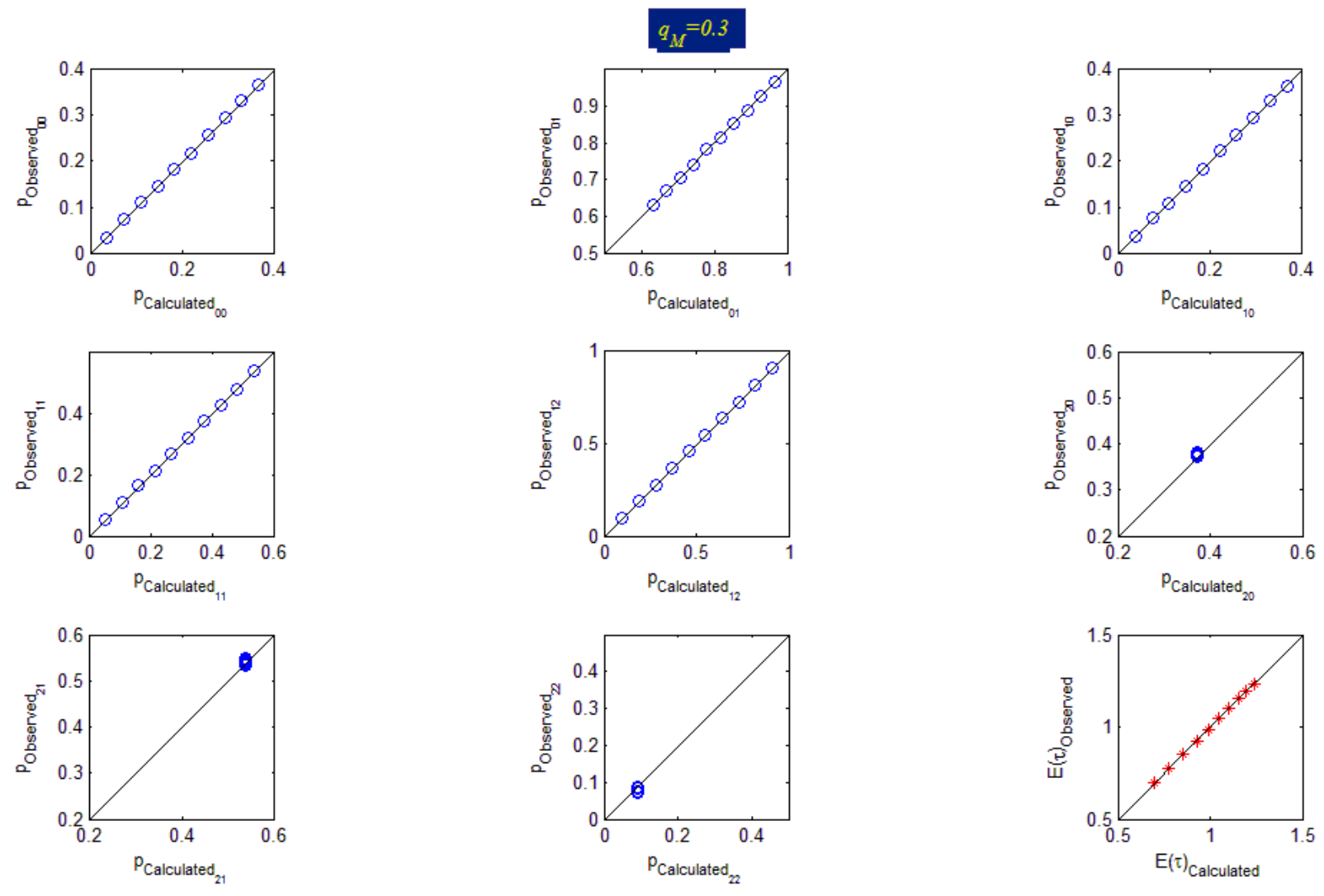

**Fig. 25.** Simulation test of $P_M$ and $E(\tau)_M$

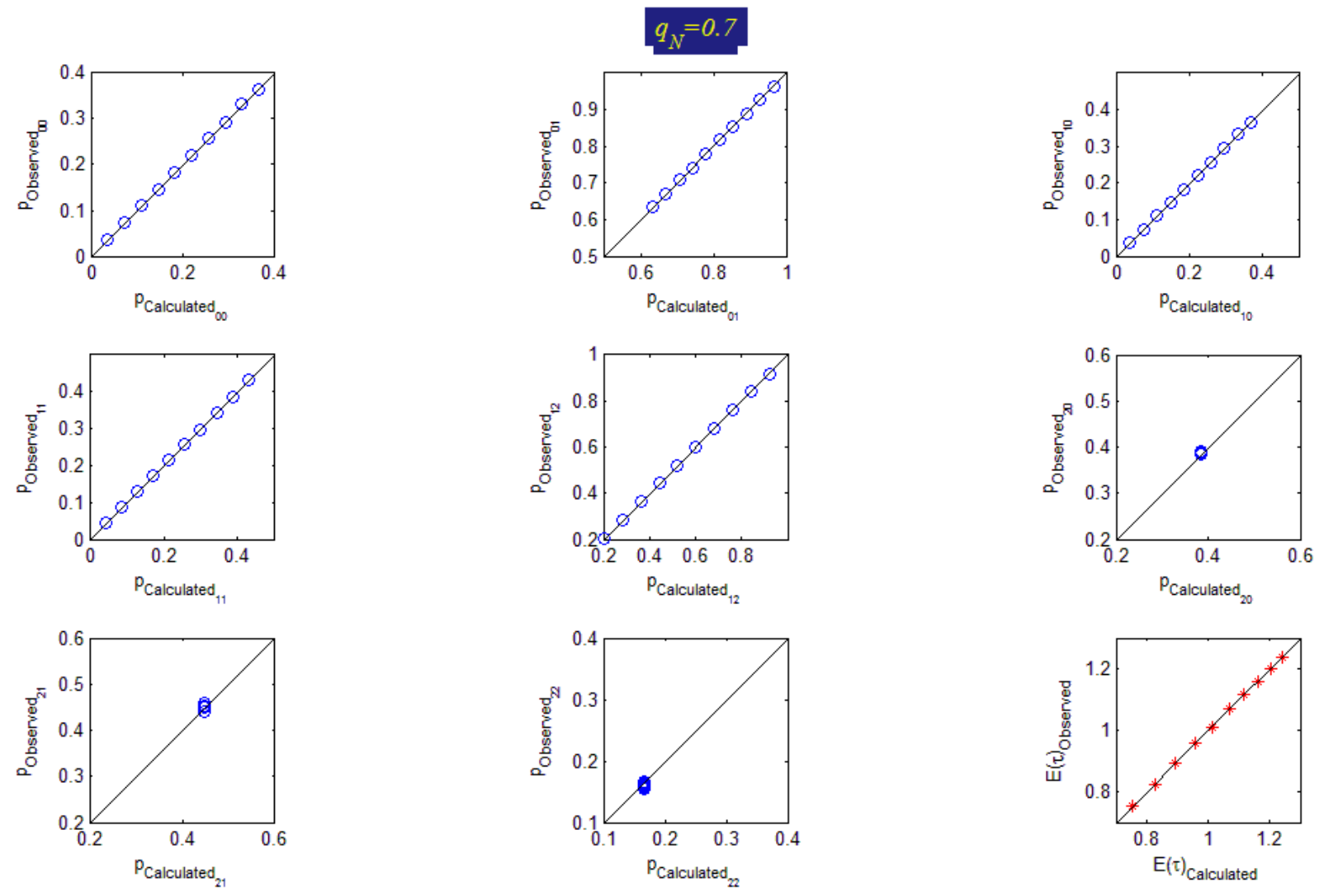

**Fig. 26.** Simulation test of $P_N$ and $E(\tau)_N$

**Fig. 27** and **Fig. 28** show the relationships between truncation of the transition probability matrix and the average delay of the system when $q_M = 0.3$ and $q_N = 0.7$ respectively. The computations in these simulations are based on (58), (59) coupled with (40)-(42). From the figure, we know that $\tau_{\min}$ is increasing with the increase of $p_{NM}$, to make the average delay of the system converge to its limitation. Also the same as the single product process, in the mixed product process, the relationship between $p_{NM}$ and the average delay of the system, $E(\tau)$, is nonlinear, and the nonlinearity becomes strong especially when $p_{NM}$ approaches to 1.

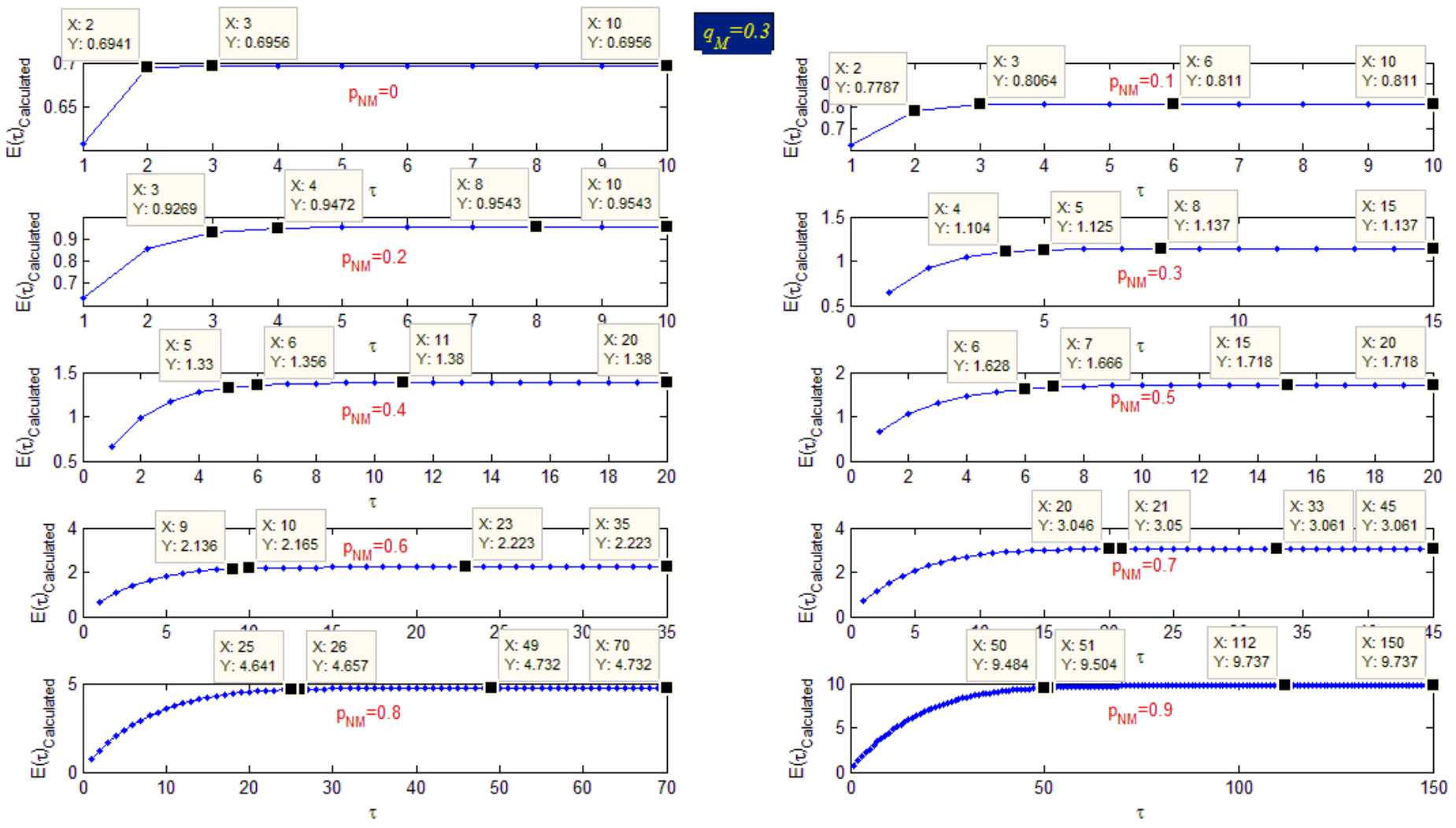

**Fig. 27.** Relationship between trancation of the transition probability matrix and average delay of the system when $q_M = 0.3$

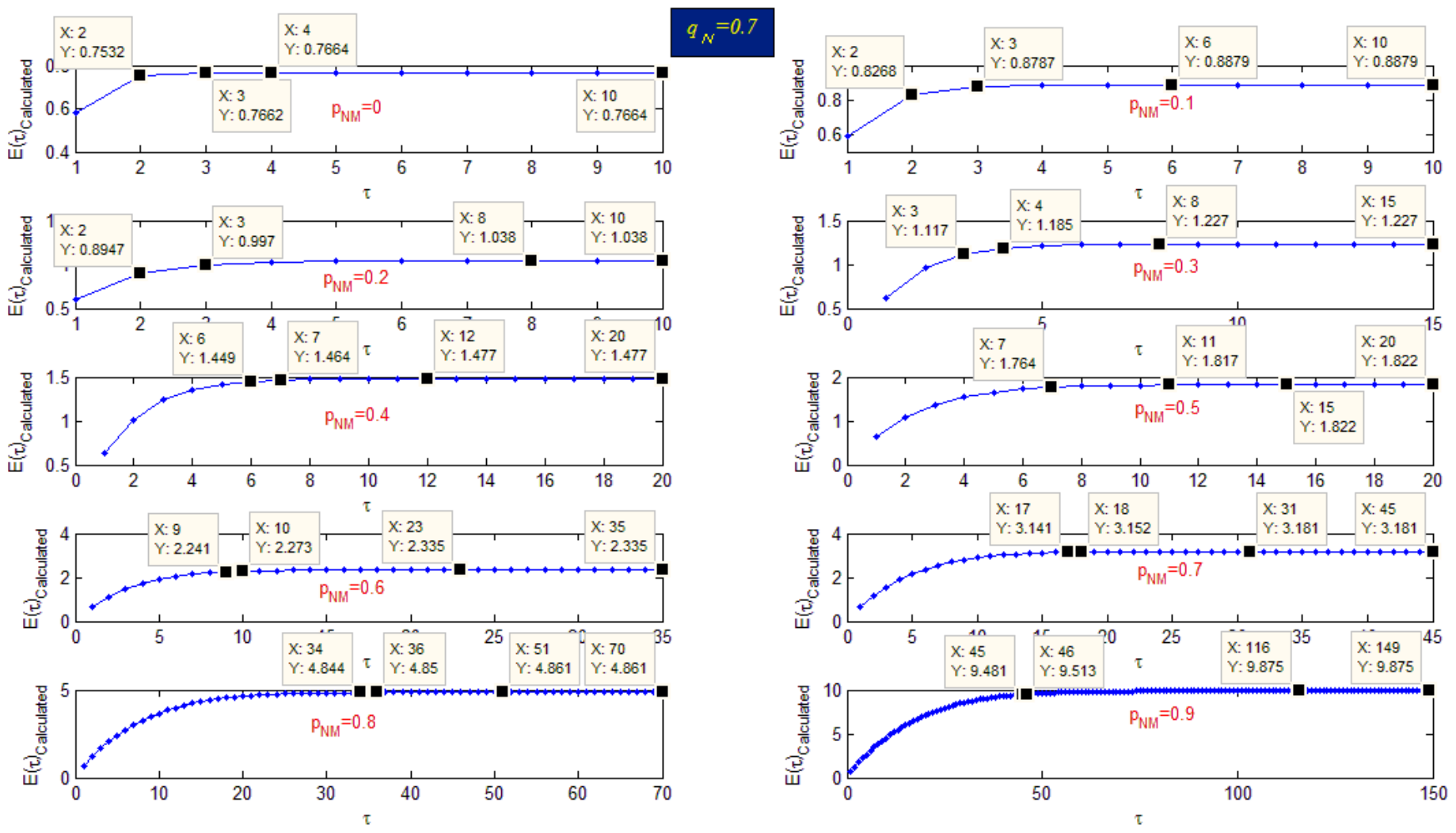

**Fig. 28.** Relationship between trancation of the transition probability matrix and average delay of the system when $q_N = 0.7$

In order to accelerate the computation, some approximations on the average delay of the system are made. **Table 3** and **Table 4** give the reasonable $\tau_p$ for each $p_{NM}$ when $q_M = 0.3$ and $q_N = 0.7$ respectively. Comparing these two tables, we know that for the same $p_{NM}$, the more the product are manufactured, the larger the average delay of the product is. For example, when $p_{NM} = 0.3$, the average delay for product $M$ is $E(\tau)_M = 1.137$ which is smaller than $E(\tau)_N = 1.227$. Comparing **Table 2**, **Table 3** and **Table 4**, we can find that for the same kind of metrology delay, each product in mixed product process enjoys less average delay than the single product process.

**Table 3:** The approximation of $E(\tau)_{Calculated}$ to $E(\tau)_{Approximated}$, $q_M = 0.3$

| $p_{NM}$ | $(\tau_{\min}, E(\tau))$ | $E(\tau)_{Approximated}$ | $(\tau_p, E(\tau)_{Calculated})$ |
|---|---|---|---|
| 0 | (3, 0.6956) | 0.70 | (3, 0.6956) |
| 0.1 | (6, 0.811) | 0.81 | (3, 0.8064) |
| 0.2 | (8, 0.9543) | 0.95 | (4, 0.95) |
| 0.3 | (8, 1.137) | 1.13 | (5, 1.125) |
| 0.4 | (11, 1.38) | 1.4 | (6, 1.356) |
| 0.5 | (15, 1.718) | 1.7 | (7, 1.666) |
| 0.6 | (23, 2.223) | 2.2 | (10, 2.165) |
| 0.7 | (33, 3.061) | 3.1 | (21, 3.05) |
| 0.8 | (49, 4.732) | 4.7 | (26, 4.657) |
| 0.9 | (112, 9.737) | 10 | (51, 9.504) |

**Table 4:** The approximation of $E(\tau)_{Calculated}$ to $E(\tau)_{Approximated}$, $q_N = 0.7$

| $p_{NM}$ | $(\tau_{\min}, E(\tau))$ | $E(\tau)_{Approximated}$ | $(\tau_p, E(\tau)_{Calculated})$ |
|---|---|---|---|
| 0 | (4, 0.7664) | 0.77 | (3, 0.7662) |
| 0.1 | (6, 0.8879) | 0.90 | (3, 0.8787) |
| 0.2 | (8, 1.038) | 1.0 | (3, 0.997) |
| 0.3 | (8, 1.227) | 1.2 | (4, 1.185) |
| 0.4 | (12, 1.477) | 1.5 | (7, 1.464) |
| 0.5 | (15, 1.822) | 1.8 | (11, 1.817) |
| 0.6 | (23, 2.335) | 2.3 | (10, 2.273) |
| 0.7 | (31, 3.181) | 3.2 | (18, 3.152) |
| 0.8 | (51, 4.861) | 4.9 | (36, 4.85) |
| 0.9 | (116, 9.875) | 10 | (46, 9.513) |

### 5.3.2. Stability Analysis for Mixed Product Process

We also assume that two kinds of products, product $M$ and product $N$, are manufactured on the same tool with the probabilities, $q_M = 0.3$ and $q_N = 0.7$. When $p_{NM} = 0$, then from **Table 3**, we know that the minimum $\tau$ for product $M$ should be 3 to obtain the limit of the average delay of the system, and that the correspondence transition probability matrix and average delay for system are

$$P_M = \begin{bmatrix} 0.3679 & 0.6321 & 0 & 0 \\ 0.3679 & 0.5328 & 0.0994 & 0 \\ 0.3679 & 0.5328 & 0.0892 & 0.0102 \\ 0.3682 & 0.5332 & 0.0892 & 0.0094 \end{bmatrix}$$

and $E(\tau)_M = 0.6956$. Form **Table 4**, we

can obtain $\tau=3$, $P_N=\begin{bmatrix} 0.3679 & 0.6321 & 0 & 0 \\ 0.3679 & 0.4290 & 0.2031 & 0 \\ 0.3679 & 0.4290 & 0.1577 & 0.0454 \\ 0.3708 & 0.4324 & 0.1589 & 0.0380 \end{bmatrix}$ and $E(\tau)_N=0.7662$ for product $N$. **Fig. 29** is the simulation results for both products. From **Fig. 29(a)** and **Fig. 29 (b)**, we know that for both products, the system with EWMA-I controller has a bigger stability region compared with the system with EWMA-II controller. From **Fig. 29(c)**, the system with EWMA-I controller, it is clear that the less manufactured product $M$ with a smaller average delay, has a bigger stability region compared with product $N$ which are more manufactured with a larger average delay. From **Fig. 29(d)**, the system with EWMA-II controller, we know that the more produced product, i.e., product $N$, has better stability region compared with product $M$, and this result is opposite to the result obtained from the system with EWMA-I controller. In addition, the result is different with what we have obtained in subsections 4.3.2 for the system with EWMA-II controller.

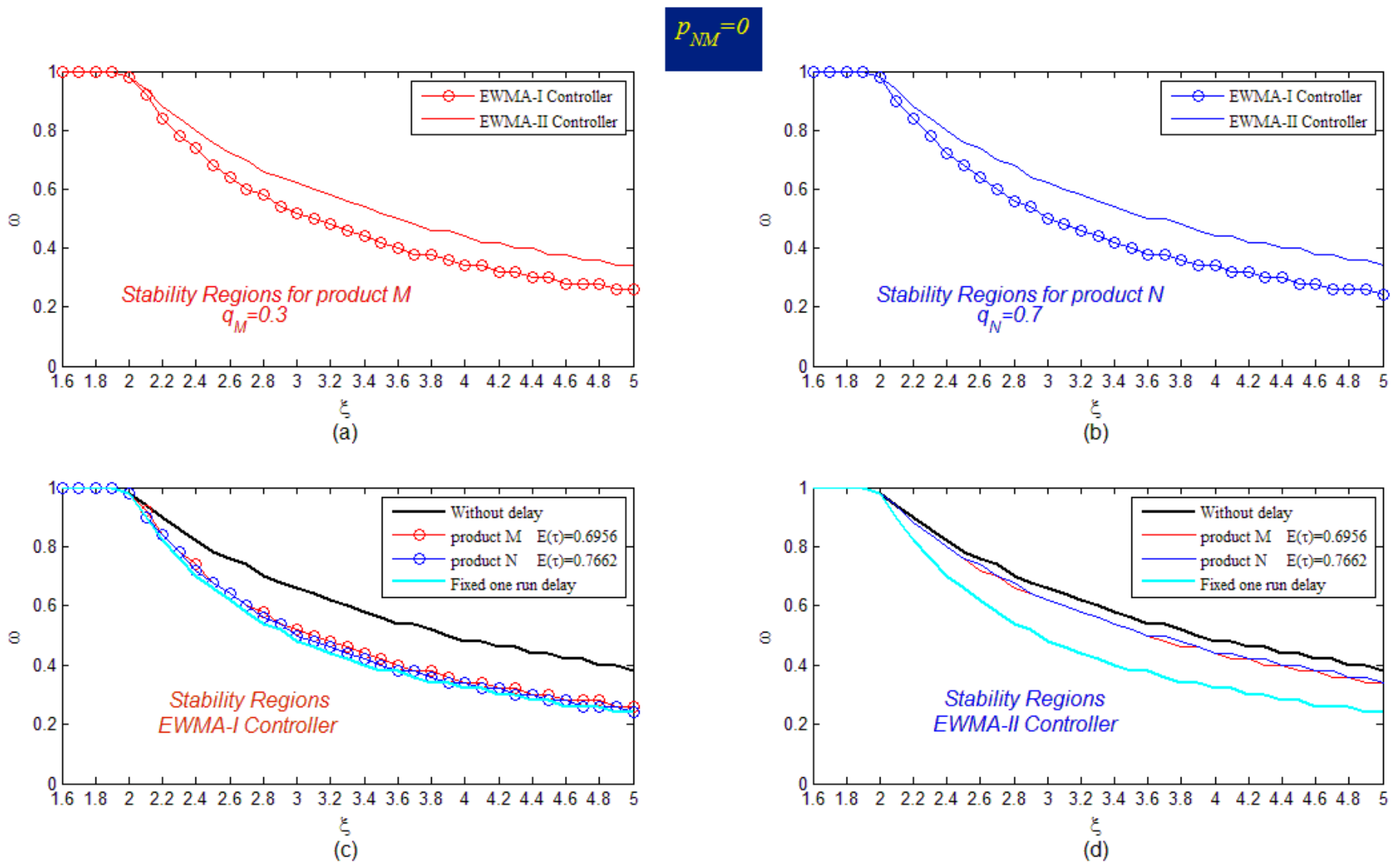

**Fig. 29.** Stability regions for both products when $p_{NM}=0$ ($q_M=0.3, q_N=0.7$)

When $P_{NM}=0.3$, from **Table 3** and **Table 4**, we know

$$P_M = \begin{bmatrix} 0.2575 & 0.7425 & 0 & 0 & 0 & 0 \\ 0.2575 & 0.3729 & 0.3695 & 0 & 0 & 0 \\ 0.2575 & 0.3729 & 0.0624 & 0.3071 & 0 & 0 \\ 0.2575 & 0.3729 & 0.0624 & 0.0066 & 0.3005 & 0 \\ 0.2575 & 0.3729 & 0.0624 & 0.0066 & 0.0005 & 0.3000 \\ 0.3679 & 0.5328 & 0.0892 & 0.0094 & 0.0007 & 0.0000 \end{bmatrix}$$

and $E(\tau)_M = 1.125$ for product $M$. And

$$P_N = \begin{bmatrix} 0.2575 & 0.7425 & 0 & 0 & 0 \\ 0.2575 & 0.3003 & 0.4422 & 0 & 0 \\ 0.2575 & 0.3003 & 0.1104 & 0.3318 & 0 \\ 0.2575 & 0.3003 & 0.1104 & 0.0264 & 0.3054 \\ 0.3683 & 0.4295 & 0.1578 & 0.0377 & 0.0067 \end{bmatrix}$$

and $E(\tau)_N = 1.185$ for product $N$. The simulation results for this case are shown in Fig. 30. From the figure, we know that the conclusions of the stability regions for both products are the same as what we concluded for $P_{NM} = 0$. Also from Fig. 30(d), it is noticed that the stability regions for both products with EWMA-II controller fall between the stability region of delay free system and that with fixed one run delay system despite that both $E(\tau)_M$ and $E(\tau)_N$ are greater than 1.

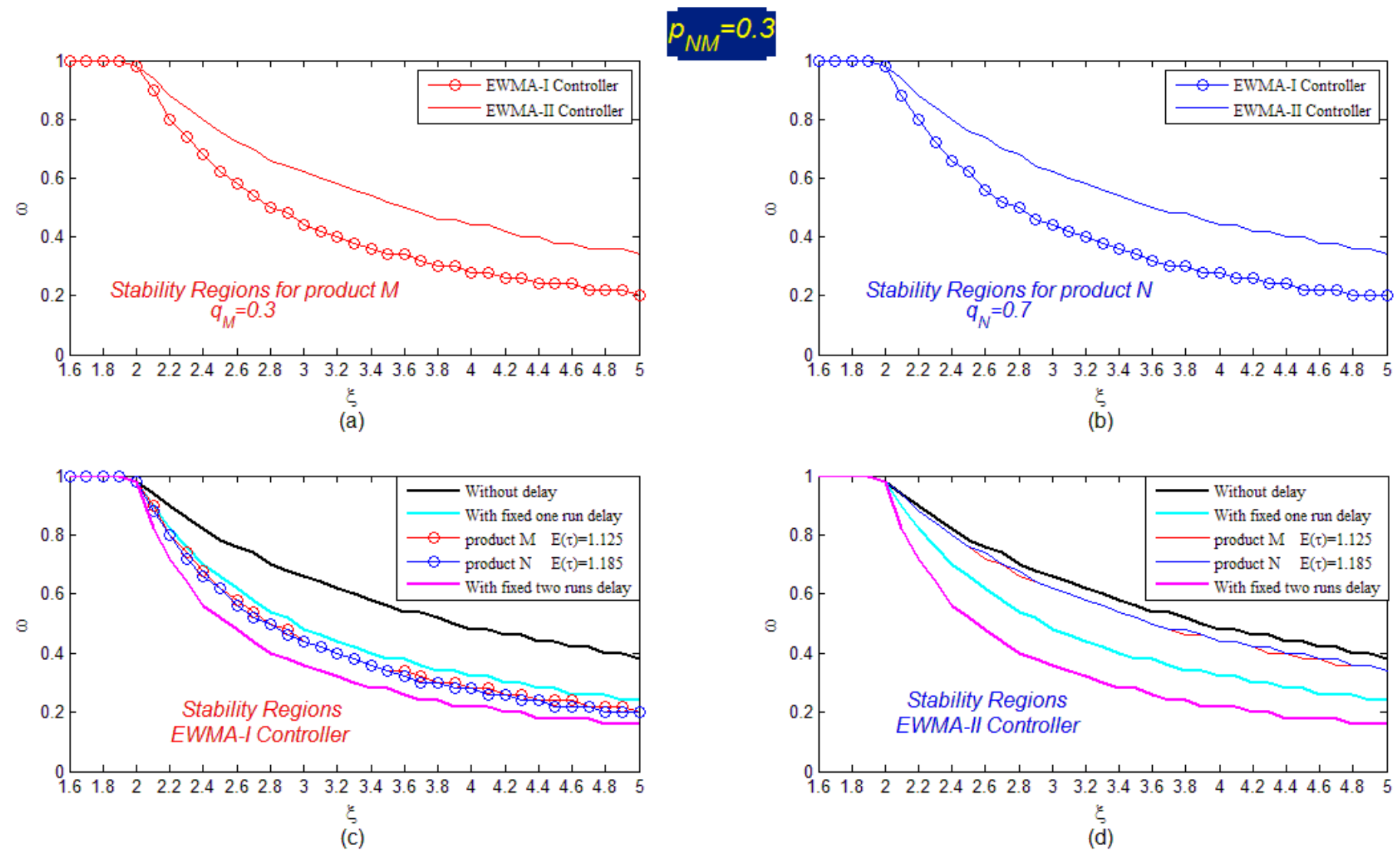

**Fig. 30.** Stability regions for both products when $p_{NM} = 0.3$ ($q_M = 0.3, q_N = 0.7$)

Detailed comparisons of the stabilities for both products under different $p_{NM}$

are shown in **Fig. 31**. From **Fig. 31(a)**, the system with EWMA-I controller, we conclude that with the increase of average delay, the stability regions decrease for both products; using $\text{ceil}(E(\tau))$ as the delay of the system will lead to conservative stability regions for each product. For the system with EWMA-II controller (see **Fig. 31(b)**), it is noticed that both $p_{NM}$ and the probabilities of the production of the products have minor effects on the stability of the system; using $\text{ceil}(E(\tau))$ as the delay of the system will lead to conservative stability regions for each product, especially for large $\text{ceil}(E(\tau))$. For the system of any kind of product with either controller is stable for any $\omega$ between 0 to 1 if $\xi < 2$.

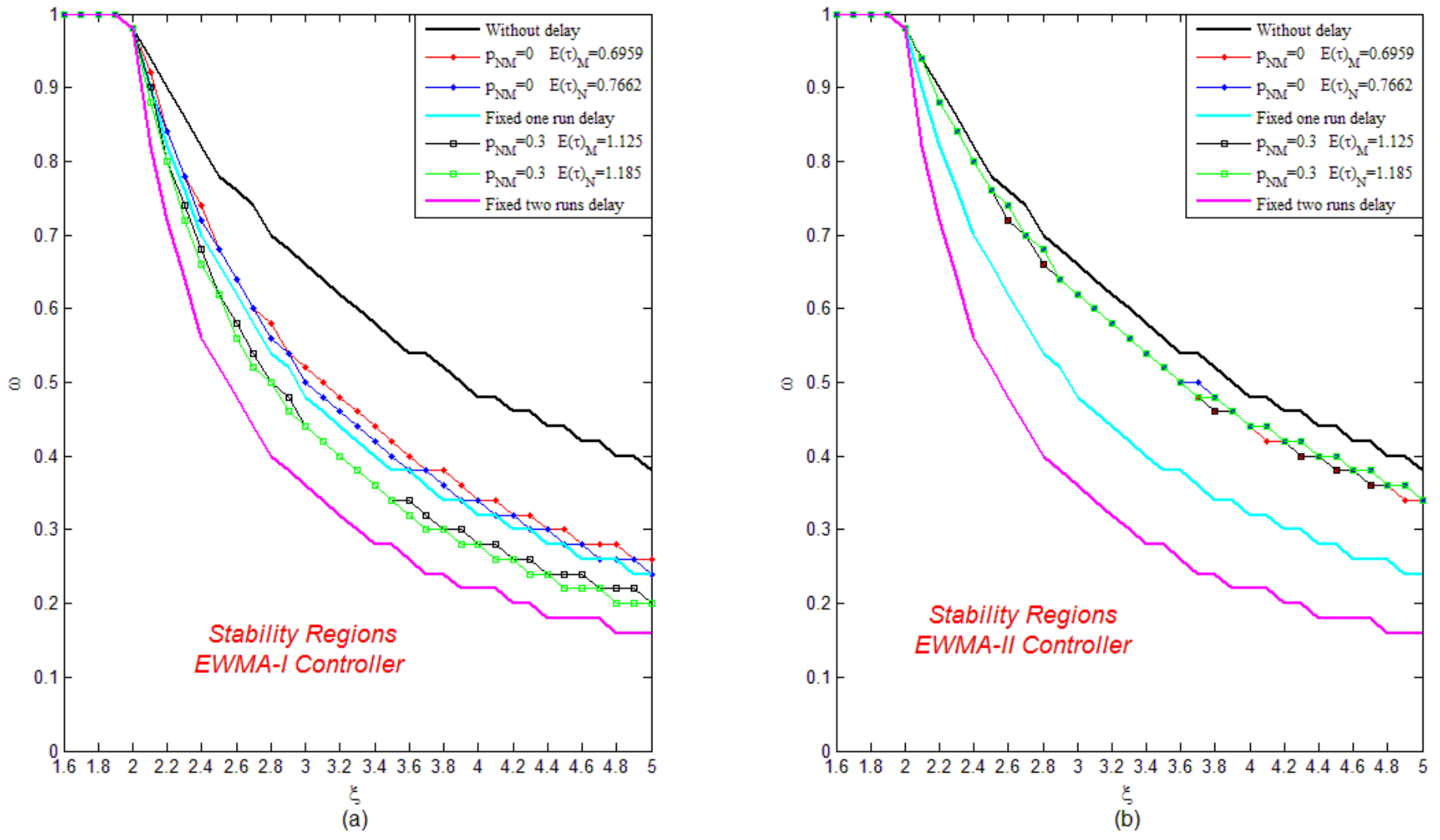

**Fig. 31.** Stability comparison for both products with different $p_{NM}$

In order to better illustrate the effect of the probability distributions of the manufacturing product on the stability of the system, we give another example which assumes that the probability of producing product $M$ is 0.05, i.e., $q_M = 0.05$, and product $N$ is 0.95, i.e., $q_N = 0.95$. $p_{NM}$ in this example is chosen as 0.3. Based on the analysis in subsection 5.3.1, we know the transition probability matrices for product $M$ and $N$ are

$$P_M = \begin{bmatrix} 0.2575 & 0.7425 & 0 & 0 & 0 & 0 \\ 0.2575 & 0.4298 & 0.3127 & 0 & 0 & 0 \\ 0.2575 & 0.4298 & 0.0124 & 0.3003 & 0 & 0 \\ 0.2575 & 0.4298 & 0.0124 & 0.0003 & 0.3000 & 0 \\ 0.2575 & 0.4298 & 0.0124 & 0.0003 & 0.0000 & 0.3000 \\ 0.3679 & 0.6141 & 0.0177 & 0.0003 & 0.0000 & 0.0000 \end{bmatrix}$$

and

$$P_N = \begin{bmatrix} 0.2575 & 0.7425 & 0 & 0 & 0 & 0 \\ 0.2575 & 0.2641 & 0.4784 & 0 & 0 & 0 \\ 0.2575 & 0.2641 & 0.1265 & 0.3519 & 0 & 0 \\ 0.2575 & 0.2641 & 0.1265 & 0.0402 & 0.3117 & 0 \\ 0.2575 & 0.2641 & 0.1265 & 0.0402 & 0.0096 & 0.3021 \\ 0.3681 & 0.3774 & 0.1808 & 0.0575 & 0.0137 & 0.0025 \end{bmatrix}$$

and the corresponding average delays are $E(\tau)_M = 1.1$ and $E(\tau)_N = 1.3$. For product $M$, the average delay of the system is less than that when $q_M = 0.3$; while for product $N$, the average delay of the system is larger compared with that when $q_N = 0.7$. Fig. 32 is the stability regions for the systems of both products under different controllers. From the figure, we have the same conclusions on the stability regions as we have previously obtained for $P_{NM} = 0$ and $P_{NM} = 0.3$, under the situations that $q_M = 0.3$ and $q_N = 0.7$.

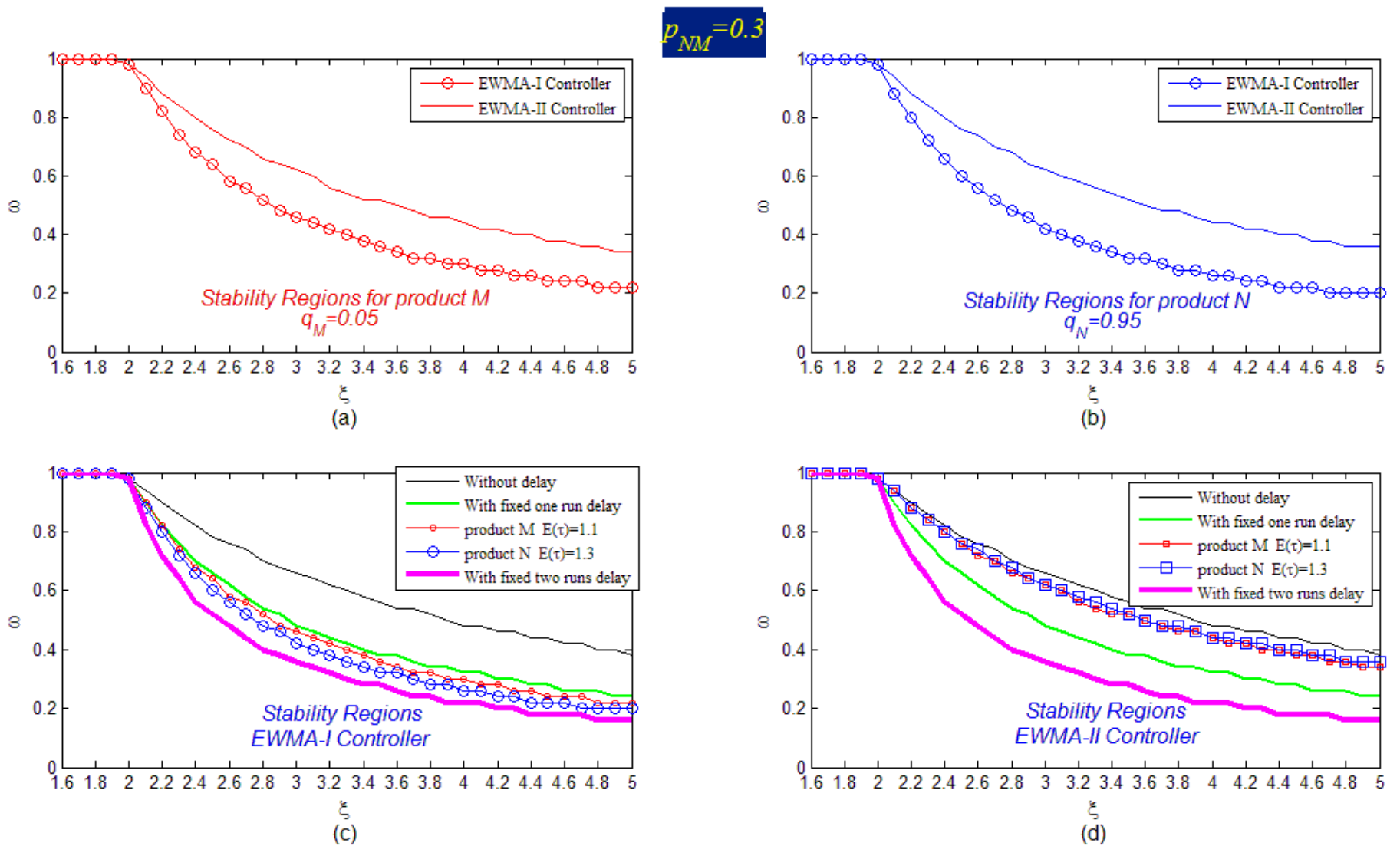

**Fig. 32**. Stability regions for both products when $q_M = 0.05$ and $q_N = 0.95$ ($p_{NM} = 0.3$)

Based on the results obtained in **Fig. 30** and **Fig. 32**, we can draw another figure, **Fig. 33**, which compares the stability regions under the different production probabilities (we denote it as $q$). Form the figure, we notice that with the increase of $q$, the stability region of the system with EWMA-I controller is decreasing. While for the system with EWMA-II controller, the increase of $q$, has minor effects on the stability of the system. In other words, under the same metrology delay, if the EWMA-I controller is adopted, then the frequently manufactured product has a smaller stability region compared with the infrequently produced product; while if the EWMA-II controller is adopted, then $q$ has little influence on the stability of the system.

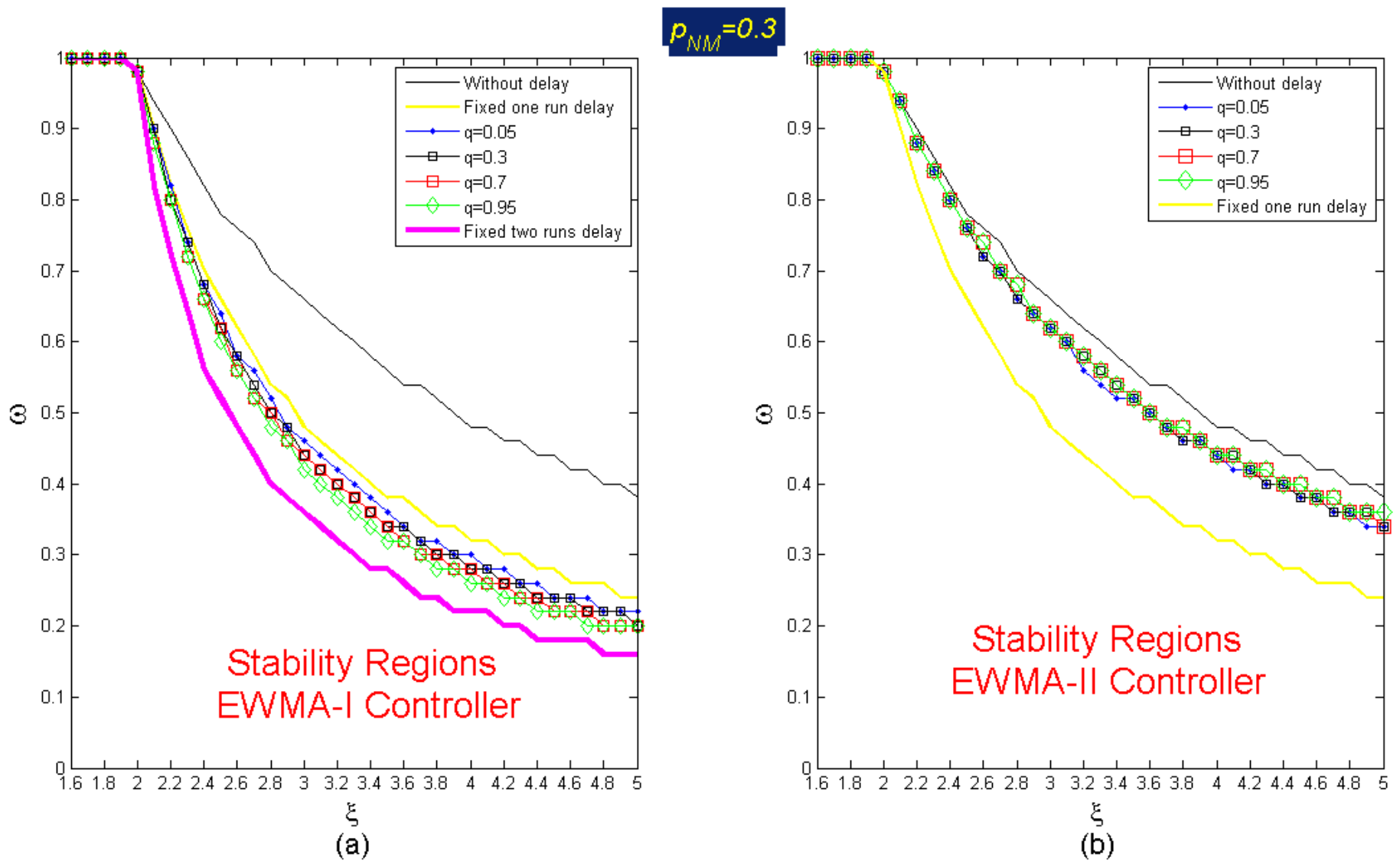

**Fig. 33.** Stability comparison for product under different production probabilities when $p_{NM} = 0.3$

***Remark 15.*** The same as what we have discussed in subsection 4.3.1, we can still affirm that in the mixed product process, if the metrology delays follow Poisson distribution, then the distributions of the production of the products have minor effects on the stability of the system of any product with EWMA-II controller; also the stability regions for system with EWMA-II controller of any kind of products lie between those of the system with fixed one run delay and those of the system without metrology delay.

## 6. Conclusion

In this paper, we proposed two kinds of EWMA controllers, i.e., EWMA-I and EWMA-II controllers for single product process and mixed product process in semiconductor manufacturing.

In the single product process, the stability problems for systems with both controllers subject to different kinds of metrology delays are studied; Routh-Hurwitz criterion as well as Lyapunov's direct method are used for obtaining the stability regions of the system with fixed metrology delay; Lyapunov's direct method is adopted to derive the sufficient and necessary conditions of the stochastic stability for the system with fixed sampling metrology and for the system subject to stochastic metrology delay. From the results of numerical computation we have the following conclusions for each kind of metrology delay in single product process:

(A) For the fixed metrology delay:

a. EWMA-I controller is equivalent to EWMA-II controller;

b. Routh-Hurwitz criterion and Lyapunov's direct method is equivalent in getting the stability regions for different fixed runs delay;

c. With the increase of metrology delay, the size of stability region will decrease.

(B) For the fixed sampling metrology:

a. For the system with EWMA-I controller, with the increase of sampling interval, the stability region is decreasing; it's wrong to use $\text{ceil}(E(\tau))$ as the delay of the system to obtain the stability region.

b. For the system with EWMA-II controller, we have less, but the latest data of the system; the stability region is not affected by the sampling interval, hence the stability of the system is the best.

(C) For the stochastic metrology delay:

a. For the system with EWMA-I controller, with the increase of average delay, the stability region will decrease; it is sometimes wrong to use $\text{ceil}(E(\tau))$ as the delay to get the stability region.

b. For the system with EWMA-II controller, the increase of average delay of the system sometimes has little influences on the stability of the system; using $\text{ceil}(E(\tau))$ as the delay of the system will lead to conservative stability regions.

(D) For all kinds of metrology delay, we have the following common conclusions:

a. When the estimated process gain is greater than half of the true process gain, the system is guaranteed closed-loop stable for any discount factor $\omega$ between 0 and 1.

b. For the same kind of metrology delay, the system with EWMA-II controller always has a better stability region compared with the stability region of the system with EWMA-I controller.

We also extended the theorems obtained for single product process to mixed product process. And we have the following results:

(A) The more the products are manufactured, i.e., the frequently run product, the larger the average delay of the system with this kind of product is. The less the products are produced, i.e., the infrequently run product, the smaller the average delay of the system with this kind of product is. In fact, if the metrology delays are the same for frequently and infrequently run products, then the infrequently run products are more likely to receive latest outputs of their previous runs compared with the frequently run products, because of that between the adjacent runs of the same kind of infrequently run products, there will be lots of runs to manufacture other kinds of products, during which this kind of infrequently run products is likely to complete their measurements; therefore the infrequently run products will enjoy a smaller average delay. The situation is the opposite for the frequently run products.

(B) If the measurement delay follows Poisson distribution, for the system with EWMA-I controller, the frequently run product has a worse stability region than the infrequently run product. However, for the system with EWMA-II controller, the distributions of the production of the products have minor effects on the stability of the system. Using $\text{ceil}(E(\tau))$ as the delay of the system will lead to conservative stability regions for the system of any kind of product, and it is much more conservative for the system with EWMA-II controller.

(C) The system of any kind of product with EWMA-II controller always has a larger stability region than the system with EWMA-I controller if the measurement delays are the same for both controllers.

(D) For the system of any kind of product with either controller, when the estimated process gain is greater than half of the true process gain, the system is guaranteed closed-loop stable for any discount factor $\omega$ between 0 to 1.